\documentclass[12pt,eqsecnum]{report}
\usepackage{cite}
\usepackage{amsfonts}
\usepackage{epsfig}
\usepackage{setspace}
\usepackage{amssymb}
\usepackage{amsmath}
\usepackage{subfigure}
\usepackage{hyperref}
\usepackage{color}
\usepackage{longtable}
\usepackage{caption}

\newcommand{\psip}{{\psi^\prime}}
\newcommand{\psid}{\dot{\psi}}

\newcommand{\Pis}{P_\psi}
\newcommand{\sub}{\left( \frac{N_r}{N} \right)}

\newcommand{\mn}{{\mu\nu}}
\newcommand{\R}{\mathcal{R}}

\newcommand{\g}{g_{(2)}}
\newcommand{\ma}[1]{\mbox{$\mathcal{#1}$}}

\newcommand{\D}{{\rm d}}

\newcommand{\dalm}{\kern1pt\vbox{\hrule height 0.9pt\hbox{\vrule width
0.9pt\hskip 2.5pt\vbox{\vskip 5.5pt}\hskip 3pt\vrule width 0.3pt}\hrule height
0.3pt}\kern1pt}

\def\b2hat{ {\hat b}_2 }

\def\none{\nonumber\\}

\def\be {\begin{equation}}
\def\ee  {\end{equation}}
\def\bea {\begin{eqnarray}}
\def\eea {\end{eqnarray}}
\def\nn {\nonumber}

\begin{document}
\newcounter{tabctr}
\pagenumbering{roman}


\begin{titlepage}
\begin{center}
\begin{Large}
Black Hole Formation in Lovelock Gravity \\
\end{Large}
\vspace*{6mm}
by \\
\vspace*{6mm}
\begin{Large}
Timothy Mark Taves \\
\end{Large}
\vspace*{30mm} 
A Thesis submitted to the Faculty of Graduate Studies of \\
\vspace*{3mm}
The University of Manitoba \\
\vspace*{3mm}
in partial fulfilment of the requirements of the degree of \\
\vspace*{18mm}
DOCTOR OF PHILOSOPHY \\
\vspace*{24mm}
\begin{large}
Department of Physics and Astronomy \\
\vspace*{3mm}
University of Manitoba \\
\vspace*{3mm}
Winnipeg\\
\end{large}
\vspace*{24mm}
Copyright \copyright\ 2013 by Timothy Mark Taves \\
\end{center}
\end{titlepage}

\pagestyle{headings}
\setcounter{enumiv}{1}


\addcontentsline{toc}{chapter}{\bf Abstract}

\begin{abstract}
\thispagestyle{plain}
\setcounter{page}{1}


Some branches of quantum gravity demand the existence of higher dimensions and the addition of higher curvature terms to the gravitational Lagrangian in the form of the Lovelock polynomials.  In this thesis we investigate some of the classical properties of Lovelock gravity.

We first derive the Hamiltonian for Lovelock gravity and find that it takes the same form as in general relativity when written in terms of the Misner-Sharp mass function.  We then minimally couple the action to matter fields to find Hamilton's equations of motion.  These are gauge fixed to be in the Painlev\'{e}-Gullstrand co--ordinates and are well suited to numerical studies of black hole formation.

We then use these equations of motion for the massless scalar field to study the formation of general relativistic black holes in four to eight dimensions and Einstein-Gauss-Bonnet black holes in five and six dimensions.  We study Choptuik scaling, a phenomenon which relates the initial conditions of a matter distribution to the final observables of small black holes.  

In both higher dimensional general relativity and Einstein-Gauss-Bonnet gravity we confirm the existence of cusps in the mass scaling relation which had previously only been observed in four dimensional general relativity.  In the general relativistic case we then calculate the critical exponents for four to eight dimensions and find agreement with previous calculations by Bland {\it et al} but not Sorkin {\it et al} who both worked in null co--ordinates. 

For the Einstein-Gauss-Bonnet case we find that the self-similar behaviour seen in the general relativistic case is destroyed.  We find that it is replaced by some other form of scaling structure.  In five dimensions we find that the period of the critical solution at the origin is proportional to roughly the cube root of the Gauss-Bonnet parameter and that there is evidence for a minimum black hole radius.  In six dimensions we see evidence for a new type of scaling.  We also show, from the equations of motion, that there is reason to expect qualitative differences between five and higher dimensions.

\end{abstract}


\newpage
\setcounter{page}{2} \addtocontents{toc}{\protect\vspace{-0.167in}}
\addcontentsline{toc}{chapter}{\bf Dedication}
\chapter*{Dedication}

\vspace{2in}

\begin{center}
For my brother, Greg.
\end{center}


\newpage
\addtocontents{toc}{\protect\vspace{-0.167in}}
\addcontentsline{toc}{chapter}{\bf Acknowledgements}
\chapter*{Acknowledgements}
This work would not have been possible if it were not for the hard work, dedication and support of my colleagues, family and friends.  In particular I would like to thank:

My supervisor and collaborator, Gabor Kunstatter.  He was always passionate about this work and dedicated to my education.  A graduate student could not ask for a better supervisor.

My other collaborators, Hideki Maeda, Robb Mann, Danielle Leonard, Nils Deppe and Jack Gegenberg for all of their hard work, enthusiasm and encouragement as well as Jonathan Ziprick for enlightening conversations.

My advisory and examining committees consisting of Tom Osborn, Eric Schippers and my external examiner, David Garfinkle for the helpful advice and taking the time to read my thesis.

My previous supervisors, Giles Santyr, Scott King and Dwight Vincent for their dedication to my education and research.

My wife, Laryssa, for being so supportive of my crazy dream of thinking for a living.  I particularly appreciate her patience with me working on evenings and weekends all of the time.  I could not have done this with some one less understanding.

My family, in particular my mother, Barb and my two brothers, Jeremy and Greg for being so supportive even though they had no idea what I was doing.

All of my friends but in particular Jon Wheelwright for running code for me and Steph Spence for making me take coffee breaks at the University of Winnipeg.  I would also like to thank Bill Todd, Alex Mirza and Kelly Dueck for their enthusiasm and support.

The University of Manitoba and NSERC for funding.


\newpage
\addtocontents{toc}{\protect\vspace{-0.167in}}
\addcontentsline{toc}{chapter}{\bf Contents}
 \tableofcontents

\newpage
\addtocontents{toc}{\protect\vspace{-0.167in}}
\addcontentsline{toc}{chapter}{\bf List of Tables}
\listoftables

\newpage
\addtocontents{toc}{\protect\vspace{-0.167in}}
\addcontentsline{toc}{chapter}{\bf List of Figures}
\begin{listoffigures}
\end{listoffigures}

\newpage

\thispagestyle{plain}
\addtocontents{toc}{\protect\vspace{-0.167in}}
\addcontentsline{toc}{chapter}{\bf List of Copyrighted Material for Which Permission Was Obtained}

\phantom{1}

\vspace{0.85in}

\noindent {\huge {\bf List of Copyrighted Material for Which Permission Was Obtained}}

\vspace{0.45in}

Chapters \ref{ch:LLADM}, \ref{ch:HDCS} and \ref{ch:CSEGBG} of this thesis are based on papers \cite{Kunstatter2013}, \cite{Taves2011} and \cite{Deppe2012} which were published either in Classical and Quantum Gravity or Physical Review.  Copyright permission was obtained to include the articles in my thesis.


\chapter{Introduction \& Background}
\pagenumbering{arabic} \setcounter{page}{1} \label{ch:Introduction}

The search to understand the role of quantum mechanics in gravity has lead to theories such as string theory and loop quantum gravity.  These theories make predictions on the Planck scale ($\approx 10^{19}GeV$) which has made direct experimental verification difficult.  This problem requires us to find alternate ways to verify candidate theories in both the theoretical and experimental realms.  To this end black holes have become a useful physical system.  Einstein's classical theory of gravity, general relativity, predicts that there is a singularity of infinite mass density at the center of a black hole, a problem which must be solved by any potentially acceptable theory of quantum gravity.  It has also been speculated that microscopic black holes may be created at the CERN collider when it is fully operational.  According to some higher dimensional theories of gravity these black holes could be created with energies as low as $1TeV$ and could therefore be used to experimentally verify predictions of quantum gravity theories.  For these reasons a deep understanding of black holes, especially microscopic black holes, may be important to gain insight into the most fundamental laws of physics.

Some quantum gravity theories, such as string theory, predict the existence of more than four dimensions, all but four of which are compactified and can only be directly observed on the Planck scale.  In certain situations string theory also predicts the existence of higher order space-time curvature terms in the gravitational Lagrangian in combinations known as the Lovelock polynomials \cite{Lovelock1970, Lovelock1971}.  The resulting theory, known as Lovelock gravity, is a natural generalization of Einstein's general relativity; it gives equations of motion with no more than second derivatives of the metric and it is also free of ghosts (imaginary mass) when linearized around a flat background.  

This thesis will not be directly concerned with string theory or any other quantum gravity theories.  Instead, we investigate the effects of higher dimensions and higher curvature corrections, in the form of Lovelock gravity, on the formation of black holes, particularly microscopic black holes.  We also concentrate on the Hamiltonian formulation for our analysis which has the potential to be useful in the study of quantum gravity in the future.  The first goal of this work is to calculate the Hamiltonian for Lovelock gravity in terms of a mass function.  This Hamiltonian is then coupled to matter and used to find the equations of motion which can be used to model the collapse of matter coupled to Lovelock gravity.  We then numerically simulate the formation of microscopic black holes.  The goal is to investigate the effects of higher dimensions and Lovelock curvature terms on Choptuik scaling, a phenomena which relates the initial conditions of a collapsing, microscopic black hole system to its final observables. 

The rest of this chapter is dedicated to a review of relevant topics.  We start with Lovelock gravity and the Painlev\'{e}-Gullstrand co-ordinate system which is used extensively in this thesis.  All work done in this thesis assumes spherical symmetry and uses Hamiltonian mechanics.  We therefore review the well known Hamiltonian formulation of spherically symmetric general relativity which was first discussed by Arnowitt, Deser and Misner (ADM) in 1962 \cite{Arnowitt2004}, although we will find it necessary to use Kucha\^r' geometrodynamics \cite{Kuchar1994} which involves treating the mass function as a canonical variable.  We then discuss the phenomena of Choptuik scaling in general relativity.  In Chapter \ref{ch:LLADM} we perform the Hamiltonian analysis of Lovelock gravity.  We find Hamilton's equations of motion for Lovelock gravity coupled to a scalar field (which are used for all subsequent, numerical calculations) and Lovelock gravity coupled to a charged scalar field and an electromagnetic field.  In Chapters \ref{ch:HDCS} and \ref{ch:CSEGBG} we use the equations of motion found in Chapter \ref{ch:LLADM} to numerically simulate black hole formation and Choptuik scaling.  In Chapter \ref{ch:HDCS} we concentrate on higher dimensional general relativity while in Chapter \ref{ch:CSEGBG} we look at the effects of the addition of the Gauss-Bonnet term (second order Lovelock term) into the action.  In Chapter \ref{ch:Conclusions} we conclude.

\section{Lovelock Gravity}
\label{sec:LLGravity}

The action of general relativity is given by

\be
I = I_{GR} + I_{matter},
\ee
where $I_{GR}$ is the Einstein-Hilbert action and $I_{matter}$ is the matter action.  The Einstein-Hilbert action in $n$ dimensions is given by

\begin{align}
\label{actiongr}
I_{GR}=&\frac{1}{2\kappa_n^2}\int \D ^nx\sqrt{-g}{\cal R},
\end{align}
where we work in units where only, $G_n$, the $n$ dimensional gravitational constant, is retained, $\kappa_n := \sqrt{8\pi G_n}$, $g$ is the determinant of the metric (we use the convention that the Minkowski metric is taken as diag$(-,+,\cdots,+)$) and ${\cal R}$ is the Ricci scalar.  This action, when coupled to matter, locally conserves energy, ie $\nabla_\mu T^\mu_{\phantom{\mu} \nu}=0$ where $T^\mu_{\phantom{\mu} \nu}$ is the energy momentum tensor derived by varying $I_{matter}$.  Also, the action is ghost free (no imaginary mass) when linearized around a flat background.  It also gives equations of motion which have at most second time derivatives of the metric.  Theories with more than second time derivatives of the metric in the equations of motion often require the use of auxiliary fields in order to apply the variational principal to find the Hamiltonian.  This introduces new degrees of freedom whose physical meaning may be difficult to interpret.  For these reasons it is desirable to work with a theory which has no more than second derivatives of the metric in the equations of motion.  When working in four dimensions the only combination of curvature terms which satisfies these conditions and contributes to the equations of motion is the Einstein-Hilbert Lagrange density.  When considering higher dimensions, however, there exists higher curvature terms, known as the Lovelock polynomials which also have these properties.  To this end consider the Lovelock action \cite{Lovelock1970,Lovelock1971} given by

\begin{align}
\label{action2}
I_{LL}=&\frac{1}{2\kappa_n^2}\int \D ^nx\sqrt{-g}\sum_{p=0}^{[n/2]}\alpha_{(p)}{\ma L}_{(p)},\\
{\ma L}_{(p)}:=&\frac{1}{2^p}\delta^{\mu_1\cdots \mu_p\nu_1\cdots \nu_p}_{\rho_1\cdots \rho_p\sigma_1\cdots \sigma_p}{\cal R}_{\mu_1\nu_1}^{\phantom{\mu_1}\phantom{\nu_1}\rho_1\sigma_1}\cdots {\cal R}_{\mu_p\nu_p}^{\phantom{\mu_p}\phantom{\nu_p}\rho_p\sigma_p},
\end{align}
where $\delta^{\mu_1...\mu_p}_{\rho_1...\rho_p}:=\delta^{\mu_1}_{[\rho_1}...\delta^{\mu_p}_{\rho_p]}\,$, ${\cal R}_{\mu_p\nu_p}^{\phantom{\mu_p}\phantom{\nu_p}\rho_p\sigma_p}$ is the Reimann curvature tensor (which obeys the conventions $[\nabla _\rho ,\nabla_\sigma]V^\mu ={{\cal R}^\mu }_{\nu\rho\sigma}V^\nu$ and ${\cal R}_{\mu \nu }={{\cal R}^\rho }_{\mu \rho \nu }$ for a given vector, $V^\mu$), $\alpha_{(p)}$ are coupling constants of dimension (length) ${}^{2(p-1)}$ and $[n/2]$ is the biggest integer less than $n/2$.  The zero$^{th}$ and first terms in Equation \ref{action2} correspond to the cosmological constant term and the Einstein-Hilbert term respectively.  The gravitational equations following from this action are given by 
\begin{align} 
{\ma G}_{\mu\nu}=\kappa_n^2 {T}_{\mu\nu}, \label{beqL}
\end{align} 
where ${T}_{\mu \nu}$ is the energy-momentum tensor for matter fields obtained from $I_{\rm matter}$ and
\begin{align} 
{\ma G}_{\mu\nu} :=& \sum_{p=0}^{[n/2]}\alpha_{{(p)}}{G}^{(p)}_{\mu\nu}, \label{generalG}\\
{G}^{\mu(p)}_{~~\nu}:=& -\frac{1}{2^{p+1}}\delta^{\mu\eta_1\cdots \eta_p\zeta_1\cdots \zeta_p}_{\nu\rho_1\cdots \rho_p\sigma_1\cdots \sigma_p}{\cal R}_{\eta_1\zeta_1}^{\phantom{\eta_1}\phantom{\zeta_1}\rho_1\sigma_1}\cdots {\cal R}_{\eta_p\zeta_p}^{\phantom{\eta_p}\phantom{\zeta_p}\rho_p\sigma_p}.
\label{LLTensor}
\end{align} 
The tensor ${G}^{(p)}_{\mu\nu}$, obtained from ${\ma L}_{(p)}$, contains up to the second derivatives of the metric and ${G}^{(p)}_{\mu\nu}\equiv 0$ is satisfied for $p\ge [(n+1)/2]$.

The action of Equation \ref{action2} combines curvature terms in such a way that it gives equations of motion which have no more than second derivatives of the metric and it is ghost free when linearized around a flat background \cite{Boulware1985}.  In addition to this it has been shown that, in the low energy limit, string theory calls for the addition of the higher order Lovelock terms to the Einstein-Hilbert term ($p=1$) in the action \cite{Gross1986,Gross1987,Metsaev1987,Metsaev1987a,Zweiebach1985}.


In this thesis we are interested in spherically symmetric systems and so we consider the $n(\ge 4)$-dimensional general metric 
\begin{eqnarray}
g_{\mu\nu}(x)dx^\mu dx^\nu=g_{AB}({\bar y})d{\bar y}^A d{\bar y}^B+R({\bar y})^2\gamma_{ab}(z)dz^adz^b,
\label{eq:structure}
\end{eqnarray}
where $g_{AB}$ is a function of one time and one space co-ordinate and $R({\bar y})$ is a scalar function.  $\gamma_{ab}$ is an $(n-2)$-dimensional metric with sectional curvature, $k=1,0,-1$.  We introduce the covariant derivatives compatible with the total metric, the 2D and the $(n-2)$D metrics as
\begin{eqnarray}
\nabla_\rho g_{\mu\nu}=0,\qquad D_F g_{AB}=0,\qquad {\bar D}_fg_{ab}=0.
\end{eqnarray}

The most general energy-momentum tensor $T_{\mu\nu}$ compatible with this spacetime symmetry governed by the Lovelock equations is given by
\begin{align}
T_{\mu\nu}\D x^\mu \D x^\nu =T_{AB}({\bar y})\D {\bar y}^A\D {\bar y}^B+p({\bar y})R^2 \gamma_{ab}\D z^a\D z^b,
\end{align}  
where $T_{AB}({\bar y})$ and $p({\bar y})$ are a symmetric two-tensor and a scalar respectively.

Lovelock gravity has been shown to obey a generalized Birkhoff theory 
\cite{Wiltshire1986,Deser2005,Zegers2005,Maeda2011}.  This allows us to write down the Hamiltonian in terms of a mass distribution in the spherically symmetric case.  This mass function has been calculated by \cite{Deruelle2004,Maeda2008,Maeda2011}.  It is a generalization of the Misner-Sharp mass function \cite{Misner1964} from general relativity and it reduces to the ADM (Arnowitt-Deser-Misner) mass \cite{Arnowitt2004} at spatial infinity in the asymptotically flat spacetime.  It is given by
\begin{align}
M :=& \frac{(n-2)V_{n-2}^{(k)}}{2\kappa_n^2}\sum_{p=0}^{[n/2]}{\tilde \alpha}_{(p)}R^{n-1-2p}[k-(DR)^2]^p,\label{qlm-L}\\
{\tilde \alpha}_{(p)}:=&\frac{(n-3)!\alpha_{(p)}}{(n-1-2p)!}, \label{alphatil}
\end{align}  
where $(DR)^2:=(D_A R)(D^A R)$ \cite{Maeda2011} and $k = 1,0,-1$ corresponds to $\gamma_{ab}$ being spherical (compact), flat or hyperbolic.
The constant $V_{n-2}^{(k)}$ represents the volume of an $n-2$ dimensional sphere in the $k=1$ case.  


We will review the well known Hamiltonian analysis in terms of this mass function for the general relativity case in Section \ref{Canonical Formalism in General Relativity1} and perform it for the Lovelock case in Chapter \ref{ch:LLADM}.  We note that the Hamiltonian analysis for full Lovelock gravity was first considered by Teitelboim and Zanelli \cite{Teitelboim1987}. Their result was rather formal in that an explicit parametrization of the phase space was not provided. (See also  \cite{Deser2012,Torii2008}.) For the case of spherical symmetry, the Hamiltonian analysis of five-dimensional Einstein-Gauss-Bonnet (i.e. quadratic Lovelock) gravity was worked out by Louko {\it et al} \cite{Louko1997} (using the methodology of Kucha\v{r}' geometrodynamics \cite{Kuchar1994}), while the Hamiltonian analysis of higher-dimensional Gauss-Bonnet gravity coupled to matter was recently done in our paper, \cite{Taves2012}.  In this thesis our analysis is done for generic Lovelock gravity in arbitrary dimensions.

\section{Painlev\'{e}-Gullstrand Co-ordinates}

In this thesis, when we must choose a co-ordinate system, we often use co-ordinates which are regular over the future horizon and are intrinsically flat on constant time slices.  They are analogous to the Painlev\'{e}-Gullstrand (PG) co-ordinates in general relativity and so we devote this section to a review of PG co-ordinates in 4D, static general relativity.  See \cite{Martel2001} and \cite{Lake1994} for more details.  Note that later on in this thesis we will refer to flat slice co-ordinates as PG co-ordinates even in the non-static, Lovelock case.

We start with the Schwarzschild metric given by 

\begin{align}
\label{eq:sch}
ds^2 =& -\left( 1-\frac{2GM}{R} \right)  dT^2 + \left( 1-\frac{2GM}{R} \right)^{-1} dR^2 + R^2d\Omega_2,
\end{align}
where $G$ is Newton's gravitational constant and $T$ is the Schwarzschild time co-ordinate.  Here, since we are concerned only with the static case, $M$ is the mass centered at the spatial origin.  Later, when we are interested in the dynamical case, $M$ will be used to represent the generalized Misner-Sharpe mass function \cite{Deruelle2004,Maeda2008,Maeda2011,Misner1964}.  $d\Omega_{2}$ is the line element of a unit 2-sphere, equal to $d\theta^2+sin^2\theta d\phi^2$.  When we work in higher dimensions we generalize to $d\Omega_{n-2}$, the line element of a unit $n-2$-sphere.

The goal is to find a co-ordinate transformation which gives a metric that is regular at the horizon, $R=2GM$.  There are many ways to do this.  We transform to the co-ordinates of a radially in-falling, massive observer.  This observer would not notice anything special when crossing the horizon.  For two events occurring along a time-like geodesic this system's time co-ordinate separation is equal to the proper time separation.

To find this transformation we start by applying the geodesic equation,

\be
\frac{\partial K}{\partial x^\alpha} - \frac{d}{d\tau} \frac{\partial K}{\partial \dot{x}^\alpha} = 0,
\ee
where $2K=g_{\alpha \beta} \dot{x}^\alpha \dot{x}^\beta$, to the metric of Equation \ref{eq:sch} to obtain the equations of motion for an ingoing, radial, massive particle,

\be
\dot{T}=\frac{E}{1-2GM/R}, \quad \dot{R}=-\sqrt{E-\left(1-2GM/R\right)}.
\ee
In this section $\tau$ is an affine parameter (which we can take as the proper time in this case) and a dot represents differentiation with respect to $\tau$.  $E$ is the observer's energy per unit mass \cite{Wald1984a}.  At spatial infinity this is equal to 

\be
E=\frac{1}{\sqrt{1-v_\infty^2}},
\ee
where $v_\infty$ is the observer's speed at spatial infinity.  The value of $v_\infty$ gives an entire family of co-ordinate systems \cite{Martel2001} but we will only be concerned with the case where $v_\infty=0$.

We are now in a position to write down the four velocity, \linebreak $u^\alpha=\left( \dot{T}, \dot{R}, 0, 0 \right)$, of a particle travelling on an ingoing, time-like geodesic,

\be
u_\alpha=\left( -1, \frac{\sqrt{2GM/R}}{1-2GM/R}, 0, 0 \right).
\ee
It is important to notice that this can be written as

\be
\label{eq:4v1}
u_\alpha=- \partial_\alpha T_{PG},
\ee
where the PG time, $T_{PG}$, is given by

\begin{align}
\label{eq:tPG}
T_{PG}=&T+\int \frac{\sqrt{2GM/R}}{1-2GM/R} dR \nn \\
=& T+4M\left( \sqrt{R/2GM} + \frac{1}{2} \ln \Bigg\lvert \frac{\sqrt{R/2GM}-1}{\sqrt{R/2GM}+1} \Bigg\rvert \right).
\end{align}
We can use Equation \ref{eq:4v1} to write

\be
\int{u_\alpha}dx^{\alpha} = -\int{\partial_{\alpha}T_{PG}}dx^{\alpha} \rightarrow -\int{g_{\alpha\beta}\dot{x}^{\beta}\dot{x}^{\alpha}d\tau} = \int dT_{PG}.
\ee
Since the dots represent the derivative with respect to proper time we can use $g_{\alpha\beta}\dot{x}^{\beta}\dot{x}^{\alpha} = -1$ along a timelike geodesic.  We can now say that for any two events the PG time separation is equal to the proper time separation.  This means that $T_{PG}$ is the time measured by an observer moving along a time-like geodesic.

Using Equation \ref{eq:tPG} we see that the PG metric is given by

\begin{align}
ds^2 =& -\left( 1-\frac{2GM}{R} \right)  dT_{PG}^2 + 2 \sqrt{\frac{2GM}{R}} dT_{PG}dR + dR^2 + R^2d\Omega^2 = \nn \\
& -dT_{PG}^2 + \left(dR + \sqrt{\frac{2GM}{R}} dT_{PG} \right)^2 + R^2d\Omega_2^2.
\end{align}
As promised it is regular at $R=2GM$ and it is intrinsically flat on constant time slices.

\section[Hamiltonian Formulation of General Relativity]{\texorpdfstring{Hamiltonian Formulation of General \\ Relativity}{Hamiltonian Formulation of General Relativity}}
\label{Canonical Formalism in General Relativity1}

\label{Canonical Formalism in General Relativity1}
In this section, we review the canonical analysis for spherically symmetric spacetimes ($k=1$) in general relativity without a cosmological constant in terms of the Arnowitt-Deser-Misner (ADM) variables.  We then perform a canonical transformation and write the action in terms of the generalized Misner-Sharp mass given in Equation \ref{qlm-L} this is a generalization of Kucha\v{r}'s analysis in four dimensions to arbitrary dimensions.  We then minimally couple the action to a matter field to find the equations of motion which govern matter collapse.

Our first task is to dimensionally reduce the Einstein-Hilbert action given in Equation \ref{actiongr}.  For the spherically symmetric spacetimes under consideration (see Equation \ref{eq:structure}), the action reduces to
\begin{align}
\label{eq:reduced action 2_GR}
I_{GR}=\frac{V_{n-2}^{(k)}}{2\kappa_n^2}\int d^2{\bar y}\sqrt{-g_{(2)}}R^{n-2} {\cal L}_{GR},
\end{align}
where $g_{(2)}:=\det(g_{AB})$ and the dimensionally reduced Einstein-Hilbert action, $ {\cal L}_{GR}$ is given from expressions (2.19) and (2.20) of \cite{Maeda2011} (with $p=1$) as 
\begin{align}
{\cal L}_{GR} =& (n-2)(n-3)\left(\frac{k-(DR)^2}{R^2}\right) - 2(n-2)\frac{D^2R}{R} + {}^{(2)}{\cal R} ,  \label{L_p_GR}
\end{align}
where ${}^{(2)}{\cal R}$ is the Ricci scalar calculated using the two-dimensional metric, $g_{AB}$ and $D^2R:=D^AD_AR$.

It should be noted here that, in general, it is not true that dimensional reduction commutes with the variational principle. That is, obtaining the equations of motion from a dimensionally reduced action generally is not the same as implementing the symmetry into equations of motion which were obtained from a non-dimensionally reduced action. In the case of spherical symmetry, however, setting the variation to zero does indeed commute with dimensional reduction \cite{Palais1979,Fels2002}.  This will be discussed in more detail in Chapter \ref{ch:LLADM}.

We write down the action of Equation \ref{eq:reduced action 2_GR} by adopting the following ADM co-ordinates $(t,x)$ as our two non-angular co-ordinates
\begin{eqnarray}
ds_{(2)}^2=g_{AB}d{\bar y}^Ad{\bar y}^B=-N(t,x)^2dt^2+\Lambda(t,x)^2(dx+N_r(t,x)dt)^2.\label{ADM}
\end{eqnarray} 
Now the canonical variables are $N$, $N_r$, $\Lambda$, and $R$ and their momentum conjugates are respectively written as $P_{N}$, $P_{N_r}$, $P_{\Lambda}$, and $P_{R}$.  In the present section, a dot and a prime denote a partial derivative with respect to $t$ and $x$, respectively.  The metric and its inverse are
\begin{align}
g_{tt}=&-(N^2-\Lambda^2N_r^2),\quad g_{tx}=\Lambda^2N_r,\quad g_{xx}=\Lambda^2,\\
g^{tt}=&-N^{-2},\quad g^{tx}=N_rN^{-2},\quad g^{xx}=N^{-2}\Lambda^{-2}(N^2-\Lambda^2N_r^2),
\end{align} 
while $\sqrt{-g_{(2)}}$ is given by  
\begin{align}
\sqrt{-g_{(2)}}=N\Lambda.
\end{align} 
For later use, we compute the following quantities:
\begin{align}
F:=&(DR)^2 =-y^2+\Lambda^{-2}{R'}^2,\label{defF}\\
\sqrt{-g_{(2)}}D^2 R=&-\partial_t(\Lambda y)+\partial_x(\Lambda N_r y+\Lambda^{-1}NR'),\label{DDR}
\end{align} 
where $y$ is defined by 
\be
y:=N^{-1}({\dot R}-N_rR').\label{defy}
\ee 

The reduced action \ref{eq:reduced action 2_GR} then becomes quite simple:
\begin{align}
I_{GR}=&\frac{\ma A_{n-2}}{2\kappa_n^2}\int d^2{\bar y}  \Biggl[2(n-2)R^{n-3}y(N_r'\Lambda +N_r\Lambda') \nonumber \\ 
& -2N\biggl((R^{n-2})''\Lambda^{-1}+(R^{n-2})'(\Lambda^{-1})'\biggl)  \nonumber \\
& + (n-2)(n-3)(1+F)N\Lambda R^{n-4}-2(n-2)R^{n-3}y {\dot \Lambda}\Biggr]. \label{action-3gr}
\end{align}
Here $\ma A_{n-2}$ is the surface area of an $(n-2)$-dimensional unit sphere, namely
\begin{align}
\ma A_{n-2} :=\frac{2\pi^{(n-1)/2}}{\Gamma((n-1)/2)}(\equiv V_{n-2}^{(1)}),
\label{unitarea}
\end{align}
where $\Gamma(x)$ is the Gamma function.

We will use the action of Equation \ref{action-3gr} to perform the Hamiltonian analysis in terms of the mass function and its conjugate momentum, ie, Kucha\v{r}' geometrodynamical phase space variables. We now briefly review some relevant aspects of Kucha\v{r}' geometrodynamics.

\subsection{Geometrodynamics}
The metric of Equation \ref{ADM} may be written in the generalized Schwarzschild form in terms of the areal coordinates as
\begin{align}
ds_{(2)}^2=-F(R,T)e^{2\sigma(R,T)}dT^2+F(R,T)^{-1}dR^2.
\label{eq:Schwarzschild metric}
\end{align}  
The generalized Misner-Sharp mass of Equation \ref{qlm-L} is then given by
\begin{align}
M(R,T)= \frac{(n-2)V_{n-2}^{(k)}}{2\kappa_n^2}\sum_{p=0}^{[n/2]}{\tilde \alpha}_{(p)}R^{n-1-2p}\biggl(k-F(R,T)\biggl)^p,\label{M-F}
\end{align}  
which reduces to
\be
M := \frac{(n-2)\ma A_{n-2}}{2\kappa_n^2}R^{n-3}(1-F)
\label{massgr}
\ee
for the general relativistic case with $k=1$.  This implicitly gives the functional form $F=F(R,M)$.

To see the relation to the ADM form of Equation \ref{ADM} we use the coordinate transformations $T=T(t,x)$ and $R=R(t,x)$, to write the metric of Equation \ref{eq:Schwarzschild metric} as
\begin{align}
ds_{(2)}^2=&-(F{\dot T}^2e^{2\sigma}-F^{-1}{\dot R}^2)dt^2+2(-F{\dot T}T'e^{2\sigma}+F^{-1}{\dot R}R')dtdx \nonumber \\
&+(-F{T'}^2e^{2\sigma}+F^{-1}{R'}^2)dx^2.
\end{align}  
Comparing with the ADM form, we identify
\begin{align}
F{\dot T}^2e^{2\sigma}-F^{-1}{\dot R}^2=&N^2-\Lambda^2{N_r}^2,\\
-F{\dot T}T'e^{2\sigma}+F^{-1}{\dot R}R'=&\Lambda^2N_r,\\
-F{T'}^2e^{2\sigma}+F^{-1}{R'}^2=&\Lambda^2 \label{Gamma1}
\end{align} 
and obtain
\begin{align}
N_r=&\frac{-F{\dot T}T'e^{2\sigma}+F^{-1}{\dot R}R'}{-F{T'}^2e^{2\sigma}+F^{-1}{R'}^2},\label{beta}\\
N=& \frac{e^{\sigma}({\dot T}R'-{\dot R}T')}{\sqrt{-F{T'}^2e^{2\sigma}+F^{-1}{R'}^2}},\label{alpha} 
\end{align} 
As discussed by Kucha\v{r} in Section IVA of \cite{Kuchar1994}, one can ensure that ${\dot T}R'-{\dot R}T'$ and hence the Lapse function $N$ are positive by an appropriate choice of $x$.  $y$ is then given from the definition of Equation \ref{defy} as 
\begin{align}
y=\frac{FT'e^{\sigma}}{\sqrt{-F{T'}^2e^{2\sigma}+F^{-1}{R'}^2}},
\end{align} 
from which we obtain
\begin{align}
T'e^{\sigma}=\frac{y\Lambda}{F},\label{T-sigma}
\end{align} 
where we used Equation \ref{Gamma1}.  Using this to eliminate $T'e^{\sigma}$ in Equation \ref{Gamma1}, we obtain
\begin{align}
F=-y^2+\frac{{R'}^2}{\Lambda^{2}}
\label{eq:F1}
\end{align} 
as required by consistency with Equation \ref{defF}
 
In the above, we derived expressions for the generalized Schwarzschild time $T$ in terms of the canonical ADM variables. As we will see in the following this determines the conjugate momentum to the Misner-Sharp mass function in a form that is appropriate for slicings that approach the Schwarzschild form at spatial infinity. Other asymptotic forms for the slicings are possible, including flat slice or generalized Painlev\'{e}-Gullstrand (PG) coordinates:
\be
ds_{(2)}^2=-{e}^{2\sigma}dT_{\rm PG}^2 + (dR + G{e}^{\sigma} dT_{\rm PG})^2,
\label{p-g-metric}
\ee
where $\sigma=\sigma(T_{\rm PG},R)$ and $G=G(T_{\rm PG},R)$.
The geometrodynamical variables appropriate for such slicings were first derived in \cite{Kunstatter2007}.  Since we have 
\be
(D R)^2=1-G^2
\ee 
for the above form of the metric, it follows that
\be
G=\pm\sqrt{1-F}.
\ee
By inspection of Equation \ref{p-g-metric} one can see that the positive sign yields an equation for ingoing null geodesics that is regular at any horizon $F=0$, so this is the choice that is suitable for describing the spacetime near a future horizon (black hole). The opposite sign must be chosen for a past horizon (white hole).  We now go through exactly the same derivation as before.
{Performing the coordinate transformations $T_{\rm PG}=T_{\rm PG}(t,x)$ and $R=R(t,x)$ in the metric of Equation \ref{p-g-metric} and comparing to the ADM form of Equation \ref{ADM} yields: }
\begin{subequations}
\label{eq:rec1}
\bea
\Lambda^2
&=& 
(R'+{e}^{\sigma} G T_{\rm PG}')^2 -{e}^{2\sigma}{T_{\rm PG}'}^2, 
\label{conformal-factor}
\\
N^2-\Lambda^2N_r^2
&=&{e}^{2\sigma} \dot{T}_{\rm PG}^2
- (\dot{R}+{e}^{\sigma} G \dot{T}_{\rm PG})^2, 
\label{eq:mixed}
\\
\Lambda^2 N_r
&=& 
(R'+{e}^{\sigma} G T_{\rm PG}')(\dot{R}+{e}^{\sigma} G \dot{T}_{\rm PG})  -{e}^{2\sigma}T_{\rm PG}'\dot{T}_{\rm PG}.
\label{shift}
\eea
\end{subequations}
Solving Equation \ref{eq:rec1} for $N$ and $N_r$, we find%
\begin{subequations}%
\label{eq:N-and-sigma-solved}%
\bea
N_r 
&=& \frac{(R'+{e}^{\sigma} G T_{\rm PG}')(\dot{R}+{e}^{\sigma} G \dot{T}_{\rm PG}) -{e}^{2\sigma}T_{\rm PG}'\dot{T}_{\rm PG}}{(R'+{e}^{\sigma} G T_{\rm PG}')^2-{{e}^{\sigma}(T_{\rm PG}')}^2} ,
\\
N 
&=& \frac{R'{e}^{\sigma}\dot{T}_{\rm PG}- \dot{R}{e}^{\sigma} T_{\rm PG}'}{\sqrt{(R'+{e}^{\sigma} G T_{\rm PG}')^2-{{e}^{\sigma}(T_{\rm PG}')}^2}} . 
\label{eq:sigma-solved}%
\eea
\end{subequations}
To complete the derivation, we use Equation \ref{defy} and the above expressions for $\Lambda$, $N$ and $N_r$ to calculate
\bea
y\Lambda &=& (1-G^2)e^{\sigma}T_{\rm PG}' + G R',
\label{eq:yLambda PG}
\eea
which yields:
\be
e^{\sigma}T_{\rm PG}' = \frac{y \Lambda}{F} \pm \frac{R'\sqrt{1-F}}{F}.
\label{eq:Tprime PG}
\ee
The second term on the right-hand side of the above guarantees that the PG time is well defined either for (with a +ve sign) future or (with a -ve sign) past horizons.  This is discussed in the Appendix \ref{app:boundary} on boundary conditions and is valid in the full Lovelock case.


\subsection{Canonical Formalism}

We now derive the expressions for $P_\Lambda$ and $P_R$.  The Lagrangian density corresponding to the action of Equation \ref{action-3gr} is
\begin{align}
{\cal L}_{GR}=& \frac{(n-2)\ma A_{n-2}}{2\kappa_n^2} \Biggl[2R^{n-3}y(N_r'\Lambda +N_r\Lambda') \nonumber \\ 
&-\frac{2}{(n-2)}N\biggl((R^{n-2})''\Lambda^{-1}+(R^{n-2})'(\Lambda^{-1})'\biggl)    \nonumber \\
&+ (n-3)(1+F)N\Lambda R^{n-4}-2R^{n-3}y {\dot \Lambda}\biggl],
\end{align}
from which we obtain $P_N=P_{N_r}=0$ and

\begin{align}
P_\Lambda=& -\frac{(n-2)\ma A_{n-2}}{\kappa_n^2}R^{n-3}N^{-1}({\dot R}-N_rR'),\label{PLambdaGR2}\\
P_R=& \frac{(n-2)\ma A_{n-2} }{2\kappa_n^2} \Biggl[2R^{n-3}N^{-1}(N_r'\Lambda +N_r\Lambda') \nonumber \\
& -2 (n-3)N^{-1}\Lambda R^{n-4}({\dot R}-N_rR')-2R^{n-3}N^{-1} {\dot \Lambda}\biggl].\label{PR}
\end{align}
With $P_\Lambda$ and $P_R$, the Hamiltonian density ${\cal H}(:={\dot \Lambda}P_{\Lambda}+{\dot R}P_{R}-{\cal L})$ is given by
\begin{align}
{\cal H}=&(N_r\Lambda P_{\Lambda})'-N_r\Lambda P_{\Lambda}'+N_rR'P_{R} \nonumber \\
& -\frac{\kappa_n^2 N}{(n-2)\ma A_{n-2}R^{n-2}}P_{\Lambda}\biggl(RP_{R}-\frac{n-3}{2}\Lambda P_{\Lambda}\biggl)  \nonumber \\
&-\frac{(n-2)\ma A_{n-2}}{\kappa_n^2}N \times \nonumber \\
& \biggl\{R^{n-3}\biggl(-R'' \Lambda^{-1}+R'\Lambda^{-2}\Lambda'\biggl)+\frac{n-3}{2} \Lambda R^{n-4}\biggl(1-\Lambda^{-2}{R'}^2\biggl)\biggl\}. \label{hamildens-gr}
\end{align} 
The first term in Equation \ref{hamildens-gr} is a total derivative and becomes a boundary term.  Since $P_N=P_{N_r}=0$, the Hamilton equations for $N$ and $N_r$ give constraint equations $H=0$ and $H_r=0$, where the super-momentum $H_{r}(:=\delta {\cal H}/\delta N_r)$ and the super-Hamiltonian $H(:=\delta {\cal H}/\delta N)$ are given by
\begin{align}
H_r=&-\Lambda P_{\Lambda}'+R'P_{R}, \label{Hr}\\
H=&-\frac{\kappa_n^2}{(n-2)\ma A_{n-2}R^{n-2}}P_{\Lambda}\biggl(RP_{R}-\frac{n-3}{2}\Lambda P_{\Lambda}\biggl)  \nonumber \\
&-\frac{(n-2)\ma A_{n-2}}{\kappa_n^2}\biggl\{R^{n-3}\biggl(-R'' \Lambda^{-1}+R'\Lambda^{-2}\Lambda'\biggl) \nonumber \\
&+\frac{n-3}{2} \Lambda R^{n-4}\biggl(1-\Lambda^{-2}{R'}^2\biggl)\biggl\}. \label{H}
\end{align} 
The action is finally written as
\begin{align}
I_{GR}=&\int dt\int dx ( {\dot \Lambda}P_{\Lambda}+{\dot R}P_{R}-N H-N_r H_{r}).\label{IMGR}
\end{align} 

It can be verified that with suitable boundary conditions the constraints $H$ and $H_r$ are first class in the Dirac sense and generate spacetime diffeomorphisms that preserve the spherically symmetric form of the metric.

It is important to note here that we have written the Hamiltonian in terms of the ADM canonical variables.  If we wanted to we could use Equations \ref{massgr}, \ref{PLambdaGR2} and \ref{PR} to group terms so that we can write the Hamiltonian in terms of the mass function $M(\Lambda, P_\Lambda, R, P_R)$, whose physical meaning is clear.  We can not repeat this procedure for the Lovelock case; the expressions for $P_\Lambda$ and $P_R$ are too difficult to invert.  We will use geometrodynamics to treat the mass function as a canonical variable.  This will let us write the Hamiltonian in terms of the mass function.  It will be important when coupling matter to the action that the gravitational Hamiltonian be written in terms of the ADM variables and so we will do an inverse canonical transformation.  This will give the Hamiltonian in terms of the ADM variables, ${\cal H} = {\cal H}(\Lambda, P_\Lambda, R, P_R, M(\Lambda, P_\Lambda, R, P_R))$.  The equation which relates the mass function to the ADM variables will be too difficult to invert but this will be unimportant for our subsequent analysis.  We now review the boundary conditions which will be important to show that the mass function and its conjugate form a pair of canonical variables.

\subsection{Boundary Condition and Boundary Terms}
To perform the geometrodynamics (ie, to show that the Misner-Sharp mass can be used as a canonical variable), the boundary condition plays a crucial role.  In this thesis we adopt the following boundary condition at spacelike infinity $x\to \pm \infty$\footnote{These boundary conditions are suited to asymptotically Schwarzschild slicings. The analogous boundary conditions for PG coordinates are given in \cite{Husain2005} and discussed in Appendix~\ref{app:boundary}.}:
\begin{align}
N\simeq& N_\infty(t)+\mathcal{O}(x^{-\epsilon_1}),\label{bc1}\\
N_r\simeq& N_r^\infty(t) x^{-(n-3)/2-\epsilon_2},\label{bc2}\\
\Lambda \simeq& 1+\Lambda_1(t)x^{-(n-3)},\label{bc3}\\
R\simeq& x+R_1(t)x^{-(n-4)-\epsilon_4},\label{bc4}
\end{align}
where $\epsilon_1$ is a positive number and $\epsilon_2$ and $\epsilon_4$ satisfy $\epsilon_2> \max[0,-(n-5)/2]$ and $\epsilon_4>\max[0,-(n-5)]$.  This boundary condition ensures that the canonical transformation is well defined (all terms in the Liouville form approach zero sufficiently rapidly), the Hamiltonian is finite and that the Misner-Sharp mass is non-zero and finite.  This is discussed in Appendix \ref{app:boundary}.
The asymptotic behaviour of $P_\Lambda$ and $P_R$ are given by 
\begin{align}
P_\Lambda\simeq& -\frac{(n-2)\ma A_{n-2}}{\kappa_n^2}N_\infty^{-1}\biggl({\dot R_1} x^{1-\epsilon_4}-N_r^{\infty} x^{(n-3)/2-\epsilon_2}\biggl),\label{bc5}\\
P_R\simeq & -\frac{(n-2)\ma A_{n-2} }{\kappa_n^2} N_\infty^{-1}\biggl[N_r^{\infty}(t) \biggl(-\frac{n-3}{2}+\epsilon_2\biggl)x^{(n-5)/2-\epsilon_2}+{\dot \Lambda}_1(t)\biggl].\label{bc6}
\end{align}
Under the boundary condition adopted, the Misner-Sharp mass converges to a finite value $M\simeq M^\infty(t)$, where $M^\infty(t)$ is related to $\Lambda_1(t)$ as
\begin{align}
\Lambda_1(t)\equiv \frac{\kappa_n^2M^\infty(t)}{(n-2)\ma A_{n-2}}.\label{bc7}
\end{align} 

Now let us consider the boundary term for the action of Equation \ref{IMGR}.  The role of the boundary term is to subtract the diverging terms at the boundary in the variation of the above action.  The action is completed by adding the boundary term, which gives a finite value in the variation.

Since the variation of $I_{GR}$ gives
\begin{align}
\delta I_{GR}=&\int dt\int dx \biggl( \partial_t (\delta {\Lambda}P_{\Lambda})-\delta \Lambda{\dot P_{\Lambda}}+ {\dot \Lambda}\delta P_{\Lambda}+ \partial_t (\delta {R}P_{R}) \nonumber \\
&-\delta R{\dot P_{R}}+{\dot R}\delta P_{R}-\delta N H-N \delta H-\delta N_r H_{r}-N_r \delta H_{r}\biggl),\label{variation1gr}
\end{align} 
we need to know the contributions from $N_r\delta H_r$ and $N\delta H$.
Using the following results;
\begin{align}
N_r\delta H_r=&-N_r\delta \Lambda P_{\Lambda}'-(N_r\Lambda\delta P_\Lambda)'+(N_r\Lambda)' \delta P_{\Lambda}+(N_r\delta RP_R)' \nonumber \\
&-\delta R(N_rP_{R})'+N_rR'\delta P_{R},\\
N\delta H=&\biggl(\mbox{irrelevant terms}\biggl) \nonumber \\
&-\frac{(n-2)\ma A_{n-2}}{\kappa_n^2}\biggl\{-\biggl((NR^{n-3}\Lambda^{-1}\delta R)'-(NR^{n-3}\Lambda^{-1})'\delta R\biggl)' \nonumber \\
&+(N'R^{n-3}\Lambda^{-1}\delta R)'-(N'R^{n-3}\Lambda^{-1})'\delta R +\nonumber \\
&(NR^{n-3}R'\Lambda^{-2}\delta \Lambda)'-(NR^{n-3}R'\Lambda^{-2})'\delta \Lambda\biggl\},
\end{align} 
we can write Equation \ref{variation1gr} in the following form:
\begin{align}
\delta I_{GR}=&\int dt\int dx \biggl(\mbox{dynamical terms}\biggl)+\int dx \biggl[\delta {\Lambda}P_{\Lambda}+\delta {R}P_{R}\biggl]_{t=t_1}^{t=t_2} \nonumber \\
&-\int dt\biggl[-N_r\Lambda\delta P_\Lambda+N_rP_R\delta R-\frac{(n-2)\ma A_{n-2}}{\kappa_n^2}\times \nonumber \\
&\biggl\{-NR^{n-3}\Lambda^{-1}\delta (R')+N'R^{n-3}\Lambda^{-1}\delta R+NR^{n-3}R'\Lambda^{-2}\delta \Lambda\biggl\}\biggl]_{x=-\infty}^{x=+\infty}.
\end{align} 
Now the boundary condition comes into play.  We assume $\delta \Lambda=\delta R=0$ at $t=t_1,t_2$ and then the second term in the above variation vanishes.  Using the boundary conditions of Equations \ref{bc1}--\ref{bc6}, we can show that only the contribution in the last integral comes from $NR^{n-3}R'\Lambda^{-2}\delta \Lambda$ as
\begin{align}
NR^{n-3}R'\Lambda^{-2}\delta \Lambda\simeq N_\infty\delta \Lambda_1= \frac{\kappa_n^2N_\infty\delta M^\infty(t)}{(n-2)\ma A_{n-2}},
\end{align} 
where we used Equation \ref{bc7}.  Finally we obtain the boundary term in a simple form:
\begin{align}
\delta I_{GR}=&\int dt\int dx \biggl(\mbox{dynamical terms}\biggl)+\int dt\biggl[N_\infty(t)\delta M^\infty(t)\biggl]_{x=-\infty}^{x=+\infty}. \label{boundaryGR}
\end{align}

\subsection{Misner-Sharp Mass as Canonical Variable}
In the ADM coordinates, the canonical variables are $\{\Lambda, P_\Lambda; R,P_R\}$.  However, the physical meanings of the variable $\Lambda$ is not so clear.  Also, as mentioned earlier, finding the Hamiltonian in terms of the ADM variable will not be possible for us in the Lovelock situation.  In this subsection, we show that the two-dimensional equivalent action is written in a rather elegant manner by introducing the Misner-Sharp mass $M$ as a canonical variable.  We introduce a new set of canonical variables $\{M, P_M; S,P_S\}$ defined by
\begin{align}
S:=&R,
\label{eq:R}\\
P_S:=&P_R-\frac{1}{R'}(\Lambda P_\Lambda'+P_MM'),
\label{eq:PM}\\
M :=& \frac{(n-2)\ma A_{n-2}}{2\kappa_n^2}R^{n-3}(1-F),\label{MS}\\
P_M:=&-T'e^{\sigma}=-\frac{y\Lambda}{F},
\label{eq:Pr1}
\end{align}
where we used Equation \ref{T-sigma} and $F$ can be written in terms of the ADM variables using Equations \ref{defy} \ref{eq:F1} and \ref{PLambdaGR2}.  We are going to show below that, under the boundary conditions of Equations \ref{bc1}--\ref{bc6}, the transformation from a set of variables $\{\Lambda, P_\Lambda; R, P_R\}$ to another set $\{M, P_M; S, P_S\}$ is a well-defined canonical transformation.

Note Equation \ref{eq:Pr1} chooses the conjugate to $M$ in terms of the Schwarzschild time $T$. As verified in Appendix \ref{app:boundary} this leads to a finite Liouville form providing one chooses boundary conditions such that the metric approaches the vacuum Schwarzschild solution sufficiently rapidly at spatial infinity. In order to use asymptotically PG slices, it is necessary to choose the conjugate to $M$ in terms of the PG time $T_{\rm PG}$. That is
\be
\tilde{P}_M = -e^\sigma T'_{\rm PG}= -
\frac{y\Lambda}{F} \mp \frac{R'\sqrt{1-F}}{F}.
\label{eq:PM PG}
\ee
In order to preserve the Liouville form it is of course necessary to transform $P_S$ as well. As verified in Appendix \ref{app:boundary}, the required term is precisely the one that preserves the form of the diffeomorphism constraint $H_r$, as required by diffeomorphism invariance ie, $\tilde{P}_S=P_S \mp M^{\prime}\sqrt{1-F}/F$.  It turns out that this term adds boundary terms to the variation that are required to make the transformation from $\{\Lambda, P_\Lambda; R,P_R\}$ to $\{M, \tilde{P}_M; S,\tilde{P}_S\}$ well defined under asymptotically PG boundary conditions.  This is also proven in Appendix \ref{app:boundary}.
For simplicity, we henceforth stick to the Schwarzschild expressions.

From the expression for the Misner-Sharp mass (Equation \ref{MS}), we obtain
\begin{align}
P_M{\dot M} =&\frac{(n-2)\ma A_{n-2}}{2\kappa_n^2}\frac{y\Lambda}{F}R^{n-3}\biggl[{\dot F}-(n-3)(1-F)\frac{\dot R}{R}\biggl],\label{P_MdotMgr}
\end{align}  
which shows (using Equations \ref{PLambdaGR2} and \ref{useful1}) $P_M{\dot M} =P_\Lambda{\dot \Lambda}+(\cdots){\dot R}+\dot{(\cdots)}+(\cdots)'$.  The Misner-Sharp mass $M$ is expressed in terms of $\{\Lambda,P_\Lambda; R,P_R\}$ as
\begin{align}
M=&\frac{\kappa_n^2P_\Lambda^2}{2(n-2)\ma A_{n-2}R^{n-3}}+\frac{(n-2)\ma A_{n-2}}{2\kappa_n^2}R^{n-3}\biggl(1-\frac{{R'}^2}{\Lambda^2}\biggl).\label{MLiou}
\end{align}  
From this expression, we can show that $M'$ is a linear combination of the constraints:
\begin{align}
M'=\Lambda^{-1}(yH_r-R'H),\label{M'2}
\end{align}  
where we used Equation \ref{PLambdaGR2} to replace $P_\Lambda$ by $y$.  This implies that in the vacuum theory $M$ is a constant on the constraint surface, as expected.

Also using Equation \ref{MLiou}, we can show that two sets of variables $\{\Lambda, P_\Lambda; R,P_R\}$ and $\{M, P_M; S,P_S\}$ satisfy the following Liouville form:
\begin{align}
P_\Lambda\dot{\Lambda}+P_R\dot{R}=P_M \dot{M}+P_S \dot{S}+ \dot{\eta}+\zeta',\label{Liouville}
\end{align}
where
\begin{align}
\eta:=&\Lambda P_\Lambda+\frac{(n-2)\ma A_{n-2}}{2\kappa_n^2}R^{n-3}R' \ln\biggl|\frac{R'+y\Lambda}{R'-y\Lambda}\biggl|,\label{deta-gr}\\ 
\zeta:=&-\frac{(n-2)\ma A_{n-2}}{2\kappa_n^2}R^{n-3}\dot{R}\ln\biggl|\frac{R'+y\Lambda}{R'-y\Lambda}\biggl|. \label{zeta'-gr}
\end{align} 
Under the boundary conditions of Equations \ref{bc1}--\ref{bc6}, the total derivative term $\zeta$ converges to zero at spacelike infinity.  Hence, the transformation from a set $\{\Lambda, P_\Lambda; R,P_R\}$ to $\{M,P_M;S,P_S\}$ is indeed a canonical transformation, namely
\begin{align}
&\int_{-\infty}^{\infty} dx(P_\Lambda \dot{\Lambda}+P_R \dot{R})-\int_{-\infty}^{\infty} dx(P_M \dot{M}+P_S \dot{S})=\dot{\omega}[\Lambda, P_\Lambda; R,P_R], \label{canonical}\\
&\omega[\Lambda, P_\Lambda; R,P_R]:=\int_{-\infty}^{\infty}dx\eta[\Lambda, P_\Lambda; R,P_R]
\end{align} 
is satisfied.  It is shown that the integrands in the above equation, namely $P_\Lambda \dot{\Lambda}+P_R \dot{R}$, $P_M \dot{M}+P_S \dot{S}$, and $\eta$ converge to zero faster than $O(x^{-1})$ at spacelike infinity under the boundary condition we adopt, and hence the above expression is well-defined.  (See Appendix \ref{app:boundary} for the proof.)

We now derive the Hamiltonian constraint and the diffeomorphism (momentum) constraint in terms of the variables $\{M,P_M;S,P_S\}$.  A straightforward calculation using the above equations verifies the following relation;
\begin{align}
{\cal L}-P_M{\dot M}-\frac{N\Lambda}{R'}M'=&-\frac{{\dot R}}{R'}P_MM'+\mbox{(t.d.)} \\
=&\frac{y\Lambda}{F}\biggl(N_r+N\frac{y}{R'}\biggl)M'+\mbox{(t.d.)}, \label{eq:GRfinal}
\end{align}
where $\mbox{(t.d.)}$ is a total derivative term (see Appendix \ref{appendix3} for the derivation in the more general Lovelock case).  From this we obtain the Hamiltonian density ${\cal H}_{\rm G}$ in the equivalent two-dimensional theory as
\begin{align}
{\cal H}_{\rm G}:=& P_M\dot{M} + P_S\dot{S}-{\cal L} \nonumber \\
   =& N^M M'+N^S P_S,   \label{eq:geometrodynamic hamiltonian}
\end{align}
where we have used the fact that $\{M,P_M;S,P_S\}$ are canonical in the first line, Equation \ref{eq:GRfinal} in the second line and defined new Lagrange multipliers $N^M$ and $N^S$ as
\begin{align}
N^M:=&- \frac{\Lambda}{R'}\biggl(N+\frac{y{\dot R}}{F}\biggl) \nonumber\\ 
=&- \frac{\Lambda}{R'}\biggl(N+\frac{y}{F}\left(Ny+N_r R'\right)\biggl) \nonumber\\ 
=&-N\left(\frac{\Lambda}{R'}-\frac{P_M y}{R'}\right) +N_r P_M, 
\label{NM} \\
N^S:=&\dot{S}= \left(Ny+N_r R'\right).\label{NS}
\end{align}

Collecting terms in $N$ and $N_r$, we can express the total Hamiltonian as 
\begin{equation}
{\cal H}_{\rm G}= N\left[\left(\frac{P_M y}{R'} -\frac{\Lambda}{R'}\right)M'+ y P_s\right] + N_r (P_M M' +P_S S').
\label{eq: HG}
\end{equation}
The coefficients of $N$ and $N_r$  are the super-Hamiltonian $H$ and the super-momentum $H_r$, respectively.  Note that $H$ can be written as
\begin{equation}
H = -\frac{\Lambda}{R'} M' + \frac{y}{R'} H_r
\end{equation} 
in agreement with Equation \ref{M'2}.

The Lagrangian density for the canonical coordinates $(M,S,N^M,N^S)$ is now written as
\begin{align}
{\cal L}= P_M\dot{M} + P_S\dot{S}- N^M M'-N^S P_S, \label{eq:geometrodynamic L}
\end{align}
which corresponds to Equation 122 of \cite{Kuchar1994}.


Lastly, let us derive the boundary term in Equation \ref{boundaryGR} with the new canonical variables.  Starting from
\begin{align}
I_{GR}=\int dt\int dx(P_M\dot{M} + P_S\dot{S}- N^M M'-N^S P_S),
\end{align}
we obtain
\begin{align}
\delta I_{GR}=&\int dt\int dx\biggl(\delta P_M\dot{M}+\partial_t(P_M\delta M)-{\dot P}_M\delta {M} \nonumber \\ 
&+ \delta P_S\dot{S} + \partial_t(P_S\delta  S)-{\dot P}_S\delta  S \nonumber \\
&- \delta N^M M'- (N^M \delta M)'+{N^M}' \delta M-\delta N^S P_S-N^S \delta P_S\biggl) \nonumber \\
=&\int dt\int dx \biggl(\mbox{dynamical terms}\biggl) \nonumber \\
&+\int dx \biggl[P_M\delta M+P_S\delta  S\biggl]_{t=t_1}^{t=t_2}-\int dt\biggl[N^M \delta M\biggl]_{x=-\infty}^{x=+\infty}.
\end{align} 
Under the boundary condition of Equations \ref{bc1}--\ref{bc6}, we obtain $\delta M \simeq \delta M^\infty(t)$ and $N^M\simeq -N_\infty(t)$ at spacelike infinity and hence we obtain the same result as Equation \ref{boundaryGR} by setting $\delta M=0$ and $\delta  S=0$ at $t=t_1,t_2$.  One important advantage of the new set of canonical variables is to greatly simplify the calculations.  We will take advantage of this simplification in the next section and in Chapter \ref{ch:LLADM}.

\section[Addition of the Massless Scalar Field:  The General Relativistic Case]{\texorpdfstring{Addition of the Massless Scalar Field:  \\ The General Relativistic Case}{Addition of the Massless Scalar Field:  The General Relativistic Case}}
\label{Scalar field GR}

In this section, we add the action for a massless scalar field to the gravity action using the ADM variables discussed in Section \ref{Canonical Formalism in General Relativity1}.  We use this to find Hamilton's equations of motion, in terms of the mass function, Equation \ref{MLiou}.  These equations govern the evolution of a scalar field under the influence of spherically symmetric, $n$D, general relativistic gravity.  We fix the gauge such that the equations of motion are in Painlev\'{e}-Gullstrand co-ordinates.  This analysis will be repeated in some detail for the more general, Lovelock case in Section \ref{Adding Matter Fields1}.  For this reason we only present highlights of the calculation here.


The action for a massless scalar field, $\psi$, is given by
\begin{align}
I_\psi = -\frac{1}{2}\int d^{n}x \sqrt{-g} (\nabla \psi)^2.
\label{scalaractionGR}
\end{align}
The equivalent two-dimensional action in the symmetric spacetime under consideration can be shown to be
\begin{align}
\label{scalaraction2GR}
I_\psi &= -\frac{{\cal A}_{n-2}}{2}\int d^2{\bar y} \sqrt{-\g} R^{n-2} (D \psi)^2 \\ \nonumber
& = -\frac{{\cal A}_{n-2}}{2}\int dxdt \frac{\Lambda R^{n-2} }{N} \left( -\psid^2 + 2N_r\psip \psid + (N^2\Lambda^{-2}-N_r^2) \psip^2 \right),
\end{align}
which can be written in terms of the phase space variables as
\begin{align}
\label{scalaraction3GR}
I_\psi=& \int dxdt \psid \Pis \nonumber \\
& -\int dxdt N \left[ \frac{1}{2\Lambda} \left( \frac{\Pis^2}{{\cal A}_{n-2} R^{n-2}} +{\cal A}_{n-2}R^{n-2} \psip^2 \right) + \Pis \psip\frac{N_r}{N}  \right],
\end{align}
where $\Pis$ is the conjugate momentum to $\psi$.

Equation \ref{scalaractionGR} does not contain any derivatives of the metric or $R$, which allows us to write the total Hamiltonian as the sum of the gravitational and matter parts.  Using Equation \ref{scalaraction3GR}, we obtain the total Hamiltonian $H_{\rm total} $ as\footnote{The matter fields' fall off conditions required for a well--defined Hamiltonian formulation are well known.  See \cite{Husain2005} for the PG case.}
\begin{align}
\label{H8GR}
H_{\rm total} = \int dx \biggl[ N(H^{\rm (G)}+H^{\rm (M)}) + N_r(H_r^{\rm (G)}+H_r^{\rm (M)}) \biggl],
\end{align}  
where the super-Hamiltonian $H=H^{\rm (G)} + H^{\rm (M)}$ and super-momentum $H_r=H_r^{\rm (G)} + H_r^{\rm (M)}$ are the given by the sums of their gravitational $(G)$ and matter $(M)$ contributions and are given by
\begin{align}
\label{HconstMGR}
H^{\rm (M)} =& \frac{1}{2\Lambda} \left( \frac{P_\psi^2}{{\cal A}_{n-2} R^{n-2}} +{\cal A}_{n-2}R^{n-2} \psip^2 \right),\\
\label{momconstMGR}
H_r^{\rm (M)} =& P_\psi \psip.
\end{align}
\begin{align}
\label{momconst1GR}
H_r^{\rm (G)} =& P_S S^\prime + P_M M^\prime = P_R R^\prime  - P_\Lambda^\prime \Lambda,\\
\label{Hconst2GR}
H^{\rm (G)} =& -\frac{\Lambda}{R^\prime}M^\prime  + \frac{y}{R^\prime}H_r^{(G)}.
\end{align}
$y$ is a function of the phase space variables, $\Lambda$, $P_\Lambda$ and $R$ via Equation \ref{PLambdaGR2} and we have used Equations \ref{defy}, \ref{eq:PM}, \ref{eq:R}, \ref{eq:Pr1}, \ref{eq:GRfinal} and \ref{M'2} to derive Equations \ref{momconst1GR} and \ref{Hconst2GR}.  It should be noted that we are now using the ADM variables and $M$ should be thought of as a function of them.  Since Equation \ref{PLambdaGR2} shows that $y$ is not a function of $N$ or $N_r$, we can see that the Hamiltonian density is the sum of Lagrange multiplier times constraints.  Since our theory is diffeomorphism invariant, $H$ and $H_r$ must be first class constraints which means that there are two gauge choices to pick (see \cite{Matschull1996} for an excellent review of constraint analysis). 
We choose
\be
\label{gauge1GR}
\chi := R - x \approx 0
\ee
and
\be
\label{gauge2GR}
\xi := \Lambda - 1\approx 0,
\ee
where the weakly equals zero symbol ($\approx$) means that the equations are only valid on the solution surface.  The first gauge choice, $\chi$ forces the spatial co-ordinate to be the areal radius while the second, $\xi$ forces the time co-ordinate to be the Painlev\'{e}-Gullstrand time.

In order to insist that $\chi$ and $\xi$ are satisfied at every time slice, we must insist that their Poisson brackets with the Hamiltonian is weakly equal to zero.  This gives the consistency conditions,
\be
\label{conscond1GR}
N_r/N +y/R^\prime \approx 0
\ee
and
\be
\label{sigma1GR}
N^\prime \sqrt{\frac{2(n-2){\ma A_{(n-2)}}R^{n-3}M}{\kappa_n^2}}  + N P_\psi \psip \approx 0.
\ee
In principal we can use Equations \ref{gauge1GR}, \ref{gauge2GR}, \ref{conscond1GR} and \ref{sigma1GR} to simplify the Hamiltonian of Equation \ref{H8GR} (as long as we use Dirac brackets to find the equations of motion).  In practice we must solve the equations of motion and constraints numerically and Equation \ref{sigma1GR} must be solved numerically at each time step to find the value of $N$.  Simplifying with Equations \ref{gauge1GR}, \ref{gauge2GR} and \ref{conscond1GR} gives

\begin{align}
\label{H8dGR}
H_{\rm total} = & \int dR N \Biggl[ -M^\prime + \frac{1}{2} \left( \frac{\Pis^2}{{\cal A}_{n-2} R^{n-2}} + {\cal A}_{n-2} R^{n-2} \psip^2 \right) \nonumber \\
& + \Pis \psip  \sqrt{\frac{2\kappa_n^2 M}{(n-2){\cal A}_{n-2}R^{n-3}} }  \Biggr]
\end{align}
where we used Equation \ref{MS} to write the Hamiltonian in terms of the mass function.

Using this Hamiltonian to find the equations of motion gives

\begin{align}
\label{psidotGR}
\dot{\psi} =& N \left( \frac{\Pis}{{\cal A}_{n-2} R^{n-2}} + \psip \sqrt{\frac{2\kappa_n^2 M}{(n-2){\cal A}_{n-2}R^{n-3}} } \right),\\
\label{PisdotGR}
\dot{P}_\psi =& \left[ N \left( {\cal A}_{n-2} R^{n-2} \psip +  \Pis \sqrt{\frac{2\kappa_n^2 M}{(n-2){\cal A}_{n-2}R^{n-3}} }  \right) \right]^\prime.
\end{align}

Given some initial data these equations can be used to evolve the scalar field forward in time.  At each time step the constraint of Equation \ref{sigma1GR} and the constraint on the Hamiltonian,

\begin{align}
& -M^\prime + \frac{1}{2} \left( \frac{\Pis^2}{{\cal A}_{n-2} R^{n-2}} + {\cal A}_{n-2} R^{n-2} \psip^2 \right) \nonumber \\
& + \Pis \psip  \sqrt{\frac{2\kappa_n^2 M}{(n-2){\cal A}_{n-2}R^{n-3}} } = 0 \label{HconstraintGR}
\end{align}
must be satisfied.

It will be important for later to notice that the equations of motion and the constraints are scale invariant; that is if $\psi(T_{PG},R)$, $\Pis(T_{PG},R)$ and $M(T_{PG},R)$ are a solution to the equations of motion and constraints then so are

\begin{align}
\psi(lT_{PG},lR), \quad l^{-(n-3)}\Pis(lT_{PG},lR), \quad l^{-(n-3)}M(lT_{PG},lR) 
\label{eq:scale invariance}
\end{align}
where $l$ is some positive number.  This will not be the case when a dimensionful parameter appears in the Lagrangian as is the case in higher order Lovelock gravity.

\section{Choptuik Scaling}
\label{sec:Choptuik}

As mentioned at the beginning of this chapter, one of the goals of this work is to numerically investigate the formation of microscopic black holes.  To this end we now review a phenomenon known as Choptuik scaling which relates observables, such as a black hole's mass, to parameters in the initial data.  Although Choptuik scaling is affected by the number of dimensions of the universe and the addition of higher order curvature terms in the action, we mostly concentrate on the case of 4D general relativity in this section.

\subsection{Phenomenology}

When simulating the evolution of a matter field coupled to gravity it is necessary to specify initial conditions, in our case the initial conditions of the scalar field and its conjugate momentum.  For many different types of matter, including scalar field matter, a given parameter in the initial data, $p$, will have a corresponding critical value $p^*$.  For $p>p^*$ a black hole will form and for $p<p^*$ the matter will disperse to infinity.  Black holes with $p$ just slightly bigger than $p^*$ are known as near critical black holes.  In 1993 Matthew Choptuik \cite{Choptuik1993} numerically explored the formation of near critical black holes.  He found that near criticality the mass of the black hole at formation, $M_{BH}$, is related to the initial parameters by the relation 

\be
M_{BH} = |p-p^*|^\gamma,
\label{CSeq}
\ee
where $\gamma\simeq0.37$ for 4D black holes.  Although he found this relation in Schwarzschild coordinates the result is independent of the coordinate system.  The result was confirmed in null coordinates by Garfinkle in 1995 \cite{Garfinkle1995}.  Note that this type of scaling, where the mass (or some other observable) goes to zero as $p$ approaches $p^*$ is seen in other areas of physics and is known as type II scaling.

In 1997 Gundlach described the origins of Choptuik scaling in terms of discrete self-similarity of the equations of motion near the critical point \cite{Gundlach1997} (similar work was also performed at this time by Hod and Piran \cite{Hod1997}).  Using this line of reasoning he then predicted the existence of a co-ordinate system dependent periodic term, $f(\ln|p-p*|)$, which should appear in the Choptuik relation as

\be
\ln(M_{BH}) = \gamma \ln|p-p^*| + f(\ln|p-p^*|).
\label{GundlachCS}
\ee

(This relation will be referred to as Chopuik scaling for the rest of this thesis.)  It turned out that upon re-inspection of Choptuik's work this periodic function was there but it was small enough in amplitude that it went unnoticed.  Although Gundlach predicted the periodic function in Equation \ref{GundlachCS} he did not state a specific form for it in any coordinate system.  In Schwarschild and null coordinates it was found that it could be very well fit to a small amplitude sine wave.  Jonathan Ziprick later found \cite{Ziprick2009a} that in PG coordinates the periodic function had cusps.

The critical exponent, $\gamma$, of Equation \ref{GundlachCS} is not a function of which coordinate system is chosen but it is a function of the number of space-time dimensions.  In 2005 Sorkin and Oren \cite{Sorkin2005} found that the value for $\gamma$ peaked near $n=10$ whereas Bland \cite{Bland2005,Bland2007} found
\be
\gamma = m_1(m_2-e^{-m_3n})
\label{Bland}
\ee
where $m_1$, $m_2$ and $m_3$ are fit parameters.  This asymptotes to $\approx0.5$ for large $n$.  This result is obviously in conflict with that of Sorkin and Oren and will be investigated further in Chapter \ref{ch:HDCS}.

\section[Quasi-Analytical Explanation of Choptuik Scaling]{\texorpdfstring{Quasi-Analytical Explanation of \\ Choptuik Scaling}{Quasi-Analytical Explanation of Choptuik Scaling}}
\label{Canonical Formalism in General Relativity1}
\label{QAECS}


Choptuik scaling, for the massless scalar field, was thought to occur due to the existence of a discretely self similar, critical surface which separates the phase space into two solutions; those which form black holes and those which end with a flat space-time.  It is thought that the existence of this self similarity is related to the scale invariance in the equations of motion such as that seen in equation \ref{eq:scale invariance}.  In Carsten Gundlach's 2003 paper \cite{Gundlach2002} on this subject he says that ``the presence of a length scale in the field equations is incompatible with exact self similarity'' although he also mentions that it is still an open question whether or not ``type II critical phenomena, where the critical solution is scale-invariant, can arise if the field equations contain a scale.''

In 1997 Gundlach \cite{Gundlach1997} verified the explanation that discrete self similarity causes Choptuik scaling by using it to calculate the critical exponent, $\gamma$ and by predicting the periodic function, $f$ in Equation \ref{GundlachCS}.  This section briefly reviews the relevant parts of that work.  We keep much of the notation used in \cite{Gundlach1997}.

\begin{figure}[ht!]
\centering
\includegraphics[width=0.8\linewidth]{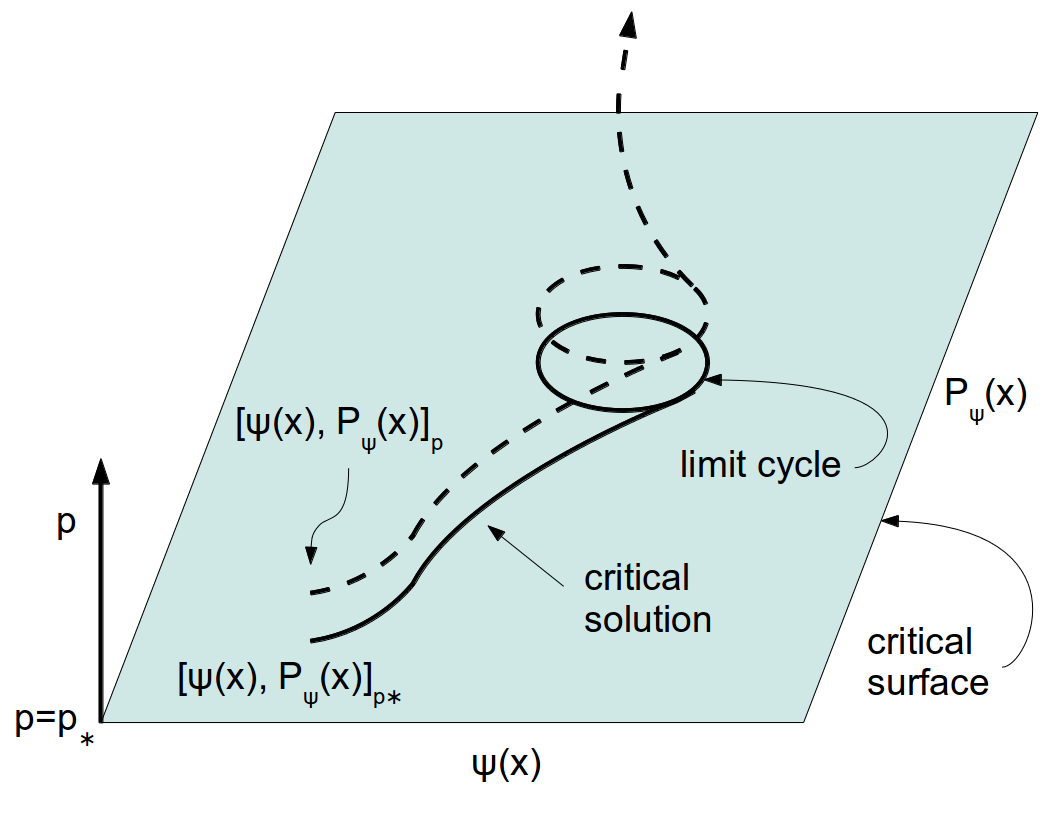}
\caption[The Critical Solution]{The critical solution (solid line) and a solution starting with initial data which is not quite on the critical surface (dashed line)}
\label{CriticalSolution}
\end{figure}

Figure \ref{CriticalSolution} shows the critical surface.  Any point on the surface corresponds to two functions, $\psi(x)$ and $P_\psi(x)$ of some space-like co-ordinate, $x$ \footnote{For concreteness we have chosen to use phase space variables for this figure but, in principal, other forms of the equations of motion could be used.}.  Any initial data on this surface ($[\psi(x), P_\psi(x)]_{p^*}$ for example) will evolve along this surface (solid line) until it reaches the limit cycle where it will oscillate forever.  This is the origin of the discrete self similarity.  We assume that the initial data contains one tunable parameter, $p$, which we call $p^*$ on the critical surface.  By adjusting $p$ we move off of the critical surface.  If we start with $p$ slightly bigger than $p^*$ in the initial data (see the point $[\psi(x), P_\psi(x)]_{p}$), ie slightly above the critical surface, then we evolve along a curve (dotted line) which approximates the critical solution and even follows the limit cycle for some time before leaving the vicinity of the critical surface to form a black hole.  If $p$ is slightly less than $p^*$ the evolution is similar except that the space-time eventually becomes flat.

In order to explore the critical solution we start by defining discrete self similarity.  A space-time is said to be discretely self similar if there exists some set of co-ordinates, $(\sigma, x^\alpha)$ such that the metric can be written as

\be
g_{\mu \nu}(\sigma, x^\alpha)=e^{2\sigma}\tilde{g}_{\mu \nu}(\sigma, x^\alpha)
\label{eq:DSS1}
\ee
where

\be
\tilde{g}_{\mu \nu}(\sigma, x^\alpha)=\tilde{g}_{\mu \nu}(\sigma+m\Delta, x^\alpha),
\label{eq:DSS2}
\ee
$m$ is an integer and $\Delta$ is some constant.  By assuming that the metric takes this form for a specific co-ordinate system on the critical surface Gundlach \cite{Gundlach1997} calculates $\Delta$ and shows agreement with results calculated using numerical collapse and dispersion simulations.  To do this calculation start with the metric in the form

\be
-\alpha(R,T)^2dT^2 + a(R,T)^2dR^2 + R^2d\Omega^2_{(2)}.
\ee
If we transform to the co-ordinates $(\sigma, z)$ given by

\be
T=e^{\sigma}\tilde{T}(\sigma, z)
\label{eq:TDSS1}
\ee
and

\be
R=e^{\sigma}\tilde{R}(\sigma, z),
\label{eq:RDSS1}
\ee
where $\tilde{T}$ and $\tilde{R}$ are periodic in $\sigma$ with period $\Delta$, we get the new line element

\begin{align}
ds^2=&e^{2\sigma}\bigg\{ -\alpha^2\left[ \left( \tilde{T} + \tilde{T},_\sigma \right)d\sigma + \tilde{T},_z dz \right]^2 + \nonumber \\
& a^2\left[ \left( \tilde{R} + \tilde{R},_\sigma \right)d\sigma + \tilde{R},_z dz \right]^2 + \tilde{R}^2 d\Omega^2 \bigg\}.
\end{align}
This metric is discretely self similar (see Equations \ref{eq:DSS1} and \ref{eq:DSS2}) provided that the old metric functions obey

\be
a(T,R)=a(e^{m\Delta}T,e^{m\Delta}R)
\label{eq:aDSS1}
\ee
and

\be
\alpha(T,R)=\alpha(e^{m\Delta}T,e^{m\Delta}R)
\label{eq:alpDSS1}
\ee
where $m$ is an integer.  In order to impose discrete self similarity into the evolution we adopt a specific co-ordinate system given by

\be
\tau := \ln\left(\frac{T}{l}\right)
\ee
and

\be
\zeta := \ln\left(\frac{R}{T}\right) - \xi_0(\tau),
\label{zeta1}
\ee
where $l$ is an arbitrary length scale and $\xi_0$ is a function with period $\Delta$.  These co-ordinates are in the form of Equations \ref{eq:TDSS1} and \ref{eq:RDSS1} with $\tau$ playing the role of $\sigma$.  In these co-ordinates Equations \ref{eq:aDSS1} and \ref{eq:alpDSS1} become

\be
a(\zeta, \tau)=a(\zeta, \tau + m\Delta)
\label{eq:aDSS2}
\ee
and

\be
\alpha(\zeta, \tau)=\alpha(\zeta, \tau + m\Delta).
\label{eq:alpDSS2}
\ee

Gundlach wrote the equations of motion (analogous to Equations \ref{psidotGR} and \ref{PisdotGR}) in terms of $\tau$ and $\xi$.  Using Fourier analysis and the assumptions given in Equations \ref{eq:aDSS2} and \ref{eq:alpDSS2} he numerically calculated $\Delta=3.4453$.  This value can be compared to that which is calculated using collapse simulations (see \cite{Ziprick2009c} for example) of $\Delta \simeq 3.44$ giving credence to the idea that a discretely self similar solution is responsible for the periodic term in the Choptuik scaling relation, Equation \ref{GundlachCS}.

In order to calculate the critical exponent, $\gamma$ of Equation \ref{GundlachCS} we must consider solutions which aren't quite on the critical surface (see the dotted line in Figure \ref{CriticalSolution}).  We define $Z$ as being short hand for $a$, $\alpha$, $\psi$ or $P_\psi$, $Z_*$ is $Z$ on the critical surface and $\delta Z$ is a small perturbation to $Z_*$.  From the equations of motion we obtain the equations governing the perturbations.  They are of the form 

\be
\delta Z,_{\zeta}=A\delta Z + B \delta Z,_\tau,
\label{eq:deltaZEOM}
\ee
where $A$ and $B$ are periodic in $\tau$ with period $\Delta$.  Note that $\delta Z$ can not be periodic (or $Z_*+\delta Z$ would be on the critical surface) and that it must be some function of the distance from the critical surface, $p-p^*$.  To achieve this Gundlach used the form

\be
\delta Z(\zeta, \tau)=\sum\limits_{i=1}^\infty C_i(p-p^*) e^{\lambda_i \tau} \delta_i Z(\zeta, \tau),
\label{eq:deltaZ1}
\ee
where $C_i$ are functions of $p$, $\lambda_i$ are constants and $\delta_i Z$ are periodic in $\tau$ with period $\Delta$.  Gundlach calculated the set of $\lambda_i$ using Equations \ref{eq:deltaZEOM} and \ref{eq:deltaZ1} and found only one growing mode ($\lambda<0$ in this case since $\tau=-\infty$ at the singularity), which he called $\lambda_1$.  He found $\lambda_1 = -2.674$ which, as we will see shortly, corresponds to a critical exponent of $\gamma = -1/\lambda_1 = 0.374$.  This agrees well with values obtained from simulations of matter collapse (see \cite{Ziprick2009c} for example).

To see the relation between Equation \ref{eq:deltaZ1} and the scaling relation, Equation \ref{GundlachCS}, consider the initial data at constant $T$ given by

\be
Z_\tau(R)=Z_*\left( \ln\frac{R}{l}, \tau \right) + \epsilon \delta_1 Z\left( \ln\frac{R}{l}, \tau \right),
\label{eq:Zinitial}
\ee
where $\epsilon$ is a small constant who's sign determines whether the solution will form a black hole or a flat space-time.  Here we keep only the $\delta_1 Z$ term since it is the only one which corresponds to a growing mode.  As we will see soon, we can relate the constant, $\tau$, to $p$ and so it plays the role of the parameter for this family of initial data.

It is important to notice that the evolution equations have no scale.  This means that the only scale in the system is $l$ in the initial data and the solution evolves as
\be
Z_\tau(R, T) = J\left(\frac{R}{l}, \frac{T}{l}, \tau \right),
\ee
where $J$ is a some function.  The mass of the black hole, which has units of length, must then be given by\footnote{Note that the black hole mass could mean the mass of the final black hole, the mass of the black hole at formation or possibly some other choice.  This will affect the form of $\mu_0$ but not its period.  In this thesis we measure the mass of the black hole at formation.}

\be
M=l e^{\mu_0(\tau)},
\label{eq:Moftau}
\ee
where $\mu_0$ is a periodic function with period $\Delta$.  Note that a similar relation exists for the Ricci scalar, which has units of length$^{-2}$, except that it is proportional to $l^{-2}$.

We now consider the evolution of a slightly super-critical system and compare to Equation \ref{eq:Zinitial} to see the relationship between the initial conditions and the black hole's mass.  For a solution near the critical surface we write Equation \ref{eq:deltaZ1} as

\be
Z(\zeta, \tau) \simeq Z_*\left( \zeta, \tau \right) + (p-p^*)\frac{\partial C_1}{\partial p}\bigg|_{p=p^*} e^{\lambda_1 \tau} \delta_1 Z\left( \zeta, \tau \right),
\label{eq:Z2}
\ee
where $Z_*$ is the critical solution.  If we define

\be
\tau_*(p):= \gamma \ln \left[ \frac{1}{\epsilon} (p-p^*)\frac{\partial C_1}{\partial p}\bigg|_{p=p^*} \right]
\label{eq:tauofp}
\ee
and

\be
R(p):=l e^{\tau_*(p) + \xi_0(\tau_*(p))},
\label{eq:Rofp}
\ee
where, as promised, $\gamma$ is given by $\gamma := -1/\lambda_1$, then Equation \ref{eq:Z2} is written as

\be
Z_p(R)=Z_*\left( \ln\frac{R}{R(p)}, \tau_*(p) \right) + \epsilon \delta_1 Z\left( \ln\frac{R}{R(p)}, \tau_*(p) \right).
\ee
This is in the form of the initial data, Equation \ref{eq:Zinitial}.  We can now use Equations \ref{eq:Moftau}, \ref{eq:tauofp} and \ref{eq:Rofp} to write

\begin{align}
M(p)&=R(p)e^{\mu_0(\tau_*(p))} \nonumber \\
&=l(p-p^*)^\gamma e^{\tau_f + \tilde{\mu}_0[\gamma\ln(p-p^*) + \tau_f]},
\label{eq:Mofp}
\end{align}
where,
\be
\tau_f:=\gamma \ln \left[ \frac{1}{\epsilon} \frac{\partial C_1}{\partial p}\bigg|_{p=p^*} \right]
\ee
and
\be
\tilde{\mu}_0:=\mu_0+\xi_0
\ee
is a periodic function.  Although Gundlach showed that $\mu_0$ and $\xi_0$ have periods of $\Delta$ (see Equations \ref{zeta1} and \ref{eq:Moftau}) he argued that $\tilde{\mu}_0$ should have a period of $\Delta/2$.  Taking the logarithm of both sides of Equation \ref{eq:Mofp} verifies the Choptuik scaling relation of Equation \ref{GundlachCS} (up to shifts in the abscissa and ordinate) and gives the period of the $f$ as $\Delta/(2\gamma) \simeq 4.61$ in agreement with the value found using numerical collapse simulations.  This supports the idea that a critical surface and discrete self similarity are responsible for Choptuik scaling.


\chapter[Hamiltonian Formalism of LL Gravity]{Hamiltonian Formalism of Lovelock Gravity}
\label{ch:LLADM}

In Section \ref{Canonical Formalism in General Relativity1} we reviewed the Hamiltonian analysis of general relativity.  We obtained the Hamiltonian in terms of a mass function and wrote the equations of motion for gravity coupled to a massless scalar field.  In this chapter we perform the non-trivial generalization of the results of Section \ref{Canonical Formalism in General Relativity1} to the Lovelock case.  This chapter is based on our paper \cite{Kunstatter2013} which is a generalization of the work that we did in \cite{Taves2012} and \cite{Kunstatter2012}.

In Section \ref{Dimensionally Reduced Action1} we derive a dimensionally reduced equivalent two-dimensional action that is the starting point of our analysis.  The Hamiltonian analysis of Lovelock gravity is presented in Section \ref{Canonical Formalism in Lovelock Gravity1}.  The contributions of scalar matter and charged scalar matter fields are discussed in Section \ref{Adding Matter Fields1}.

\section{Dimensionally Reduced Action}
\label{Dimensionally Reduced Action1}
\subsection{Covariant Form}
Our first task is to dimensionally reduce the Lovelock action, given by Equation \ref{action2} of Section \ref{sec:LLGravity}, 

\begin{align}
I_{LL}=&\frac{1}{2\kappa_n^2}\int \D ^nx\sqrt{-g}\sum_{p=0}^{[n/2]}\alpha_{(p)}{\ma L}_{(p)},\nonumber \\
{\ma L}_{(p)}:=&\frac{1}{2^p}\delta^{\mu_1\cdots \mu_p\nu_1\cdots \nu_p}_{\rho_1\cdots \rho_p\sigma_1\cdots \sigma_p}{\cal R}_{\mu_1\nu_1}^{\phantom{\mu_1}\phantom{\nu_1}\rho_1\sigma_1}\cdots {\cal R}_{\mu_p\nu_p}^{\phantom{\mu_p}\phantom{\nu_p}\rho_p\sigma_p}, \nonumber
\end{align}
so that it incorporates spherical symmetry.  The dimensionally reduced action is given by

\begin{align}
\label{eq:reduced action 2}
I_{LL}=\frac{V_{n-2}^{(k)}}{2\kappa_n^2}\int d^2{\bar y}\sqrt{-g_{(2)}}R^{n-2}\sum^{[n/2]}_{p=0}\alpha_{(p)} {\cal L}_{(p)},
\end{align}
where $g_{(2)}:=\det(g_{AB})$ and the dimensionally reduced $p$th-order Lovelock term $ {\cal L}_{(p)}$ is given from Equations 2.19 and 2.20 of \cite{Maeda2011} as 
\begin{align}
{\cal L}_{(p)} =& \frac{(n-2)!}{(n-2p)!} \biggl[(n-2p)(n-2p-1)\left(\frac{k-(DR)^2}{R^2}\right)^{p} \nonumber \\
&- 2p(n-2p)\frac{D^2R}{R}\left(\frac{k-(DR)^2}{R^2}\right)^{p-1} \nonumber \\
& + 2p(p-1) \frac{(D^2R)^2 - (D^AD_BR)(D^BD_AR)}{R^2} \left(\frac{k-(DR)^2}{R^2}\right)^{p-2} \nonumber \\
&+ p {}^{(2)}{\cal R} \left(\frac{k-(DR)^2}{R^2}\right)^{p-1} \biggl],  \label{L_p}
\end{align}
where ${}^{(2)}{\cal R}$ is the Ricci scalar calculated using $g_{AB}$, $D^2R:=D^AD_AR$ and we recall that $k = 1,0,-1$ corresponds to $\gamma_{ab}$ being spherical (compact), flat or hyperbolic.  At a glance, there is a non-minimal coupling between $(DR)^2$ and the terms containing second derivatives in ${\cal L}_{(p)}$.  In this form it is not clear how to use this Lagrangian to perform the canonical analysis.  As proven in Appendix \ref{appendix1}, we can write it, up to total divergences, without this coupling: 
\begin{align}
\label{eq:lovelock simplified}
 {\cal L}_{(p)} =& \frac{(n-2)!}{(n-2p)!} \Biggl[pk^{p-1} {}^{(2)}{\cal R} R^{2-2p} \nonumber \\
&+ pR^{2-n}\frac{D^A(R^{n-2p})D_A((DR)^2)}{(DR)^2}\biggl\{k^{p-1}-(k-(DR)^2)^{p-1}\biggl\}  \nonumber \\
& + (n-2p)(n-2p-1)\biggl\{\left(k-(DR)^2\right)^{p} +2pk^{p-1}(DR)^2\biggl\}R^{-2p} \Biggr]. 
\end{align}
This is a key result of this work and the starting point of our canonical analysis.

Note that in the following we work exclusively with the equations of motion derived from the reduced action of Equation \ref{eq:reduced action 2}. In general it is not true that dimensional reduction commutes with the variational principle. That is, the space of extrema of a dimensionally reduced action in principle may not coincide with the space of symmetric solutions of the unreduced action. However, in a very elegant and powerful set of papers \cite{Palais1979,Fels2002} (see also \cite{Deser2003a}), it has been rigorously proven that if the symmetry group is a compact Lie group, as in our case, then for any local metric theory of gravity in arbitrary space-time dimensions, with or without matter, variation does indeed commute with dimensional reduction. The spherically symmetric equations of motion obtained from the full, unreduced Lovelock action with matter were explicitly written down in \cite{Maeda2011}.  The proof that the solution space is the same in both cases nonetheless requires the more detailed analysis of \cite{Fels2002,Palais1979}. Unfortunately, this analysis is only valid in the compact case, so that more work needs to be done in order to prove that the dimensionally reduced action is sufficient when $k=0,-1$. This is one of the reasons that we only consider the $k=1$ case for the rest of this work.

\subsection{ADM Form}
We are now going to write down the action of Equation \ref{eq:lovelock simplified} using the ADM coordinates $(t,x)$ as we did in Section \ref{Canonical Formalism in General Relativity1} for the general relativistic case.  Recall from Equation \ref{ADM} that the ADM metric is given by
\begin{eqnarray}
ds_{(2)}^2=-N(t,x)^2dt^2+\Lambda(t,x)^2(dx+N_r(t,x)dt)^2. \nonumber
\end{eqnarray} 
As we did in Equations \ref{defF}, \ref{DDR} and \ref{defy} for the general relativistic case it is useful to use the following definitions and relationships
\begin{align}
F:=&(DR)^2 \nonumber \\
=&-y^2+\Lambda^{-2}{R'}^2,\nonumber\\
\sqrt{-g_{(2)}}D^2 R=&-\partial_t(\Lambda y)+\partial_x(\Lambda N_r y+\Lambda^{-1}NR') \nonumber,
\end{align} 
where $y$ is again defined by 
\be
y:=N^{-1}({\dot R}-N_rR'). \nonumber
\ee 
For the Lovelock case we also need the following relationship:
\be
D^A(R^{n-2p})D_A((DR)^2)= 
  (n-2p) R^{n-2p-1} \left(  - \frac{1}{N}y \dot{F}
+ \left(\frac{R'}{\Lambda^2}
   +\frac{N_r}{N} y\right) F'\right).
\ee
Using these results, the action of Equation (\ref{eq:lovelock simplified}) is written in the following simple form:
\begin{align}
I_{LL}=&\frac{(n-2)V_{n-2}^{(k)}}{2\kappa_n^2}\sum^{[n/2]}_{p=0}\int d^2{\bar y}\sqrt{-g_{(2)}}\frac{{\tilde \alpha}_{(p)} }{(n-2p)} \Biggl[pk^{p-1}{}^{(2)}{\cal R}R^{n-2p} \nonumber \\
&- p(n-2p)\frac{R^{n-2p-1}}{N\Lambda}\{k^{p-1}-(k-F)^{p-1}\} \nonumber \\
&\times\biggl\{\Lambda y\frac{\dot F}{F}-(\Lambda N_r y+\Lambda^{-1}NR')\frac{F'}{F}\biggl\} \nonumber \\
& + (n-2p)(n-2p-1)\biggl\{\left(k-F\right)^{p} +2pk^{p-1}F\biggl\}R^{n-2-2p} \Biggr]. \label{action-3}
\end{align}
The first term is two-dimensional gravity non-minimally coupled to a scalar field $R$, which is essentially the same as the general relativistic case.  This term can be explicitly written down in terms of the canonical variables using
\begin{align}
\sqrt{-g_{(2)}}R^{n-2p}{}^{(2)}{\cal R}=&-2N^{-1}\biggl((R^{n-2p})'N_r-\partial_t(R^{n-2p})\biggl)(N_r'\Lambda +N_r\Lambda'-{\dot \Lambda}) \nonumber \\
&-2N\biggl((R^{n-2p})''\Lambda^{-1}+(R^{n-2p})'(\Lambda^{-1})'\biggl) \nonumber \\
&+\partial_t(\cdots)+\partial_x (\cdots).\label{intR}
\end{align} 
Based on the action of Equation \ref{action-3}, we will perform the canonical analysis in the subsequent sections using geometrodynamical phase space variables as we did in Section \ref{Canonical Formalism in General Relativity1} for the general relativistic case.

\section{Canonical Formalism in Lovelock Gravity}
\label{Canonical Formalism in Lovelock Gravity1}
In this section, we show that the important results of Section \ref{Canonical Formalism in General Relativity1} can be generalized to full Lovelock gravity.  In particular the transformation from the ADM variables $\{\Lambda, P_\Lambda; R,P_R\}$ to $\{M, P_M; S,P_S\}$ is a well-defined canonical transformation using definitions of $P_M$, $S$, and $P_S$ that are the same as those in general relativity, Equations \ref{eq:R}, \ref{eq:PM} and \ref{eq:Pr1}, and $M$ defined by Equation \ref{qlm-L}.

\subsection{ADM variables}
First we derive the ADM conjugate momenta $P_\Lambda$ and $P_R$.  The Lagrangian density from the action of Equation \ref{action-3} is 
\begin{align}
{\cal L}_{LL}=& \frac{(n-2){\cal A}_{n-2}}{2\kappa_n^2}\sum^{[n/2]}_{p=0}{\tilde \alpha}_{(p)}  \Biggl[2pR^{n-2p-1}y(N_r\Lambda)' -\frac{2pN}{n-2p}\biggl((R^{n-2p})'\Lambda^{-1}\biggl)' \nonumber \\
& + (n-2p-1)\biggl\{\left(1-F\right)^{p} +2pF\biggl\}N\Lambda R^{n-2-2p}-2pR^{n-2p-1}y {\dot \Lambda}  \nonumber \\
&+ pR^{n-2p-1}\biggl\{1-(1-F)^{p-1}\biggl\}\biggl\{(\Lambda N_r y+\Lambda^{-1}NR')\frac{F'}{F}-\Lambda y\frac{\dot F}{F} \biggl\}\biggl].\label{Lag-0}
\end{align}
Using the binomial expansion and integration by parts many times, we can rewrite the above Lagrangian density in the following form up to total derivatives.  The derivation is presented in Appendix \ref{appendix2}.
\begin{align}
&{\cal L}_{LL}= \frac{(n-2){\cal A}_{n-2}}{2\kappa_n^2}\sum^{[n/2]}_{p=0}{\tilde \alpha}_{(p)}  \Biggl[2pR^{n-2p-1}y(N_r\Lambda)'-\frac{2pN}{n-2p}\biggl((R^{n-2p})'\Lambda^{-1}\biggl)' \nonumber \\
& + (n-2p-1)\biggl\{\left(1-F\right)^{p} +2pF\biggl\}N\Lambda R^{n-2-2p}-2pR^{n-2p-1}y {\dot \Lambda}\biggl] \nonumber \\
&-\frac{(n-2){\cal A}_{n-2}}{2\kappa_n^2}\sum^{[n/2]}_{p=2}{\tilde \alpha}_{(p)}\biggl[\sum_{w=0}^{p-2}\frac{p!(-1)^{p-1-w}}{w!(p-1-w)!}\biggl\{ \Lambda N_r yR^{n-2p-1}F^{p-2-w}(\Lambda^{-2}{R'}^2)' \nonumber \\
&+\sum_{j=0}^{p-2-w}\frac{2(p-2-w)!(-1)^{p-2-w-j}}{j!(p-2-w-j)!}\frac{(N_r R^{n-2p-1}\Lambda^{1-2j}{R'}^{2j})'y^{2(p-w-j)-1}}{2(p-w-j)-1} \nonumber \\
&-\frac{(\Lambda^{-1}NR'R^{n-2p-1})'F^{p-1-w}}{p-1-w}\biggl\} +\sum_{w=1}^{p-1}\frac{2p!(-1)^w}{w!(p-1-w)!}R^{n-2p-1} F^{w-1} y\Lambda^{-2}{\dot \Lambda}{R'}^2\nonumber \\
&-\sum_{w=1}^{p-1}\frac{2p!}{w(p-1-w)!}\sum_{j=0}^{w-1}\frac{(-1)^{2w-1-j}}{j!(w-1-j)!} \biggl\{\frac{\partial_t(R^{n-2p-1}\Lambda^{1-2j}){R'}^{2j}y^{2(w-j)+1} }{2(w-j)+1}\nonumber \\
&-j\sum_{q=0}^{2(w-j)+1}\frac{(2w-2j)!(-1)^q}{q!(2w-2j+1-q)!}\\ \nonumber
& \times \frac{{\dot R}^{w-j+1}(R^{n-2p-1}\Lambda^{1-2j}N^{-2(w-j)-1}{R'}^{2j-1+q}N_r^q)'}{2(w-j+1)} \\ \nonumber
&-\sum_{q=0}^{2(w-j)-1}\frac{(2w-2j-1)!(-1)^q}{q!(2w-2j-q)!}{\dot R}^{2w-2j-q} \\ \nonumber
& \times (R^{n-2p-1}\Lambda^{-1-2j}N^{-2(w-j)+1}{R'}^{2j+1+q}N_r^q)'\biggl\}\biggl] \label{Lag-1}.
\end{align}
From this Lagrangian density, we obtain
\begin{align}
&P_\Lambda= -\frac{(n-2){\cal A}_{n-2}}{\kappa_n^2}\biggl[\sum^{[n/2]}_{p=0}{\tilde \alpha}_{(p)}pR^{n-2p-1}y \\ \nonumber
& +\sum^{[n/2]}_{p=2}{\tilde \alpha}_{(p)}R^{n-2p-1}y\frac{{R'}^2}{\Lambda^{2}} \sum_{w=1}^{p-1}\frac{p!(-1)^w}{w!(p-1-w)!} \\ \nonumber
& \times \biggl\{F^{w-1}- \sum_{j=0}^{w-1}\frac{(-1)^{w-1-j}(w-1)!}{j!(w-1-j)!} \frac{(1-2j)y^{2(w-j)}}{2(w-j)+1}\frac{{R'}^{2j-2}}{\Lambda^{2j-2}}\biggl\} \biggl]\label{PLambda3}
\end{align}
and
\begin{align}
&P_R= \frac{(n-2){\cal A}_{n-2}}{\kappa_n^2}\sum^{[n/2]}_{p=0}{\tilde \alpha}_{(p)} pR^{n-2p-1} \nonumber \\
& \times  \Biggl[\frac{(N_r\Lambda)'-{\dot \Lambda}}{N}+(n-2p-1)\biggl\{\left(1-F\right)^{p-1} -2\biggl\}\frac{y\Lambda}{R}\biggl] \nonumber \\
&-\frac{(n-2){\cal A}_{n-2}}{2\kappa_n^2}\sum^{[n/2]}_{p=2}{\tilde \alpha}_{(p)} p \nonumber \\
& \times \biggl[\sum_{w=0}^{p-2}\frac{(p-1)!(-1)^{p-1-w}}{w!(p-1-w)!}N^{-1}\biggl\{2(\Lambda^{-1}NR'R^{n-2p-1})'F^{p-2-w}y\nonumber \\
&+ \sum_{j=0}^{p-2-w}\frac{2(p-2-w)!(-1)^{p-2-w-j}}{j!(p-2-w-j)!}(N_r R^{n-2p-1}\Lambda^{1-2j}{R'}^{2j})'y^{2(p-w-j-1)} \nonumber \\
&+\Lambda N_r R^{n-2p-1}F^{p-3-w}(\Lambda^{-2}{R'}^2)'\biggl(F-2(p-2-w) y^2\biggl) \biggl\} \nonumber \\
&+\sum_{w=1}^{p-1}\frac{2(p-1)!(-1)^w}{w!(p-1-w)!}\biggl\{R^{n-2p-1} \Lambda^{-2}{\dot \Lambda}{R'}^2N^{-1}F^{w-2}\biggl(F-2(w-1) y^2\biggl)\nonumber \\
&- \sum_{j=0}^{w-1}\frac{(-1)^{w-1-j}}{j!(w-1-j)!}\nonumber \\
& \times \biggl(\partial_t(R^{n-2p-1}\Lambda^{1-2j})\frac{{R'}^{2j}y^{2(w-j)}}{N} +\frac{n-2p-1}{2(w-j)+1}\frac{R^{n-2p-2}{R'}^{2j}y^{2(w-j)+1}}{\Lambda^{2j-1}} \nonumber \\
&-\sum_{q=0}^{2(w-j)+1}\frac{2j(2w-2j)!(-1)^q}{q!(2w-2j+1-q)!}{\dot R}^{2(w-j)+1}\nonumber \\
& \times(R^{n-2p-1}\Lambda^{1-2j}N^{-2(w-j)-1}{R'}^{2j-1+q}N_r^q)' \nonumber \\
&-\sum_{q=0}^{2(w-j)-1}\frac{(2w-2j-1)!(-1)^q}{q!(2w-2j-1-q)!}{\dot R}^{2w-2j-q-1}\nonumber \\
& \times(R^{n-2p-1}\Lambda^{-1-2j}N^{-2(w-j)+1}{R'}^{2j+1+q}N_r^q)'\biggl)\biggl\}\biggl].
\end{align}

In general relativity, $P_\Lambda=P_\Lambda[{\dot \Lambda},y({\dot R})]$ and $P_R=P_R[{\dot \Lambda},y({\dot R})]$ can be algebraically solved to give a unique set of ${\dot \Lambda}={\dot \Lambda}[P_\Lambda,P_R]$ and ${\dot R}={\dot R}[P_\Lambda,P_R]$.  In higher-order Lovelock gravity, by contrast, it is not possible to obtain a unique expression in general because of the fact that $P_\Lambda=P_\Lambda[{\dot \Lambda},y({\dot R})]$ and $P_R=P_R[{\dot \Lambda},y({\dot R})]$ are higher-order polynomials of $y$.  As a result, it is difficult to obtain the explicit forms of the super-momentum $H_r$ and the super-Hamiltonian $H$, such that
\begin{align}
{\cal L}= {\dot \Lambda}P_{\Lambda}+{\dot R}P_{R}-N H-N_r H_{r}
\end{align} 
in terms of the ADM variables. However, it is not necessary to do so at this stage.  Things are greatly simplified by using the generalized Misner-Sharp mass as a new canonical variable.  As we will show, the super-momentum and the super-Hamiltonian with the new set of canonical coordinates are the same as those in general relativity and then the boundary terms at spatial infinity can be easily derived.

\subsection{Generalized Misner-Sharp Mass as a Canonical Variable}
We introduce a new set of canonical variables $\{M,P_M;S, P_S\}$ defined in the same way as in general relativity, namely by using Equations \ref{eq:R}--\ref{eq:Pr1} but with Equation \ref{qlm-L} used for the mass function.  Then, we prove that $\{\Lambda,P_\Lambda;R, P_R\}$ and $\{M,P_M;S, P_S\}$ again satisfy the Liouville form, Equation \ref{Liouville}, with the following total variation and the total derivative terms. The derivation is presented in Appendix \ref{app:Liouville}.
\begin{align}
\eta:=&\frac{(n-2){\cal A}_{n-2}}{2\kappa_n^2}\biggl[\sum_{p=1}^{[n/2]}{\tilde \alpha}_{(p)}pR^{n-1-2p}\biggl(2y\Lambda-R' \ln\biggl|\frac{R'+y\Lambda}{R'-y\Lambda}\biggl|\biggl) \nonumber \\
&-\sum_{p=2}^{[n/2]}{\tilde \alpha}_{(p)}R^{n-1-2p} \sum_{w=1}^{p-1}\frac{p!}{w(p-1-w)!}\nonumber \\
& \times\sum_{j=0}^{w-1}\frac{2(-1)^{2w-1-j}{R'}^{2j}y^{2(w-j)+1}}{j!(w-1-j)![2(w-j)+1]\Lambda^{2j-1}}\biggl], \label{deta} \\
\zeta:=&\frac{(n-2){\cal A}_{n-2}}{2\kappa_n^2}\biggl[\sum_{p=1}^{[n/2]}{\tilde \alpha}_{(p)}pR^{n-1-2p}\ln\biggl|\frac{R'+y\Lambda}{R'-y\Lambda}\biggl|\nonumber \\
&+\sum_{p=2}^{[n/2]}{\tilde\alpha}_{(p)}R^{n-1-2p}\sum_{w=1}^{p-1}\frac{2p!(-1)^wyR'}{w!(p-1-w)!\Lambda}\nonumber \\
&\times\biggl\{F^{w-1}+\sum_{j=0}^{w-1}\frac{2j(w-1)!(-1)^{w-1-j}{R'}^{2j-2}y^{2(w-j)}}{j!(w-1-j)![2(w-j)+1]\Lambda^{2j-2}}\biggl\}\biggl]\delta R. \label{zeta'}
\end{align} 

In Appendix \ref{appendix3}, it is proven that Equation \ref{eq:GRfinal} still holds in full Lovelock gravity. This immediately implies that the Hamiltonian density in the equivalent two-dimensional theory takes the same form as that in general relativity, Equation \ref{eq:geometrodynamic hamiltonian}, where the definitions of the new Lagrange multipliers $N^M$ and $N^S$ are the same as those in general relativity (Equations \ref{NM} and \ref{NS}).  Finally, the Lagrangian density for the canonical coordinates $\{M,P_M;S,P_S\}$ can be again written as
\begin{align}
{\cal L}= P_M\dot{M} + P_S\dot{S}- N^M M'-N^S P_S \label{eq:geometrodynamic L2}
\end{align}
and the super-Hamiltonian and super-momentum constraints are the same as in the general relativistic case, Equation \ref{eq: HG}.
 
In comparison to the rather complicated starting point in Equation \ref{L_p}, this equivalent Lagrangian density is extremely simple and the physical meaning of the canonical variables are very clear.  Remarkably, the coupling constants $\alpha_{(p)}$ do not appear explicitly in Equation \ref{eq:geometrodynamic L2}.  They are in fact hidden in the definition of the mass function.  This makes it possible to treat any class of Lovelock gravity in a similar way to how we would treat general relativity.

\subsection{Fall-off Rate at Infinity and Boundary Terms}
In order to prove that the transformation from $\{\Lambda, P_\Lambda; R,P_R\}$ to \linebreak $\{M,P_M;S,P_S\}$ is canonical and well-defined, we have to discuss the asymptotic behaviour of the variables.  We adopt the same boundary conditions as in general relativity, Equations \ref{bc1}--\ref{bc4}.  With these conditions, one can verify that the generalized Misner-Sharp mass, Equation \ref{qlm-L} behaves near spacelike infinity as
\begin{align}
M \simeq \frac{(n-2)A_{n-2}{\tilde \alpha}_{(1)}\Lambda_1(t)}{\kappa_n^2}.
\end{align}  
This is the same as in general relativity and hence we set $\Lambda_1$ as in Equation \ref{bc7} (where ${\tilde \alpha}_{(1)}=1$) in order that $M\simeq M^\infty(t)$ at infinity.

It can then be shown that the leading terms of $P_\Lambda$, $P_R$, $\zeta$, $\eta$, $P_S$, $P_M$, $N^M$, and $N^S$ are the same as those in the general relativistic case under the boundary conditions of Equations \ref{bc1}--\ref{bc4}.  As a consequence, the proof carries over from general relativity and all the terms in the Liouville form of Equation \ref{Liouville} are well behaved at spacelike infinity.  This is sufficient to prove that the transformation from $\{\Lambda, P_\Lambda,; R,P_R\}$ to $\{M,P_M;S,P_S\}$ is indeed a well-defined canonical transformation and that the Hamiltonian $\int^\infty_{-\infty}dx(N^M M'+N^S P_S)$ is also finite.

We now have the following two-dimensional action with a new set of canonical variables;
\begin{align}
I_{LL}=\int dt\int dx(P_M\dot{M} + P_S\dot{S}- N^M M'-N^S P_S),
\end{align}
with the same asymptotic behaviour as in general relativity. The boundary term for the above action that makes the variational principle well defined is then also the same as in general relativity:
\begin{align}
\delta I_{LL}=\int dt\int dx \biggl(\mbox{dynamical terms}\biggl)+\int dt\biggl[N_\infty(t)\delta M^\infty(t)\biggl]_{x=-\infty}^{x=+\infty}.
\end{align} 

Given the above, we can now write down the super-Hamiltonian and super-momentum constraints for full Lovelock gravity. In terms of the geometrodynamical variables they are the same expressions as in general relativity:
\bea
H &=& \left(\frac{P_M y}{R'} -\frac{\Lambda}{R'}\right)M'
+ y P_s\, ,
\label{eq:super H} \\
H_r &=& P_M M' +P_S S'.
\label{eq:super Hr}
\eea
The expressions in terms of ADM variables are considerably more complicated and can in principle be obtained once again by substitution from Equations \ref{eq:R}, \ref{eq:PM} and \ref{eq:Pr1}, with $M$ defined by Equation \ref{qlm-L}.


\section{Adding Matter Fields}
\label{Adding Matter Fields1}
In this section, we introduce matter fields in the argument with the ADM variables discussed in the previous sections.  Here we write the super-momentum and super-Hamiltonian for gravity as $H_r^{\rm (G)}$ and $H^{\rm (G)}$ in order to distinguish from the total super-momentum and super-Hamiltonian including matter contributions.  The following argument is valid in full Lovelock gravity.

It can be shown from Equations \ref{defy}, \ref{eq:PM}, \ref{eq:R}, \ref{eq:Pr1} and \ref{eq:GRfinal} that the gravitational Hamiltonian $H_{\rm G}$ is given by:
\be
\label{HGADM}
H_{\rm G} = \int dx (NH^{\rm (G)} + N_rH_r^{\rm (G)}),
\ee
where
\begin{align}
\label{momconst1}
H_r^{\rm (G)} =& P_S S^\prime + P_M M^\prime = P_R R^\prime  - P_\Lambda^\prime \Lambda,\\
\label{Hconst1}
H^{\rm (G)} =& -\frac{\Lambda}{R^\prime}M^\prime  + \frac{y}{R^\prime}H_r^{(G)}.
\end{align}
We have used Equations \ref{M'2} to derive \ref{Hconst1}.  Here it is important to think of $M$ and $y$ as a function of the ADM variables.  Since Equations \ref{M-F}, \ref{defF} and \ref{PLambda3} show that $M$ and $y$ are not a functions of $N$ or $N_r$, we can see that the Hamiltonian density is the sum of Lagrange multipliers times constraints.

\subsection{Massless Scalar Field}
\label{Scalar field}
First we consider a massless scalar field $\psi$ as a matter field, for which the matter part of the action, $I_{\rm matter}=I_\psi$, is given by 

\begin{align}
I_\psi = -\frac{1}{2}\int d^{n}x \sqrt{-g} (\nabla \psi)^2.
\end{align}
The equivalent two-dimensional action in the symmetric spacetime under consideration is given by 
\begin{align}
\label{scalaraction2}
I_\psi &= -\frac{{\cal A}_{n-2}}{2}\int d^2{\bar y} \sqrt{-\g} R^{n-2} (D \psi)^2 \\ \nonumber
& = -\frac{{\cal A}_{n-2}}{2}\int dxdt \frac{\Lambda R^{n-2} }{N} \left( -\psid^2 + 2N_r\psip \psid + (N^2\Lambda^{-2}-N_r^2) \psip^2 \right).
\end{align}
This gives the momentum conjugate $\Pis$ to $\psi$ as
\be
\label{Pip}
\Pis =  \frac{{\cal A}_{n-2}\Lambda R^{n-2}}{N} \left( \psid - N_r \psip \right),
\ee
with which we can write the matter action as
\begin{align}
\label{scalaraction3}
I_\psi=& \int dxdt \psid \Pis \nonumber \\
&- \int dxdt N \left[ \frac{1}{2\Lambda} \left( \frac{\Pis^2}{{\cal A}_{n-2} R^{n-2}} +{\cal A}_{n-2}R^{n-2} \psip^2 \right) + \Pis \psip\frac{N_r}{N}  \right].
\end{align}

Equation \ref{scalaraction2} does not contain any derivatives of the metric or $R$, which means that adding the scalar action to the gravitational action of Equation \ref{eq:geometrodynamic hamiltonian} does not change $P_\Lambda$ or $P_R$.  This allows us to write the total Hamiltonian as the sum of the gravitational and matter parts.  Using Equations \ref{HGADM}, \ref{momconst1}, \ref{Hconst1} and \ref{scalaraction3}, we obtain the total Hamiltonian $H_{\rm total} $ as
\begin{align}
\label{H8}
H_{\rm total} = & \int dx N\Bigg[-\frac{\Lambda}{R^\prime}M^\prime  + \frac{y}{R^\prime}H_r^{(G)}  + \frac{N_r}{N} H_r^{(G)} \nonumber \\
& +\frac{1}{2\Lambda} \left( \frac{P_\psi^2}{{\cal A}_{n-2} R^{n-2}} + {\cal A}_{n-2} R^{n-2} \psip^2 \right) +P_\psi \psip\frac{N_r}{N}   \Bigg] \none
= & \int dx \biggl[ N(H^{\rm (G)}+H^{\rm (M)}) + N_r(H_r^{\rm (G)}+H_r^{\rm (M)}) \biggl],
\end{align}
where $y$ is a function of the phase space variables, $\Lambda$, $P_\Lambda$ and $R$ via Equation \ref{PLambda3}.  The super-Hamiltonian $H^{\rm (M)}$ and super-momentum $H_r^{\rm (M)}$ for $\psi$ are given by
\begin{align}
\label{HconstM}
H^{\rm (M)} =& \frac{1}{2\Lambda} \left( \frac{P_\psi^2}{{\cal A}_{n-2} R^{n-2}} +{\cal A}_{n-2}R^{n-2} \psip^2 \right),\\
\label{momconstM}
H_r^{\rm (M)} =& P_\psi \psip.
\end{align}
Since our theory is diffeomorphism invariant the Poisson bracket of the Hamiltonian constraint, $H = H^{\rm (G)} + H^{\rm (M)}$ with the total momentum constraint, $H_r = H_r^{\rm (G)} + H_r^{\rm (M)}$ must be weakly equal to zero.  This implies that both constraints are first class.  

Because there are two first class constraints, there are two gauge choices to pick.  We choose our first gauge as
\be
\label{gauge1}
\chi := R - x \approx 0.
\ee
Any gauge choice must be second class with at least one of the first class constraints.  Equation \ref{gauge1} doesn't commute with $H_r^{(G)}$, from Equation \ref{momconst1} which satisfies this condition.  Our gauge choice above forces the spatial coordinate to be the areal radius which means that $R$ is no longer a phase space variable, it is now a coordinate.  In order to insist that $\chi$ is satisfied at every time slice, we must insist that $\dot{\chi}=\{\chi,H_{total}\} \approx 0$, which requires that $N_r/N +y/R^\prime \approx 0$.  We use this relation to write one Lagrange multiplier in terms of the other.  This leaves us with one Lagrange multiplier which reflects the fact that there is only one gauge fix left to choose.

We can now plug the gauge choice given by Equation \ref{gauge1} and its consistency condition into the Hamiltonian as long as we use Dirac brackets to evaluate the equations of motion in the end \cite{Matschull1996}.  Note that the remaining phase space variables, $\Lambda$, $P_\Lambda$, $\psi$ and $P_\psi$, all commute with $\chi$ and so the Poisson bracket is the same as the Dirac bracket.  Plugging $\chi = \dot{\chi} = 0$ into Equation \ref{H8} gives
\be
\label{H8b}
H_{\rm total} = \int dR N \Biggl[-\Lambda M^\prime + \frac{1}{2\Lambda} \left( \frac{\Pis^2}{{\cal A}_{n-2}R^{n-2}} + {\cal A}_{n-2}R^{n-2} \psip^2 \right) - y \Pis \psip \Biggr].
\ee
In the last term we replaced $N_r/N$ by $-y$ as required.  Since the mass Equation \ref{qlm-L} is written as
\begin{align}
\label{Massfunc}
M=\frac{(n-2){\cal A}_{n-2}}{2\kappa_n^2}\sum_{p=0}^{[n/2]} \tilde{\alpha}_{(p)}R^{n-1-2p}\left(1-\Lambda^{-2} + y^2\right)^p,
\end{align}
we can write $y$($=N_r/N$) in terms of the mass function.  For this reason we leave the factor of $N_r/N$ in the Hamiltonian with the understanding that it is the solution to Equation \ref{Massfunc}.

For our second gauge choice we choose
\be
\label{gauge2}
\xi := \Lambda - 1\approx 0
\ee
which is second class with the remaining constraint (the square brackets in Equation \ref{H8b}).  By the same reasoning used for the first gauge choice we can set $\xi$ strongly to zero (namely since $\Lambda$ commutes with $\psi$ and $P_\psi$) which gives the Hamiltonian
\be
\label{H8d}
H_{\rm total} = \int dR N \Biggl[ -M^\prime + \frac{1}{2} \left( \frac{\Pis^2}{{\cal A}_{n-2} R^{n-2}} + {\cal A}_{n-2} R^{n-2} \psip^2 \right) + \Pis \psip  \frac{N_r}{N}  \Biggr]
\ee
and the mass function
\begin{align}
\label{Nsig}
M= \frac{(n-2){\cal A}_{n-2}}{2\kappa_n^2}\sum_{p=0}^{[n/2]} \tilde{\alpha}_{(p)}R^{n-1-2p}\left(\frac{N_r}{N}\right)^{2p}.
\end{align}
To see the significance of this gauge choice, notice from Equation \ref{qlm-L} that $g^{11} \to 1-2\kappa_n^2M/[(n-2){\ma A_{(n-2)}}\tilde{\alpha}_{(1)}R^{n-3}]$ in the general relativistic case when we strongly set $\xi$ and $\chi$ to zero.  This gives the metric in the non-static version of Painlev\'{e}-Gullstrand coordinates:
\begin{align}
ds_{(2)}^2=& -N^2 \left( 1-\frac{2\kappa_n^2M}{(n-2){\ma A_{(n-2)}}\tilde{\alpha}_{(1)}R^{n-3}} \right)  dT_{PG}^2 \nonumber \\
& + 2 N \sqrt{\frac{2\kappa_n^2M}{(n-2){\ma A_{(n-2)}}\tilde{\alpha}_{(1)}R^{n-3}}} dT_{PG}dR + dR^2.
\end{align}

To ensure that the second gauge condition is conserved in time we must insist that $d(\Lambda - 1)/dt = \{\Lambda - 1, H\} = 0 \rightarrow \delta H/\delta P_\Lambda = 0$.  Although we have chosen to write $N_r/N$ in terms of the mass function, it can also be written in terms of $P_\Lambda$.  All of the $P_\Lambda$ dependence in the Hamiltonian is in $N_r/N$.  Therefore we can write
\begin{align}
\frac{\delta H_{\rm total}}{\delta P_\Lambda} & = \frac{\delta}{\delta P_\Lambda} \int dR \left( N^\prime M + N P_\psi \psip  \frac{N_r}{N}  \right) \nonumber \\ 
& = N^\prime \frac{\partial M}{\partial (N_r/N)} \frac{\partial (N_r/N)}{\partial P_\Lambda} + N P_\psi \psip\frac{\partial (N_r/N)}{\partial P_\Lambda},  \label{consistency conditon 2}
\end{align}
from which the consistency condition is given as
\be
\label{sigma1}
N^\prime \frac{\partial M}{\partial (N_r/N)} + N P_\psi \psip = 0,
\ee
where it is understood that we use Equation \ref{Nsig} to find $\partial M/\partial (N_r/N)$ and write $N_r/N$ in terms of the mass function $M$.  Notice that the actual relation between $N_r/N$ and $P_\Lambda$ is not needed.

Using Hamilton's equations and Equation \ref{H8d}, we find
\begin{align}
\label{psidot}
\dot{\psi} =& N \left( \frac{\Pis}{{\cal A}_{n-2} R^{n-2}} + \psip\frac{N_r}{N}  \right),\\
\label{Pisdot}
\dot{P}_\psi =& \left[ N \left( {\cal A}_{n-2} R^{n-2} \psip +  \Pis \frac{N_r}{N}  \right) \right]^\prime,
\end{align}
where the dot now represents differentiation with respect to the PG time.  These equations, along with the consistency conditions (Equations \ref{Nsig} and \ref{sigma1}) and the Hamiltonian constraint
\be
\label{Cconstraint}
-M^\prime + \frac{1}{2} \left( \frac{\Pis^2}{{\cal A}_{n-2}R^{n-2}} + {\cal A}_{n-2} R^{n-2} \psip^2 \right) + \Pis \psip  \frac{N_r}{N}= 0,
\ee
determine the evolution of a collapsing scalar field.

\subsection{Charged Scalar Field}
\label{sec:CSF}
In this subsection, we consider a Electromagnetic field, $A_\mu$, coupled to a charged complex massless scalar field $\psi = (\psi_1 + i\psi_2)/\sqrt{2}$, where $\psi_1$ and $\psi_2$ are real functions.  We write the action for this matter as $I_{\rm matter}=I_{\rm EM}$:
\be
\label{EMaction1}
I_{\rm EM}= \int d^nx \sqrt{-g}\left[-\left( \partial^\mu + ieA^\mu \right) \psi^* \left( \partial_\mu - ieA_\mu \right)  \psi- \frac{1}{4} F^{\mu \nu} F_{\mu \nu} \right],
\ee
where $e$ is the charge and the Faraday tensor $F_{\mu \nu}$ is defined in terms of $A_\mu$ as $F_{\mu \nu}=\partial_\mu A_\nu-\partial_\nu A_\mu$.  Under the symmetry assumption in the present paper both for gravity and matter, the equivalent two-dimensional action is given by 
\begin{align}
\label{EMaction2}
I_{\rm EM}=&{\cal A}_{n-2}\int dtdx \sqrt{-g_{(2)}} R^{n-2} \nonumber \\
& \times \left[-\left( \partial^B + ieA^B \right) \psi^* \left( \partial_B - ieA_B \right)  \psi -\frac{1}{4} F^{AB} F_{AB}\right].
\end{align}
Adopting the ADM coordinates, we obtain
\begin{align}
\label{EMaction3}
I_{\rm EM} =&\frac{{\cal A}_{n-2}}{2} \int dtdx \frac{R^{n-2}\Lambda}{N} \nonumber \\
&\times\Bigg[(\dot{\psi}_1^2 + \dot{\psi}_2^2) - 2N_r(\dot{\psi_1}\psi_1^\prime + \dot{\psi_2}\psi_2^\prime) +(N_r^2-N^2\Lambda^{-2})(\psi_1^{\prime 2} + \psi_2^{\prime 2}) \nonumber \\
& -2e \biggl\{ (A_0-N_rA_1)(\dot{\psi}_2 \psi_1 - \dot{\psi}_1 \psi_2) \nonumber \\
& - \biggl(N_r(A_0-N_rA_1) + N^2\Lambda^{-2}A_1\biggl)(\psi_2^\prime \psi_1 - \psi_1^\prime \psi_2) \biggl\}\nonumber \\ 
& +e^2 \biggl((A_0-N_rA_1)^2-N^2\Lambda^{-2}A_1^2\biggl)(\psi_1^2 + \psi_2^2) + \Lambda^{-2} (\dot{A_1} - A_0^\prime)^2 \Bigg],
\end{align}
where $A_\mu dx^\mu=A_0(t,x)dt+A_1(t,x)dx$.
From the above action we find the conjugate momenta:
\begin{align}
\label{Ppsi1}
P_{\psi 1} =& \frac{{\cal A}_{n-2}R^{n-2}\Lambda}{N}\left[ \dot{\psi}_1 - N_r \psi_1^\prime +e(A_0 - N_r A_1) \psi_2 \right],\\
\label{Ppsi2}
P_{\psi 2} =&\frac{{\cal A}_{n-2}R^{n-2}\Lambda}{N}\left[ \dot{\psi}_2 - N_r \psi_2^\prime -e(A_0 - N_r A_1) \psi_1 \right],\\
\label{PA0}
P_{A0} =& 0,\\
\label{PA1}
P_{A1} =& \frac{{\cal A}_{n-2}R^{n-2}(\dot{A_1} - A_0^\prime)}{N\Lambda},
\end{align}
which give the Hamiltonian for the present matter field:
\begin{align}
\label{HEM1}
H_{\rm EM}=&\int dx \Bigg[ \frac{N}{2{\cal A}_{n-2}\Lambda R^{n-2}}(P_{\psi 1}^2 + P_{\psi 2}^2) + e(A_0-N_rA_1)(P_{\psi 2} \psi_1 - P_{\psi 1} \psi_2) \nonumber \\
& + N_r(P_{\psi 1} \psi_1^\prime + P_{\psi 2} \psi_2^\prime) + \frac{NR^{n-2}}{2{\cal A}_{n-2}\Lambda}\biggl\{ (eA_1\psi_1-\psi_2^\prime)^2+(eA_1\psi_2+\psi_1^\prime)^2 \biggl\} \nonumber \\
& + \frac{N\Lambda}{2{\cal A}_{n-2}R^{n-2}}P_{A1}^2 + P_{A1}A_0^\prime \Bigg].
\end{align}
Since the action (\ref{EMaction3}) contains no derivatives of the metric or $R$, the addition of $I_{\rm EM}$ to the gravitational action does not alter the Hamiltonian analysis and allows us to write the total Hamiltonian as
\be
\label{HEMtotal1}
H_{\rm total} = \int dx \biggl[N(H^{\rm (G)} + H^{\rm (EM)}) + N_r(H_r^{\rm (G)} + H_r^{\rm (EM)}) + A_0H_{A0}^{\rm (EM)} \biggl],
\ee
where $H^{\rm (G)}$ and $H_r^{\rm (G)}$ are given by Equations \ref{Hconst1} and \ref{momconst1} and $H^{\rm (EM)}$, $H_r^{\rm (EM)}$ and $H_{A0}^{\rm (EM)}$ are given by
\begin{align}
H^{\rm (EM)}=& \frac{P_{\psi 1}^2 + P_{\psi 2}^2}{2{\cal A}_{n-2}\Lambda R^{n-2}} +\frac{{\cal A}_{n-2}R^{n-2}}{2\Lambda}\biggl[ (eA_1\psi_1-\psi_2^\prime)^2+(eA_1\psi_2+\psi_1^\prime)^2 \biggl] \nonumber \\
&+ \frac{\Lambda P_{A1}^2}{2 {\cal A}_{n-2}R^{n-2}},\label{HEM}\\
\label{HrEM}
H_r^{\rm (EM)}=& -eA_1(P_{\psi 2} \psi_1 - P_{\psi 1} \psi_2) + (P_{\psi 1} \psi_1^\prime + P_{\psi 2} \psi_2^\prime),\\
\label{HA0EM}
H_{A0}^{\rm (EM)}=&e(P_{\psi 2} \psi_1 - P_{\psi 1} \psi_2)-P_{A1}^\prime,
\end{align}
where we used integration by parts and the asymptotic condition $P_{A1} A_0\to 0$ at infinity to derive Equation \ref{HA0EM}.  The consistency condition on Equation \ref{PA0} is $\{P_{A0},H_{\rm total}\} = 0$, which gives $e(P_{\psi 2} \psi_1 - P_{\psi 1} \psi_2)-P_{A1}^\prime=0$.  This condition is already added into the Hamiltonian with $A_0$ as its Lagrange multiplier.  Since $P_{A0}$ is weakly equal to zero we can use the equation of motion for $P_{A0}$ to show that $H_{A0}^{\rm (EM)}$ is weakly equal to zero and is, therefore, a constraint in the same way as the constraints multiplying $N$ and $N_r$.

This Hamiltonian is composed of three first class constraints which means that there are three gauge choices to make.  Our first two gauge choices will be the same as in Section \ref{Scalar field}.  Using similar reasoning we can write the Hamiltonian as
\begin{align}
H_{\rm total} =& \int dR \biggl[ N\biggl\{ -M^\prime +\frac{P_{\psi 1}^2 + P_{\psi 2}^2}{2{\cal A}_{n-2}R^{n-2}}\nonumber \\
& +\frac{{\cal A}_{n-2}R^{n-2}}{2}\biggl( (eA_1\psi_1-\psi_2^\prime)^2 + (eA_1\psi_2+\psi_1^\prime)^2 \biggl) +\frac{P_{A1}^2}{2{\cal A}_{n-2}R^{n-2}}\nonumber \\
& + \frac{N_r}{N} \biggl( -eA_1(P_{\psi 2} \psi_1 - P_{\psi 1} \psi_2) + (P_{\psi 1} \psi_1^\prime + P_{\psi 2} \psi_2^\prime) \biggl) \biggl\} \nonumber \\ 
& + A_0\biggl(e(P_{\psi 2} \psi_1 - P_{\psi 1} \psi_2)-P_{A1}^\prime\biggl)\biggl]. \label{HEMtotal1b}
\end{align}
Just as in Section \ref{Scalar field} the consistency condition on the first gauge fix requires us to write $N_r/N$ as a function of $M$ using Equation \ref{Nsig}.  The consistency condition on the second gauge choice, analogous to Equation \ref{sigma1}, is given by
\be
\label{EM consistency condition 2}
N^\prime \frac{\partial M}{\partial (N_r/N)} + N \biggl( -eA_1(P_{\psi 2} \psi_1 - P_{\psi 1} \psi_2) + (P_{\psi 1} \psi_1^\prime + P_{\psi 2} \psi_2^\prime) \biggl) = 0.
\ee
For our third gauge we choose
\be
\label{A0gauge}
\epsilon:=A_1 \approx 0,
\ee
which is the coulomb gauge with the constant, $A_1$ chosen to be zero.  This condition is second class with the remaining constraint,
\be
\label{EMConstraint}
e(P_{\psi 2} \psi_1 - P_{\psi 1} \psi_2)-P_{A1}^\prime \approx 0,
\ee
as required for a valid gauge choice.  Note that Equations \ref{A0gauge} and \ref{EMConstraint} removes $A_1$ and its conjugate momentum $P_{A1}$ from the set of phase space variables.  We can therefore set $\epsilon$ strongly to zero in the Hamiltonian as we did for the first two gauge choices.  This gives the following Hamiltonian:
\begin{align}
H_{\rm total} =& \int dR \biggl[ N\biggl\{ -M^\prime +\frac{P_{\psi 1}^2 + P_{\psi 2}^2}{2{\cal A}_{n-2}R^{n-2}} +\frac{{\cal A}_{n-2}R^{n-2}}{2}(\psi_2^{\prime2} + \psi_1^{\prime2} ) \nonumber \\
&+\frac{P_{A1}^2}{2{\cal A}_{n-2}R^{n-2}} + \frac{N_r}{N} (P_{\psi 1} \psi_1^\prime + P_{\psi 2} \psi_2^\prime) \biggl\} \nonumber \\
&+ A_0\biggl(e(P_{\psi 2} \psi_1 - P_{\psi 1} \psi_2)-P_{A1}^\prime\biggl)\biggl], \label{HEMtotal2}
\end{align}
where it is understood that $P_{A1}$ is the solution of Equation \ref{EMConstraint}.  The consistency condition on Equation \ref{A0gauge} is given by
\begin{align}
\{\epsilon , H_{\rm total} \} \approx 0 &\to A_0^\prime + \frac{NP_{A1}}{{\cal A}_{n-2}R^{n-2}} \approx 0 \nonumber \\
&\to A_0^\prime \approx  -\frac{eN}{{\cal A}_{n-2}R^{n-2}} \int dR (P_{\psi 2} \psi_1 - P_{\psi 1} \psi_2),\label{EM consistency condition 3}
\end{align}
which puts a condition on the final Lagrange multiplier and must be satisfied at every time slice.  This is the last consistency condition on $\epsilon$.

With the fully gauge fixed Hamiltonian of Equation \ref{HEMtotal2} we may write down Hamilton's equations of motion in terms of the remaining phase space variables, $\psi_1$, $P_{\psi1}$, $\psi_2$ and $P_{\psi2}$.  The equations of motion are given by
\begin{align}
\label{EMpsi1EOM}
\dot{\psi_1} =& N\left(\frac{P_{\psi1}}{{\cal A}_{n-2}R^{n-2}} + \frac{N_r}{N} \psi_1^\prime\right) - eA_0\psi_2,\\
\label{EMpsi2EOM}
\dot{\psi_2} =& N\left(\frac{P_{\psi2}}{{\cal A}_{n-2}R^{n-2}} + \frac{N_r}{N} \psi_2^\prime\right) + eA_0\psi_1,\\
\label{EMP1EOM}
\dot{P}_{\psi1} =& \left[N \left({\cal A}_{n-2}R^{n-2} \psi_1^\prime + \frac{N_r}{N} P_{\psi1} \right) \right]^\prime - eA_0 P_{\psi2},\\
\label{EMP2EOM}
\dot{P}_{\psi2} =& \left[N \left({\cal A}_{n-2}R^{n-2} \psi_2^\prime +\frac{N_r}{N} P_{\psi2} \right) \right]^\prime + e A_0 P_{\psi1}.
\end{align}
As in the massless scalar field case, the dot represents differentiation with respect to the PG time.  It must be remembered that at every time slice the equations of motion must be supplemented by the consistency conditions of Equations \ref{Nsig}, \ref{EM consistency condition 2} and \ref{EM consistency condition 3}, as well as the Hamiltonian constraint:
\begin{align}
\label{EMNconstraint}
&-M^\prime + \frac{P_{\psi 1}^2 + P_{\psi 2}^2}{2{\cal A}_{n-2}R^{n-2}} + \frac{{\cal A}_{n-2}R^{n-2}}{2}(\psi_2^{\prime2} + \psi_1^{\prime2})\nonumber \\
& + \frac{P_{A1}^2}{2{\cal A}_{n-2}R^{n-2}} + \frac{N_r}{N} (P_{\psi 1} \psi_1^\prime + P_{\psi 2} \psi_2^\prime) = 0,
\end{align}
where it is understood that $P_{A1}$ is the solution of Equation \ref{EMConstraint}.

\chapter[Higher Dimensional Choptuik Scaling]{Higher Dimensional Choptuik Scaling in General Relativity}
\label{ch:HDCS}


So far in this thesis we have been concerned with developing the Hamiltonian analysis of Lovelock gravity.  One set of important results of this analysis are the equations of motion for a collapsing, massless scalar field given by Equations \ref{psidot} and \ref{Pisdot} and the constraints, Equations \ref{Nsig}, \ref{sigma1} and \ref{Cconstraint}.  It is useful to use these equations to numerically investigate black hole formation since they are in the Painlev\'{e}-Gullstrand gauge which are regular at horizon formation.  These equations of motion also allow us to investigate collapse in any number of dimensions, by changing $n$ and with any number of Lovelock curvature terms, by changing $p$.  The effects of higher curvature terms such as the Gauss-Bonnet term would only be felt in regions of high curvature.  For this reason it is of particular interest to investigate the formation of small black holes, since the horizon forms near the singularity, via the phenomenon of Choptuik scaling discussed in Section \ref{sec:Choptuik}.  The rest of this thesis is dedicated to the numerical investigation of Choptuik scaling.  In this chapter we look at mass scaling in higher dimensional general relativity and explore the effect of higher order curvature terms in the next chapter.  This chapter is based on our paper \cite{Taves2011}.

Recall from Equation \ref{GundlachCS} that near criticality the scaling relation is given by

\be
\ln(M_{BH}) = \gamma \ln|p-p^*| + f(\ln|p-p^*|)
\label{Chop}
\ee
where $p$ is a parameter in the initial data with critical value $p^*$, $\gamma$ is the critical exponent and $f$ is a periodic function with period, $T$. $s\gamma$ and $T$ are universal in the sense that they are independent of the type of initial data and the quantity being measured ($s$ is the power of the length dimensions of the quantity being measured).  They do depend on the type of matter that is collapsing, however. The specific form of the oscillating function, $f$, is known not to be universal. The most common quantities used to test this relationship are the mass of the final black hole (or equivalently the radius of the event horizon) and the maximum value of the curvature at the origin in sub-critical collapse.  In this chapter we work with the mass of the black hole at formation, $M_{BH}$ but we will consider subcritical cases in the next chapter.  Previous calculations for the collapse of a spherically symmetric massless scalar field in four spacetime dimensions in Schwarzschild, double null and Painlev\'{e}-Gullstrand (PG) coordinates have all given $\gamma \approx 0.37$ and $T \approx 4.6$ \cite{Choptuik1993,Garfinkle1995,Gundlach1997,Hod1997,Ziprick2009a,Bland2005,Ziprick2009c,Hamade1996}.  
 
In Schwarzschild and double null coordinates the form of the periodic function, $f$, is well approximated by a small--amplitude sine function.  By contrast, in PG coordinates the periodic function in four dimensions showed distinctive cusps \cite{Ziprick2009c,Ziprick2009a}.  This difference in the form of $f$ in PG coordinates is perhaps surprising, but not inconsistent. The PG calculation did not measure the final black hole mass after all matter had fallen through the horizon, which is independent of slicing and would exhibit the same behaviour in all coordinates. Instead, what was plotted was the radius of the apparent horizon on formation, a quantity that does depend on the slicing.  It is nonetheless a geometrical quantity that exhibits Choptuik scaling, as shown in \cite{Ziprick2009c,Ziprick2009a}.  What is less clear is whether the cusp-like features of the scaling function is a peculiarity of PG coordinates in 4 dimensions or generic in some sense.

Most previous calculations of Choptuik scaling have involved four dimensional black holes. The first higher dimensional analysis was done by Garfinkle and Duncan \cite{Garfinkle1999}. More recently, a program was initiated \cite{Birukou2002} whose purpose was to calculate the critical exponent and echoing period for spherical collapse in arbitrary spacetime dimensions.  Preliminary results were obtained in 5 and 6 spacetime dimensions.  Subsequently, accurate results were obtained up to $n=14$ \cite{Bland2005,Bland2007}.  These results provide strong evidence that the critical exponent was a monotonic function of $n$ that converged asymptotically to 1/2. Also around that time, Oren and Sorkin\cite{Sorkin2005} produced results for the critical exponent up to $n=11$ which they used to speculate that the critical exponent was not a monotonic function, but instead achieved a maximum near $n=10$. Moreover, it was argued on the basis of the behaviour of black string solutions that a critical dimension near $n=10$ might not be unexpected. It should be noted that both sets of calculations were done using double null coordinates, using different parametrizations of the fields.

The higher dimensional calculations referred to above were done in double null coordinates, whereas the purpose of the present chapter is to look at spherically symmetric scalar field collapse in higher dimensions using PG coordinates. It is perhaps worth highlighting the relative merits of the two different sets of coordinates. The main advantage of double null coordinates for studying critical behaviour has to do with the fine spatial resolution that is possible due to the convergence of outgoing null rays near horizon formation. However, this same convergence makes it difficult, if not impossible to allow numerical simulations to run all the way up to horizon formation (see numerics section of \cite{Garfinkle1995}).  For this reason the mass/horizon radius of the black hole in null coordinates must be approximated by choosing in advance how close one wishes to get to horizon formation.  This problem is also encountered in Schwarzschild coordinates. PG coordinates provide spatial slices that are regular across future horizons, so that the code can be run up to and even beyond horizon formation. Away from criticality, this allows one to map out the structure of the trapping horizon as done in \cite{Ziprick2009c,Ziprick2009a}. Near criticality, we can precisely (to numerical accuracy) determine the radius of the horizon on formation. Note that this quantity is different from the mass of the final black hole. It marks the location that the PG slicing intersects the trapping horizon and therefore verifies the scaling relationship in a different geometrical quantity thereby providing independent measures of the critical exponent and echoing scale. This also explains why a different oscillatory function is possible. The main disadvantage of PG coordinates is that one does not get an automatic improvement in the spatial resolution from the convergence of null rays. 


As with previous calculations, the resulting numerics get more difficult as the number of spacetime dimensions is increased. We obtain reliable results in 4, 5, 6 and 7 dimensions for the scaling law obeyed by the areal radius of the apparent horizon on formation.  We investigate the periodic function, $f$, in this scaling relation and confirm the cusp-like nature in higher dimensions.  In addition, the higher dimensional critical exponents and periods are calculated and compared to previous work.  Our result in 7 dimensions is consistent within error to that of \cite{Bland2005} and inconsistent with \cite{Sorkin2005}.

\section{Methodology and Results}
\label{HDCSResults}

\subsection{Computational Details}

The equations of motion and constraints that we use are the general relativistic (ie $p=1$) versions of Equations \ref{psidot}, \ref{Pisdot}, \ref{Nsig}, \ref{sigma1} and \ref{Cconstraint} with the cosmological constant set to zero.  These are given by Equations \ref{psidotGR}, \ref{PisdotGR}, \ref{sigma1GR} and \ref{HconstraintGR} and are rewritten below for ease of reference.

\begin{align}
\dot{\psi} =& N \left( \frac{\Pis}{{\cal A}_{n-2} R^{n-2}} + \psip \sqrt{\frac{2\kappa_n^2 M}{(n-2){\cal A}_{n-2}R^{n-3}} } \right)\nonumber \\
\dot{P}_\psi =& \left[ N \left( {\cal A}_{n-2} R^{n-2} \psip +  \Pis \sqrt{\frac{2\kappa_n^2 M}{(n-2){\cal A}_{n-2}R^{n-3}} }  \right) \right]^\prime \nonumber
\end{align}

\begin{align}
N^\prime \sqrt{\frac{2(n-2){\ma A_{(n-2)}}R^{n-3}M}{\kappa_n^2}}  + N P_\psi \psip = 0 \nonumber
\end{align}

\begin{align}
& -M^\prime + \frac{1}{2} \left( \frac{\Pis^2}{{\cal A}_{n-2} R^{n-2}} + {\cal A}_{n-2} R^{n-2} \psip^2 \right) \nonumber \\
& + \Pis \psip  \sqrt{\frac{2\kappa_n^2 M}{(n-2){\cal A}_{n-2}R^{n-3}} } = 0 \nonumber
\end{align}

In order to solve the equations of motion we used the initial conditions:  

\begin{eqnarray}
\psi &=& A R^2 \exp \left[ -\left( \frac{R-R_0}{B} \right)^2 \right] \nonumber \\
P_\psi &=& 0
\label{eq:initialpsi}
\end{eqnarray}
where $A$, $B$ and $R_0$ are parameters that can in principle be varied to study mass scaling. These initial conditions are the same as those used in \cite{Ziprick2009c,Ziprick2009a}. We verified universality of our results by varying the amplitude, $A$, and the width, $B$, of the initial pulse. 

The equations of motion were solved using a forth order Runge-Kutta scheme with derivatives being calculated using the finite difference method.

In order to maintain stability we used an adaptive time step, $\Delta T_{PG}(T_{PG})$
\be
\Delta T_{PG}(T_{PG}) = min_R \left[ \left(\frac{dR}{dT_{PG}}\right)^{-1} \Delta R(R) \right],
\label{timespacing}
\ee
where $\Delta R(R)$ is the spacing of the spatial lattice and $\frac{dR}{dT_{PG}}$ is the local speed of an ingoing null geodesic.

Much of the interesting behaviour of the collapse near criticality occurs near the origin.  This requires close spacing of the spatial mesh near $R=0$.  It is not computationally realistic, however, to use this spacing along the entire spatial slice, which needs to be long enough so that none of the mass leaves during the simulation. The use of close spacing for the entire slice would dramatically increase the simulation time.  For this reason the spacing near the origin was set to $10^{-5}$ and then smoothly increased to $10^{-2}$ over the first 100 of the 1200 total lattice points.

\subsection{Results}

To find Choptuik's mass scaling relation we first found the critical values, $A^*$ and $B^*$, using a bisection method. 
We then varied the parameters $A$ and $B$ in our initial data function, Equation \ref{eq:initialpsi}, and found the mass of the black hole at formation, $M_{BH}$, as a function of the initial parameters.  An apparent horizon forms when an outgoing null geodesic becomes momentarily stationary. For equations of motion in PG gauge this is signalled by the condition
\be
(DR)^2=0 \rightarrow 2GM(R_{AH}) - \frac{(n-2)^2R_{AH}^{n-3}}{8(n-3)} =0,
\label{AHGR}
\ee
where $R_{AH}$ is the areal radius of the horizon on formation. The mass of the black hole at horizon formation is then given by the value of the mass function $M(R_{AH})$ at that point. We emphasize again that this is not the ADM mass of the final static black hole that is left behind once all the matter has fallen through the horizon. 

\begin{figure}[ht!]
\centering
\subfigure[4 Dimensions]{
\includegraphics[width=0.4\linewidth]{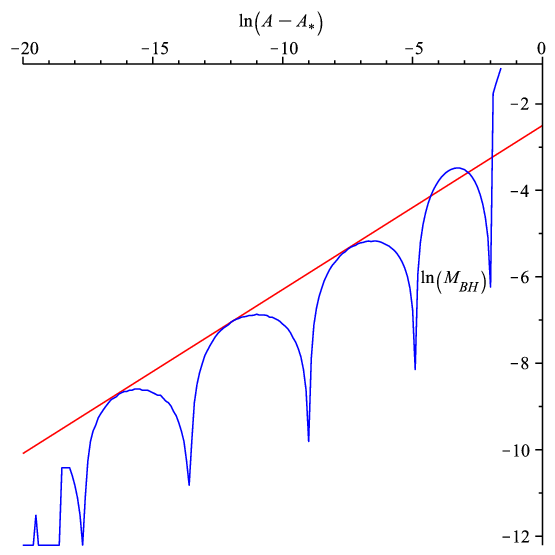}
\label{4DChop}
}
\hspace{0.25in}
\subfigure[5 Dimensions]{
\includegraphics[width=0.4\linewidth]{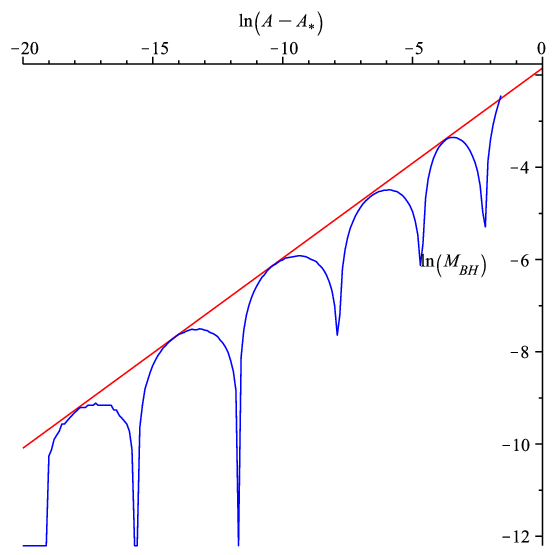}
\label{5DChop}
}
\subfigure[6 Dimensions]{
\includegraphics[width=0.4\linewidth]{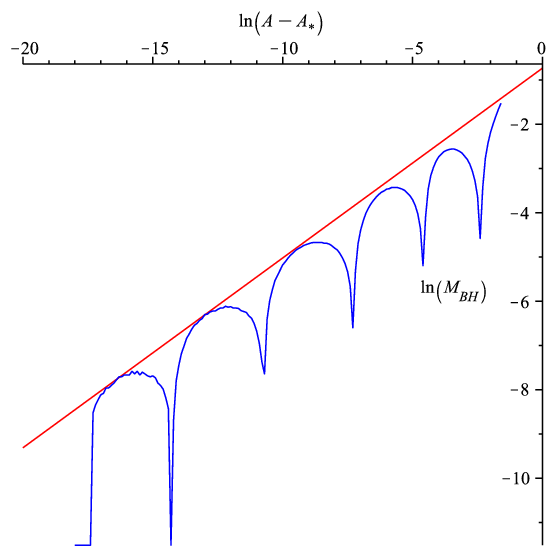}
\label{6DChop}
}
\hspace{0.25in}
\subfigure[7 Dimensions]{
\includegraphics[width=0.4\linewidth]{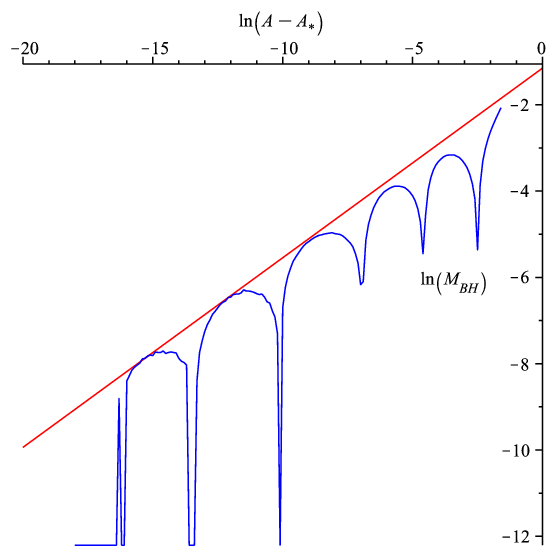}
\label{7DChop}
}
\caption[Mass scaling of $A$ parameter]{Mass Scaling in 4, 5, 6 and 7 dimensions with the $A$ parameter varied}
\label{mass scaling plots}
\end{figure}

The mass scaling plots can be seen in Figure \ref{mass scaling plots}.  A straight line, which osculates the curve, has been plotted with the data to illustrate the linear term in the mass scaling relation, Equation \ref{Chop}.  Notice that these plots confirm the existence in 5, 6 and 7 dimensions of the cusps in the periodic function that were originally noted in four dimensions in \cite{Ziprick2009c,Ziprick2009a}.  It is also important to note from Figure \ref{mass scaling plots} that as we move away from the near critical region (i.e. as $A-A^*$ gets large) the cusps systematically move away from the critical straight line.  In 4 dimensions the results move above the critical line, in five dimensions they more or less stay on the straight line, but in higher dimensions the cusps move downward. This illustrates graphically that in higher dimensions the critical exponent will be underestimated if one is not sufficiently close to criticality.  

Points in the data that were close to the straight lines ($< 2 \%$) of Figure \ref{mass scaling plots} were chosen and were used to find the critical exponent using linear regression.  An estimate of error was found by slightly varying which points were chosen to calculate the slope.  The period of $f$ in Equation \ref{Chop} was calculated by measuring the distance between the cusps of the plots.  A measure of error for these cusps was taken as the largest deviation from the average.  All measurements were made using the first three periods of the data where the near criticality approximation is valid.  This analysis was done for both of the cases where $A$ and $B$ were varied in the initial data.  The results can be seen in Table \ref{gammadelta}.  These results show good agreement with the results of Bland {\it et al} \cite{Bland2005} and Ziprick and Kunstatter \cite{Ziprick2009c,Ziprick2009a}.  

\begin{figure}[ht!]
\centering
\includegraphics[width=0.4\linewidth]{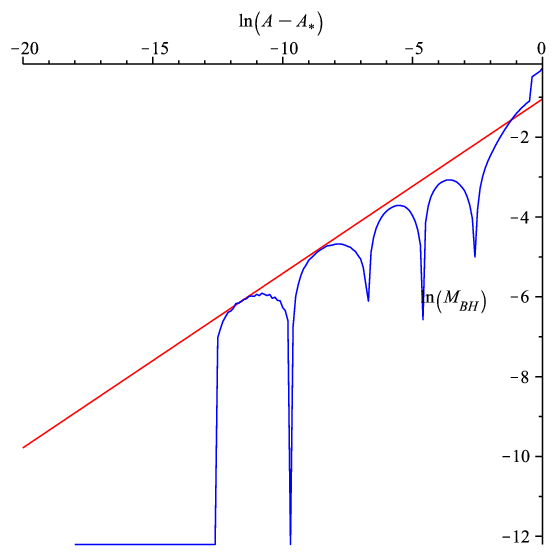}
\caption[8D Mass scaling of A parameter]{Mass Scaling in 8 dimensions ($A$ varied)}
\label{8DChop}
\end{figure}

In order to give a clearer sense of why our results are limited to seven dimensions, we present eight dimensional data in Figure \ref{8DChop}.  In lower dimensions we were able to obtain super-critical data down to  $\ln(A-A^*) \approx -17$. In eight dimensions we were limited to $\ln(A-A^*)<-12.6$ without decreasing the lattice spacing near the origin, which in turn would dramatically increase the simulation time.  The effect of this is seen in Figure \ref{8DChop}:  we are unable to get close enough to criticality for the slope to be approximately constant over three periods.  For this reason we were unable to reliably calculate the critical exponent for eight dimensions with the same accuracy as lower dimensional simulations.  If we nonetheless obtain a slope from the first two periods of the simulation (as opposed to three) then we find $\gamma = 0.44 \pm 0.02$ and $T = 3.0 \pm 0.1$.  This result lies between that of Bland {\it et al} \cite{Bland2005} ($\gamma = 0.4459 \pm 0.0054$, $T = 3.11 \pm 0.1$) and Sorkin and Oren \cite{Sorkin2005} ($\gamma = 0.436 \pm 2\%$, $T = 3.1 \pm 3\%$).  It should be noticed that given the arguments above and the negative curvature that we see in Figure \ref{8DChop} our current value is almost certainly an underestimate. That is, a third cusp closer to criticality would most likely raise the value of the slope, bringing the critical exponent closer to that of \cite{Bland2005}. 

\begin{table}[ht]
\caption{Critical Exponent, $\gamma$ and Period, $T$ for Four to Seven Dimensions, $D$}
\centering
\begin{tabular}{c c c c}
\hline\hline 
D &  & $\gamma$ & T \\ 
\hline
4 & A varied & 0.378 $\pm$ 0.002 & 4.4 $\pm$ 0.3 \\
  & B varied & 0.379 $\pm$ 0.002 & 4.4 $\pm$ 0.2 \\
  & from \cite{Bland2005} & 0.374 $\pm$ 0.002 & 4.55 $\pm$ 0.1 \\
  & from \cite{Ziprick2009c,Ziprick2009a} & 0.375 $\pm$ 0.004 & 4.6 $\pm$ 0.1 \\
  & from \cite{Sorkin2005} & 0.372 $\pm$ 1\% & 4.53 $\pm$ 2\% \\
\hline
5 & A varied & 0.413 $\pm$ 0.002 & 3.9 $\pm$ 0.8 \\
  & B varied & 0.416 $\pm$ 0.002 & 3.7 $\pm$ 0.4 \\
  & from \cite{Bland2005} & 0.412 $\pm$ 0.004 & 3.76 $\pm$ 0.1 \\
  & from \cite{Sorkin2005} & 0.408 $\pm$ 2\% & 4.29 $\pm$ 2\% \\
\hline
6 & A varied & 0.429 $\pm$ 0.003 & 3.4 $\pm$ 0.3 \\
  & B varied & 0.428 $\pm$ 0.002 & 3.3 $\pm$ 0.2 \\
  & from \cite{Garfinkle1999}&0.424 &3.03\\
  & from \cite{Bland2005} & 0.430 $\pm$ 0.003 & 3.47 $\pm$ 0.1 \\
  & from \cite{Sorkin2005} & 0.422 $\pm$ 2\% & 4.05 $\pm$ 2\% \\
\hline
7 & A varied & 0.440 $\pm$ 0.005 & 3.1 $\pm$ 0.1 \\ 
  & B varied & 0.440 $\pm$ 0.006 & 3.1 $\pm$ 0.4 \\
  & from \cite{Bland2005} & 0.441 $\pm$ 0.004 & 3.36 $\pm$ 0.1 \\
  & from \cite{Sorkin2005} & 0.429 $\pm$ 2\% & 3.80 $\pm$ 2\% \\
\hline\hline
\end{tabular}
\label{gammadelta}
\end{table}

\chapter[Choptuik Scaling in EGB Gravity]{Choptuik Scaling in Einstein-Gauss-Bonnet Gravity}
\label{ch:CSEGBG}


In Section \ref{sec:LLGravity} we saw that the Lovelock action retains many of the desirable properties of the Einstein--Hilbert action: it gives equations of motion which are second order in derivatives of the metric, ghost free when linearized about a flat background and obeys a Birkhoff theorem that yields a one parameter family of spherically symmetric black hole solutions.  This makes Lovelock gravity a natural, higher dimensional generalization of general relativity.

Recall from Equation \ref{action2} that the Lovelock action, $I$, written in terms of the Lovelock polynomials, ${\cal L}_{(p)}$ is given by
\be
I=\frac{1}{2\kappa_n^2}\int d^nx\sqrt{-g}\sum^{[n/2]}_{p=0}\alpha_{(p)} {\cal
L}_{(p)}, \nonumber
\label{eq:lovelock action}
\ee
\be
{\cal L}_{(p)}=
\frac{p!}{2^p}\delta^{\mu_1...\mu_p\nu_1..\nu_p}_{
\rho_1...\rho_p\sigma_1..\sigma_p}
    \R_{\mu_1\nu_1}{}^{\rho_1\sigma_1}...\R_{\mu_p\nu_p}{}^{\rho_p\sigma_p}, \nonumber
 \label{eq:lovelock lagrangian}
\ee
The first two terms, ${\cal L}_{(0)}$ and ${\cal L}_{(1)}$, correspond to the cosmological constant and Einstein-Hilbert term, respectively, while  ${\cal L}_{(2)}$ is the Gauss-Bonnet term.
Here we focus on the simplest non-trivial extension of general relativity, namely Einstein-Gauss-Bonnet (EGB) gravity, containing only the Einstein term and the $p=2$, Gauss-Bonnet term\footnote{The addition of a cosmological constant should not affect the short distance behaviour that is the subject of this chapter.}.  Since the Gauss-Bonnet term is proportional to curvature squared its effects would be most visible in regions of high curvature such as near the horizon of a small black hole.  It is the purpose of this chapter to investigate Choptuik scaling of Einstein-Gauss-Bonnet gravity minimally coupled to a massless scalar field and is based on our paper \cite{Deppe2012}.  This is in some ways similar to the work presented in Chapter \ref{ch:HDCS} but addresses many new issues not seen in general relativity.


It has been known for quite some time that the spherically symmetric collapse of a massless scalar field minimally coupled to general relativity exhibits critical behaviour \cite{Choptuik1993} as discussed in Section \ref{sec:Choptuik} and demonstrated in Chapter \ref{ch:HDCS}.
The presence of a dimensionful constant in the action in general changes the above scenario, as verified for Yang-Mills collapse \cite{Choptuik1996}, massive scalar field collapse \cite{Brady1997} and massive gauge field collapse \cite{Garfinkle2003}.  In massive scalar field collapse, for initial data whose width is smaller than the Compton wavelength of the scalar field, the usual second order phase transition is found, whereas in the other limit the phase transition exhibits a mass gap and is first order. It is clearly of interest to study the effects on Choptuik scaling of the Gauss-Bonnet parameter and higher order Lovelock coupling constants. Golod and Piran \cite{Golod2012} recently presented such an analysis for the spherical collapse of massless scalar matter coupled to Einstein-Gauss-Bonnet gravity in five dimensions using double null co-ordinates. They found, as expected, that the Gauss-Bonnet term dominates the dynamics at short distances and destroys the discrete self-similarity characteristic of Choptuik scaling.  Their work concentrated on the regime where the Gauss-Bonnet terms strongly dominated the dynamics.

The purpose of the present chapter is to investigate further the critical collapse of a spherically symmetric, massless scalar field minimally coupled to five and six dimensional Einstein-Gauss-Bonnet gravity. We work in flat slice, or generalized Painlev\'{e}-Gullstrand, co-ordinates since they have several advantages over double null co-ordinates in the present context as discussed in Chapter \ref{ch:HDCS}.  It will be especially important in this chapter that the horizon scaling function in PG co-ordinates has cusps which has the advantage of making the potential appearance of the periodicity in an equation such as Equation \ref{GundlachCS} more obvious.  As we will explain in the next section, the nature of the dynamical equations suggest that qualitative differences can occur in different numbers of spacetime dimensions. It is for this reason that we investigate both five and six spacetime dimensions.  


We confirm some of the results in five dimensions of \cite{Golod2012}, extend the analysis to six dimensions and obtain some surprising new results in both five and six dimensions. For all initial data and choice of parameter $p$ that we examined there exists a critical value $p^{*}$ that separates black hole formation from dispersion.  As expected, when the horizon forms far from the singularity the general relativistic term dominates and the standard Choptuik critical scaling relation is found. Things change as one gets close enough to criticality to enter the region in which the Gauss-Bonnet terms dominate the dynamics. Near criticality the scalar field at the origin oscillates with a constant period $T$ that converges as $(p-p^{*})\to0$ to a value that depends on the Gauss-Bonnet parameter as previously shown \cite{Golod2012}.  We find a different relationship between $T$ and the Gauss-Bonnet parameter than in \cite{Golod2012}, albeit for smaller values of the Gauss-Bonnet parameter.

In addition, we explore in detail the scaling in the Gauss-Bonnet dominated region.  We find qualitatively different behaviour in five and six dimensions. In five dimensions there is evidence for a radius gap; in the super-critical region the radius of the apparent horizon on formation asymptotes to a constant value as criticality is approached from above. The maximum value of the trace of the energy-momentum tensor at the origin also appears to approach a constant value as criticality is approached from below. 

In six dimensions, the behaviour is qualitatively different. In the Gauss-Bonnet region the radius of the apparent horizon formation obeys a relationship similar to Equation \ref{GundlachCS} but with different exponent and period. The maximum of the trace of the energy-momentum tensor at the origin also exhibits this same scaling relation with another scaling exponent, and small, but irregular oscillations.

The rest of this chapter is organized as follows.  In Section \ref{EoM} we describe the equations of motion which we derived using Hamiltonian formalism in \cite{Taves2012} and Chapter \ref{ch:LLADM}.  In Section \ref{Numerics and Methods} we discuss the numerical implementation of the solution and describe the general methods used to obtain results.  In Section \ref{Results} we give our results.

\section{Equations of Motion}
\label{EoM}

As stated above, we start with the action for a massless scalar field $\psi$ minimally coupled to the Einstein-Gauss-Bonnet action.  For the Einstein-Gauss-Bonnet case the action can be expanded as
 \be
  I= \frac{1}{2 \kappa_n^2}\int d^nx \sqrt{-g}\left(\R + \alpha_{(2)}\left[ \R^2 - 4\R_\mn \R^\mn + \R_{\mu \nu \rho \sigma}\R^{\mu \nu \rho \sigma}   \right]   + \kappa_n^2 \left(\nabla\psi\right)^2
  \right)
  \label{eq:GB action}
  \ee
(recall  $\kappa_n := \sqrt{8\pi G_n}$).  The equations of motion and the constraints are once again given by Equations \ref{psidot} \ref{Pisdot}, \ref{sigma1} and \ref{Cconstraint}.  They are, up to slight changes in notation,
\be
\label{psidot_new}
\dot{\psi} = N \left( \frac{\Pis}{R^{n-2}} + \left( \frac{N_r}{N} \right)
\psip \right),
\ee
\be
\label{Pisdot_new}
\dot{P}_\psi = \left[ N \left( R^{n-2} \psip + \left( \frac{N_r}{N}
\right) \Pis \right) \right]^\prime,
\ee
\be
\label{sigma_code}
N^\prime = -\frac{kGN \Pis \psip}{R^{n-3}} \Big/ \left( \sub \left( 1
+ \frac{\tilde{\alpha}}{R^2} \sub^2 \right) \right),
\ee
and
\be
\label{Hconstraint_new}
M^\prime = \frac{1}{2} \left( \frac{\Pis^2}{R^{n-2}} + R^{n-2} \psip^2 \right) + \sub \Pis \psip.
\ee
where the $n$ dimensional gravitational constant, $G$ is defined as $2kG = 2 \kappa_n^2/(n-2)A_{n-2}$  \cite{Ziprick2009a}, $A_{n-2}$ is the surface area of an $n-2$ dimensional sphere, $k := 8(n-3)/(n-2)^2$, and $\tilde{\alpha}_{(2)}:=((n-3)!/(n-5)!)\alpha_{(2)}$ (see Equation \ref{alphatil}).  For the Einstein-Gauss-Bonnet case the constraint, Equation \ref{Nsig} is given by 
\bea
M:= \frac{1}{2kG}\left[
R^{n-3}\left(\frac{N_r}{N}\right)^2 + 
\tilde{\alpha}_{(2)}R^{n-5}\left(\frac{N_r}{N}\right)^4 \right].
\label{eq:M EL2}
\eea
This can be solved algebraically and yields
\be
\frac{N_r}{N} =\sqrt{\frac{R^2}{\tilde{\alpha}} \left(  \sqrt{1 +
\frac{2\tilde{\alpha}}{R^2} \frac{2kG M}{R^{n-3}}} -1 \right)}.
\label{Nsig_code}
\ee
where we have defined $\tilde{\alpha}:=2(n-4)(n-3) \alpha_{(2)}=2\tilde{\alpha}_{(2)}$ for purposes which will become obvious later.  The sign of the inner square root in the above has been chosen to give the correct general relativistic limit as $\tilde{\alpha}\to0$.

Using Equation \ref{Nsig_code} to replace  $N_r/N$ by $M$ in Equation \ref{Hconstraint_new} provides a differential equation that can be solved for $M$ and hence $N_r/N$ in terms of the scalar field and its conjugate momenta on each spacial slice.  Note that Equations \ref{Nsig_code} and \ref{sigma_code} are the only ones in the set of equations of motion and constraints which differentiates Einstein-Gauss-Bonnet from any other form of Lovelock gravity.  For the case of general relativity $N_r/N$ is given by
\be
\frac{N_r}{N} =\sqrt{\frac{2kG M}{R^{n-3}}}.
\label{NsigGR}
\ee

Given the solution for $M$, one can look for apparent horizons by searching for locations where $(DR)^2=0$. For the Lovelock case this gives
\be
AH:=1- \left(\frac{N_r}{N}\right)^2=0
\label{eq:AH}
\ee
where for ease of reference we refer to $AH$ as the horizon function.  This equation is the generalization of Equation \ref{AHGR} which we used in the general relativistic case.  For the Einstein-Gauss-Bonnet case one can also use Equation \ref{eq:M EL2} and the above to obtain:
\be
\label{MAH}
M(R_{AH})=\frac{1}{4kG} \left( \tilde{\alpha} R_{AH}^{n-5} + 2R_{AH}^{n-3} \right).
\ee
Note that in 5D the first term is constant so that there is an algebraic lower bound on the black hole mass as the radius of the horizon goes to zero.

Our goal is to solve the time evolution equations for the scalar field and its conjugate momentum, with $N$ and $N_r/N$ on each time slice determined using Equations \ref{Nsig_code}, \ref{sigma_code} and \ref{Hconstraint_new} and then use Equation \ref{eq:AH} to look for the formation and location of an apparent horizon.  We will also be interested in calculating the value of some scalar invariant which would allow us to explore Choptuik scaling in the sub-critical regime.  When investigating Choptuik scaling in general relativity it is common to use the trace of the energy momentum tensor, $T_{\mu \nu}$ (which is proportional to the Ricci scalar in the general relativistic case) for this purpose.  This is the invariant that we use in this chapter.  It should be mentioned here that in Einstein-Gauss-Bonnet gravity the equations of motion (see equation (2.11) of \cite{Maeda2011}) give

\begin{align}
\kappa_n^2 T^\mu_{\phantom{\mu}\mu}=-\frac{(n-2)}{2}\R - \alpha_{(2)}\frac{n-4}{2}\left[\R^2 - 4\R_\mn \R^\mn + \R_{\mu \nu \rho \sigma}\R^{\mu \nu \rho \sigma}   \right]
\label{EMTensor}
\end{align}
which demonstrates that, although $T^\mu_{\phantom{\mu}\mu}$ is still a scalar invariant, it is not the Ricci scalar as in the general relativistic case.

It is important to point out that the actual time evolution equation as implemented in the code was obtained by expanding the derivative in Equation \ref{Pisdot_new} and replacing the derivatives of $M'$ and $N'$ using Equations \ref{Nsig_code} and \ref{Hconstraint_new}.  This gives
\begin{align}
\label{Pisdot_code}
& \dot{P}_\psi = N \Bigg\{ \Bigg[ Gk \left( \frac{\Pis^3}{2R^{2n-5}} -
\frac{\psip^2\Pis R}{2} \right) \Big/ \sub + \\ \nonumber
& - \frac{(n-3) \Pis}{2R} \sub - \frac{\tilde{\alpha}(n-5)\Pis}{4R^3} \sub^3
\Bigg]
\Big/ \left( 1 + \frac{\tilde{\alpha}}{R^2} \sub^2 \right) \\ \nonumber
& + (n-2)R^{n-3}\psip + R^{n-2}\psi^{\prime \prime} + \Pis^\prime \sub \Bigg\}.
\end{align}
Note that in five space-time dimensions the last term proportional to $1/R^3$ in the square brackets above vanishes. One might therefore expect behaviour for $n>5$ that is qualitatively different from $n=5$. It is for this reason that it is important to study higher dimensions.  In the present paper we restrict consideration to five and six dimensions.

\section{Numerics and Methods}
\label{Numerics and Methods}

The code used for this work is based on that used for Chapter \ref{ch:HDCS}.  The numerics for this project were more difficult and so it is worth mentioning some of the details of the numerical method here.  The system is evolved using c++ code as follows:

\begin{enumerate}
\itemsep -1mm
\item Initialize the spatial lattice.  We set the lattice spacing to $10^{-5}$ (unless otherwise stated) for the first 100 points near the origin and then slowly increase it to $10^{-2}$ at the 1200$^{th}$ and final lattice point.

\item Set up initial conditions.  We initialized $\Pis$ to zero and $\psi$ to be either a Gaussian $\psi_G$ or hyperbolic tangent $\psi_H$ as follows 
\be
\label{init}
\psi_G = AR^2 \exp\left[-\left(\frac{R-R_0}{B}\right)^2\right]
\, ; \qquad
\psi_H = A \tanh\left[\frac{R-R_0}{B}\right] 
\ee
where $A$, $B$ and $R_0$ are parameters.

\item At $R=0$ set $M=0$ and $N=1$ (which corresponds to setting the time co--ordinate to be the proper time at $R=0$) and use a subroutine to calculate $N_r/N$ using Equation \ref{Nsig_code}.  Integrate $N$ and $M$ forward in $R$ using Equations \ref{Nsig_code}, \ref{sigma_code} and \ref{Hconstraint_new}. This is done using an fourth order Runge-Kutta (RK4) method.  Spatial derivatives are calculated using a central difference routine except at the boundaries where forward and backward difference are used.

\item Integrate $\psi$ and $\Pis$ forward in time using Equations \ref{psidot_new}, \ref{Pisdot_new} and \ref{Nsig_code} employing an RK4 method. Stability is maintained by insisting that the size of the time step, $\Delta
t(t)$, is determined by
\be
\label{stability}
\Delta T_{PG}(T_{PG}) < \min_R \left\{ \left(\frac{dR}{dT_{PG}} \right)^{-1} \Delta R(R)
\right\},
\ee
where $\Delta R(R)$ is the lattice spacing and $\frac{dR}{dT_{PG}}$ is the maximum value of either the ingoing or outgoing local speed of light as was the case in Chapter \ref{ch:HDCS}.

\item Monitor the apparent horizon function, $AH:=(DR)^2 =1-\sub^2$.  At any point where $AH=0$, there is an apparent horizon.  When $AH$ forms a local minimum it signals that an apparent horizon is soon to form, so the time steps are diminished by a factor of ten.

\item Calculate the quantities of interest such as the mass density and the trace of the energy momentum tensor, $T^\mu_{\phantom{\mu}\mu}$.  Note that $T^\mu_{\phantom{\mu}\mu}$ is calculated using the phase space variables.  Starting with $T_{\mu \nu}=\nabla_\mu \psi \nabla_\nu \psi - \frac{1}{2}g_{\mu \nu} |\nabla \psi|^2$ and using Equation \ref{Pip} we find $T^\mu_{\phantom{\mu}\mu}=-\Pis^2/R^{2(n-2)}+\psip^2$.

\item Repeat steps 3-6 until the formation of an apparent horizon or until the field has dispersed. 

\end{enumerate}

For comparison purposes it was important that the code could simulate collapse without the Gauss-Bonnet term, ie in the general relativistic case.  
It is not possible to take this limit when numerically calculating $N_{r}/N$ using Equation \ref{Nsig_code} so an if statement was added to the routine which calculates $N_r/N$ in order to return $N_r/N=\sqrt{2kGM/R^{n-3}}$ when $\tilde{\alpha} = 0$.

When $\tilde{\alpha}$ is not zero a problem arises in the calculation of $N_r/N$ when $4\tilde{\alpha} kGM/R^{n-1}$ is sufficiently less than one.  When this term is added to unity in the inner square root in Equation \ref{Nsig_code}, digits are lost and thus double precision can not be claimed.  For this reason a 16$^{th}$ order Taylor expansion of the inner square root in Equation \ref{Nsig_code} was used in the case that $2\tilde{\alpha} 2kGM/R^{n-1}<0.1$.  Quadruple precision allowed for the investigation of overflow and underflow, as well as subtraction and addition round off errors.

The code was capable of parallel processing, and many simulations were run on multiple processors using twelve processor years on the WestGrid and SHARCNET computing clusters.  When generating data for mass and $T^\mu_{\phantom{\mu}\mu}$ scaling plots the speed up was linear with the number of processors used, whereas for the binary search used to find critical values the speed up was logarithmic.

\begin{figure}[ht!]
\centering
\subfigure[5D]{
\includegraphics[width=0.4\linewidth]{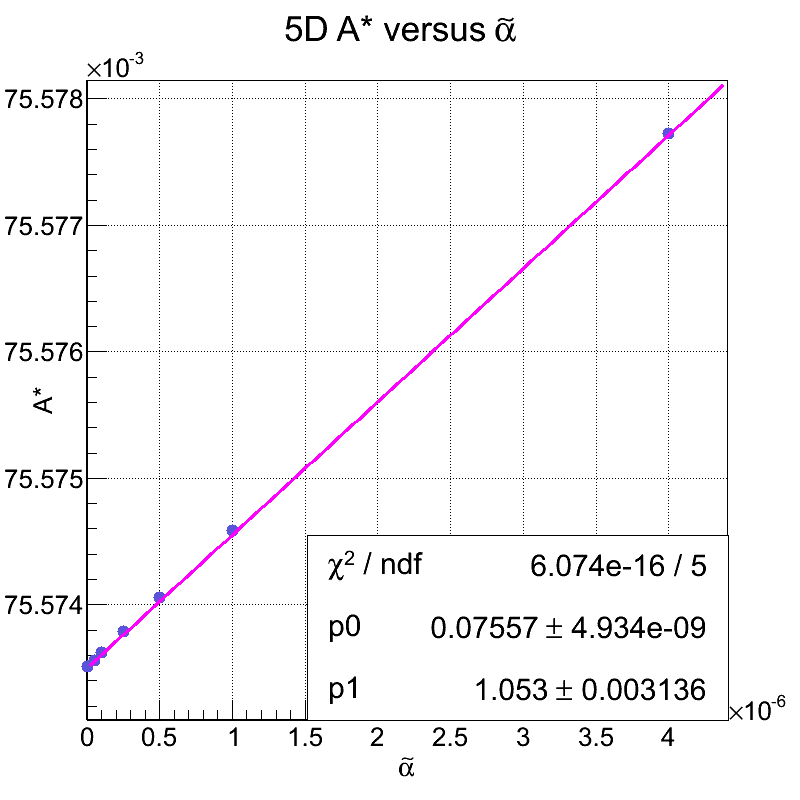}
\label{Astar5}
}
\hspace{0.25in}
\subfigure[6D]{
\includegraphics[width=0.4\linewidth]{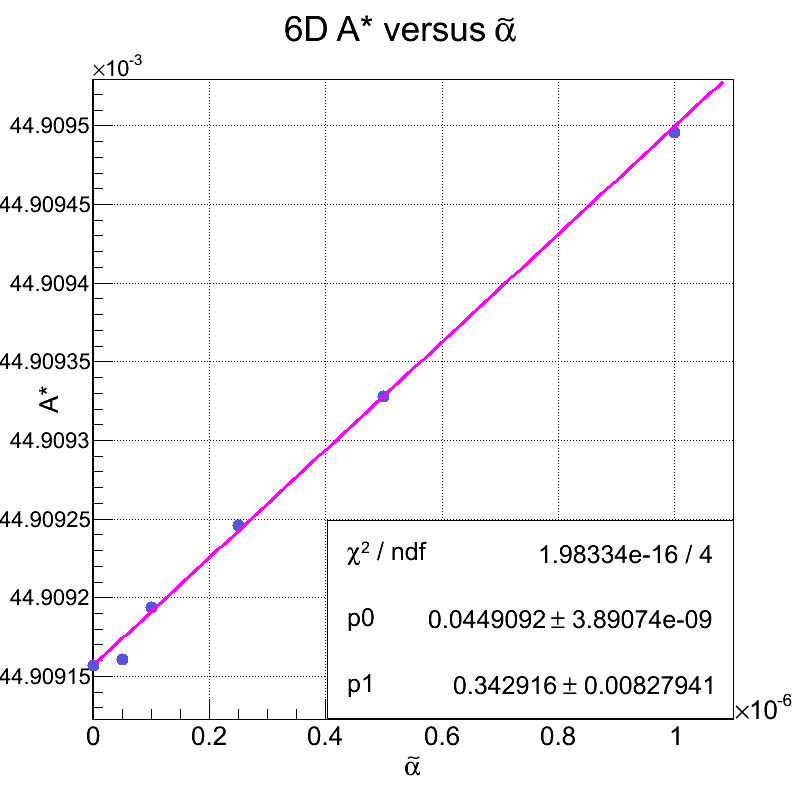}
\label{Astar6}
}
\caption{$A^*$ as a function of $\tilde{\alpha}$}
\label{Astar}
\end{figure}

We first performed a binary search to find the critical value of $A$ in Equation \ref{init}.  $\psi(T_{PG}, R=0)$ and $M$ were then checked at late times to confirm that they blew up for $A$ slightly bigger than $A^*$ and remained finite for $A$ slightly smaller than $A^*$.  We were able to get consistent results to 12 significant figures.  The $A^*$ values for different values of $\tilde{\alpha}$ can be seen in Figure \ref{Astar}.  Interestingly the points are very well fit to straight lines.  The above procedure, of course, also gives $B^*$ and $R_0^*$, which could also be varied.  Using our values for $A^*$ we calculated $\psi$ at the origin as a function of PG time and used these plots to find the period of oscillation near criticality, as a function of $\tilde{\alpha}$.

We simulated matter bounce and dispersal for 280 simulations (the number 280 was chosen to optimize graph resolution and computing time) with $A<A^*$ and recorded the maximum value of the trace of the energy momentum tensor $T^\mu_{\phantom{\mu}\mu}$ at $R=0$ for each simulation.  Plotting $T^\mu_{\phantom{\mu}\mu \phantom{\mu} max}$ as a function of $A^*-A$ with a log-log scale gives the energy momentum scaling plot.  Similarly we simulated collapse for 280 simulations with $A>A^*$ and recorded the radius of the initial apparent horizon, $R_{AH}$.  This procedure was repeated in five and six dimensions checking for scaling with both the $A$ and $B$ parameters in both the gaussian and tanh initial data of Equation \ref{init} to check for universality.  Using gaussian initial data radius and $T^\mu_{\phantom{\mu}\mu}$ scaling plots were created for $\tilde{\alpha} = 0, 10^{-8}, 10^{-7}, 5\times10^{-7}, 10^{-6}$ in the 5D case and $\tilde{\alpha} = 0, 10^{-7}, 5\times10^{-7}, 10^{-6}, 10^{-7}, 10^{-4}$  in the 6D case to investigate the effects of the Gauss-Bonnet terms on the critical exponent, period and the existence of a mass gap.



\section{Results}
\label{Results}
\subsection{Scalar Field Oscillations}
\begin{figure}[ht!]
\centering
\subfigure[5D,$\psi(0,T_{PG})$ near criticality, GR]{
\includegraphics[width=0.4\linewidth]{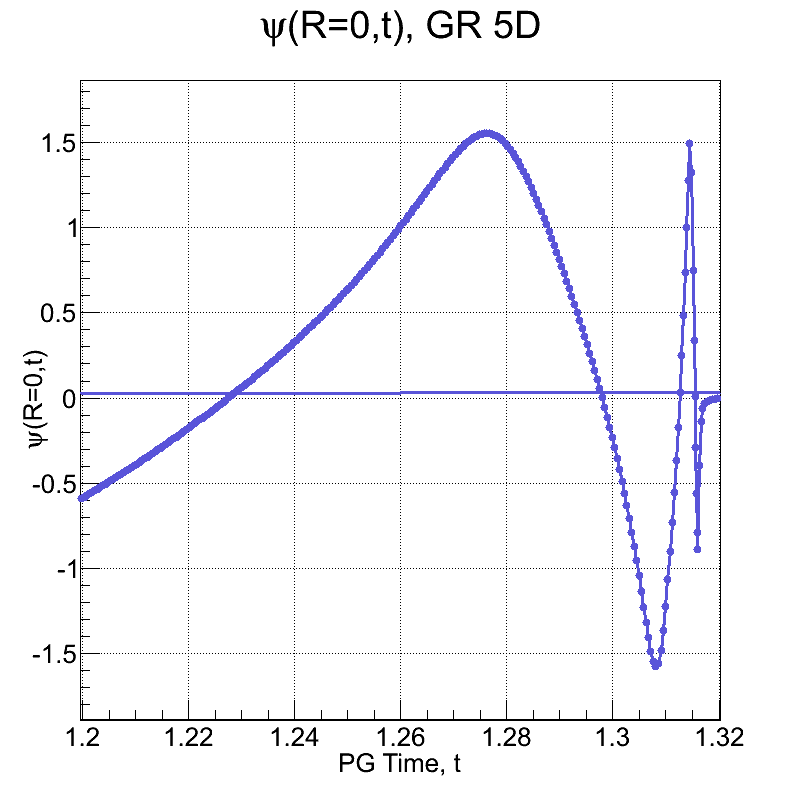}
\label{5Dscalar1_gr}
}
\hspace{0.25in}
\subfigure[5D,$\psi(0,T_{PG})$ near criticality, $\tilde{\alpha}=10^{-6}$]{
\includegraphics[width=0.4\linewidth]{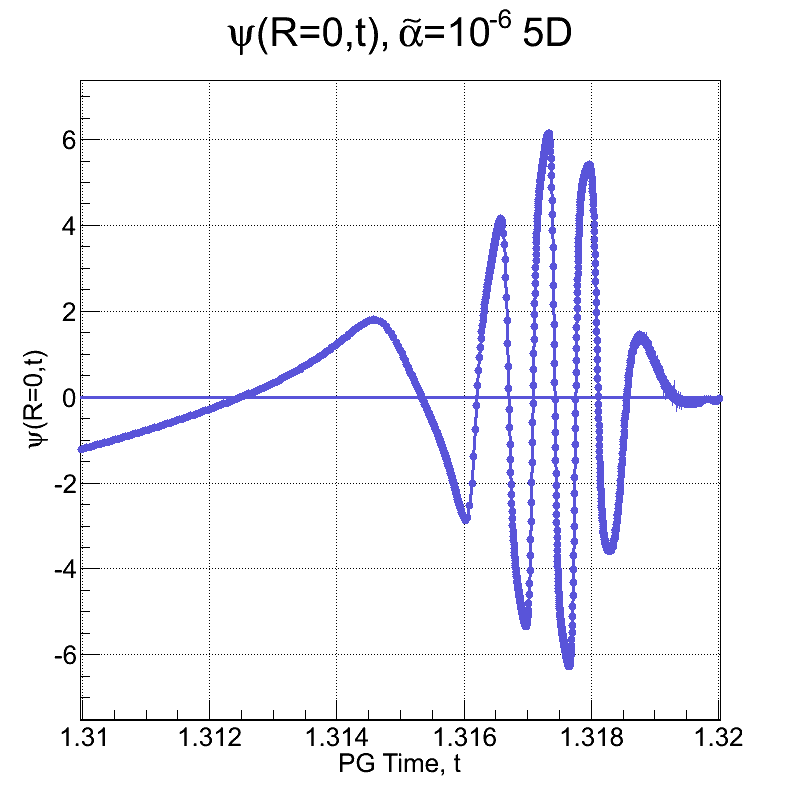}
\label{5Dscalar1a}
}
\subfigure[5D, Period of $\psi(0,T_{PG})$ near criticality, $\tilde{\alpha}=10^{-7}$, showing convergence]{
\includegraphics[width=0.4\linewidth]{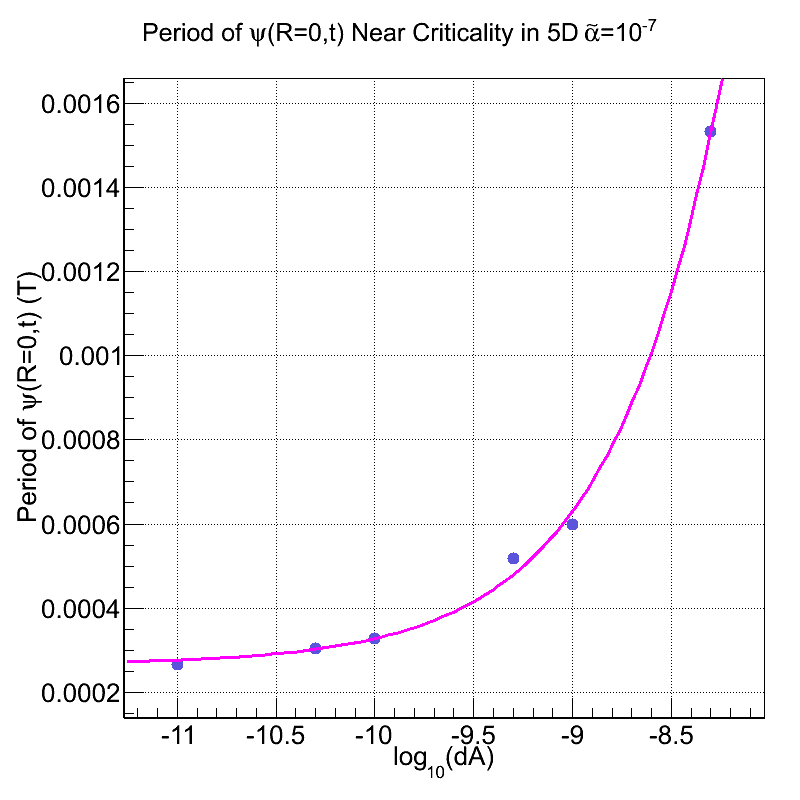}
\label{5Dscalar1b}
}
\hspace{0.25in}
\subfigure[Period of $\psi(0,T_{PG})$ in 5D near criticality as a function of Gauss-Bonnet parameter]{
\includegraphics[width=0.4\linewidth]{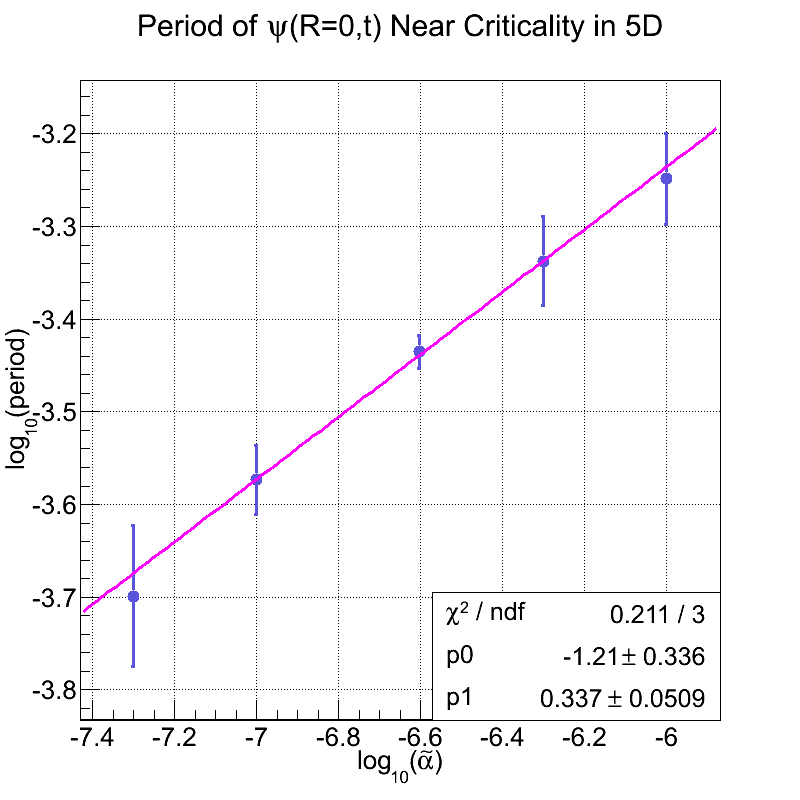}
\label{5Dscalar1c}
}
\caption{Scalar field oscillations}
\label{5Dscalar}
\end{figure}

\begin{figure}[ht!]
\centering
\subfigure[6D,$\psi(0,T_{PG})$ near criticality, GR.]{
\includegraphics[width=0.4\linewidth]{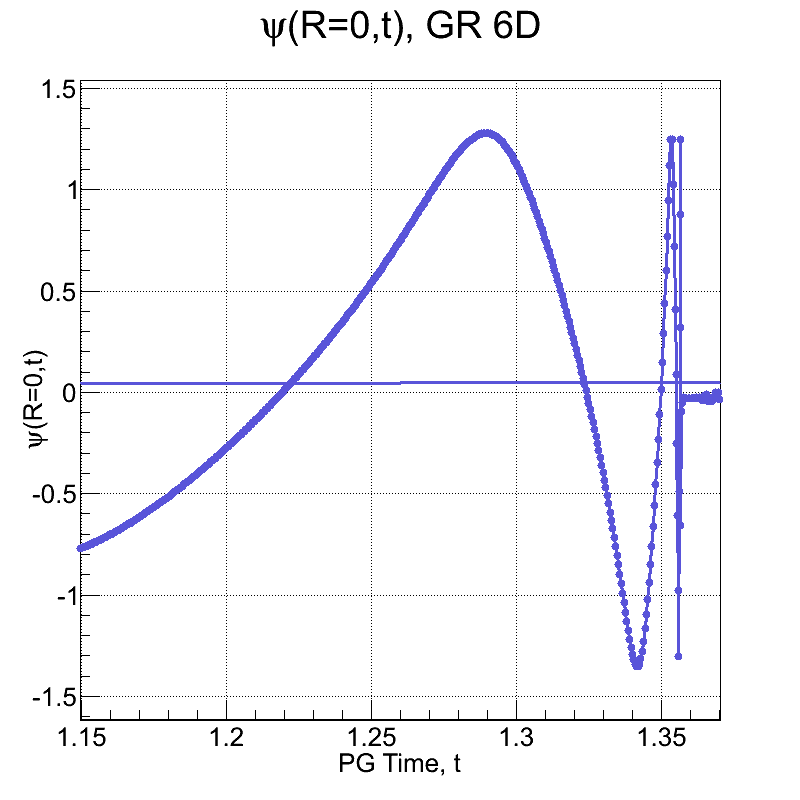}
\label{6DscalarGR}
}
\hspace{0.25in}
\subfigure[6D,$\psi(0,T_{PG})$ near criticality, $\tilde{\alpha}=10^{-6}$.]{
\includegraphics[width=0.4\linewidth]{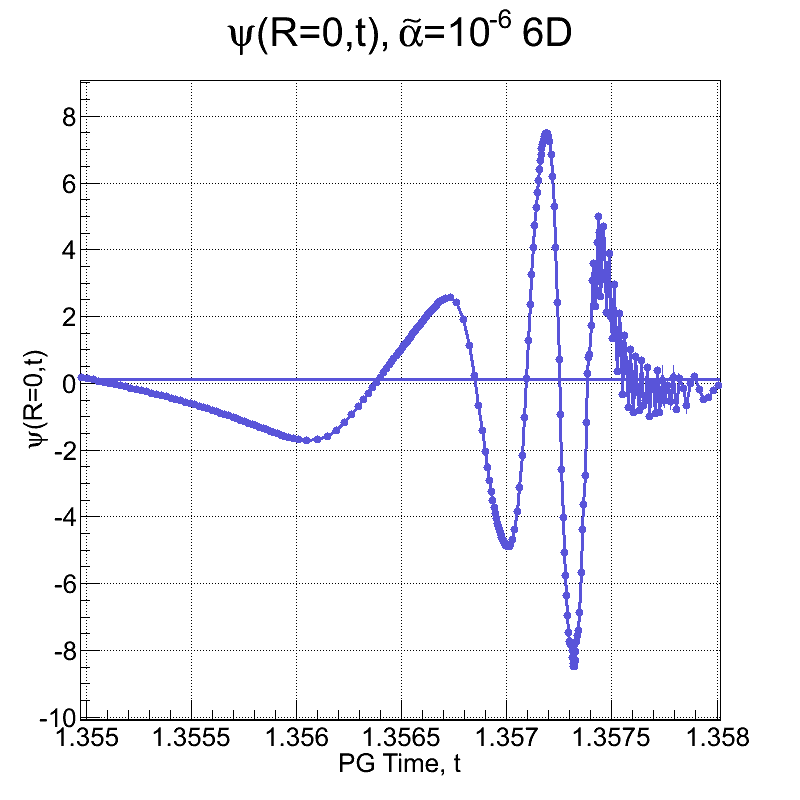}
\label{6Dscalar1a}
}
\subfigure[6D, period of $\psi(0,T_{PG})$ near criticality, $\tilde{\alpha}=10^{-7}$, showing convergence.]{
\includegraphics[width=0.4\linewidth]{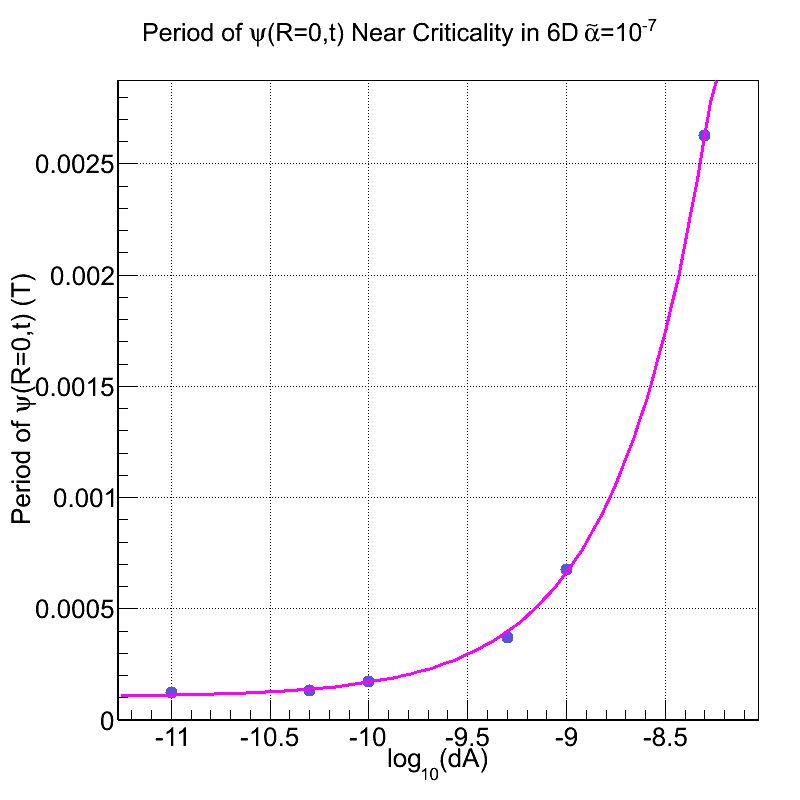}
\label{6Dscalar1b}
}
\hspace{0.25in}
\subfigure[Period of $\psi(0,T_{PG})$ in 6D near criticality as a function of Gauss-Bonnet parameter]{
\includegraphics[width=0.4\linewidth]{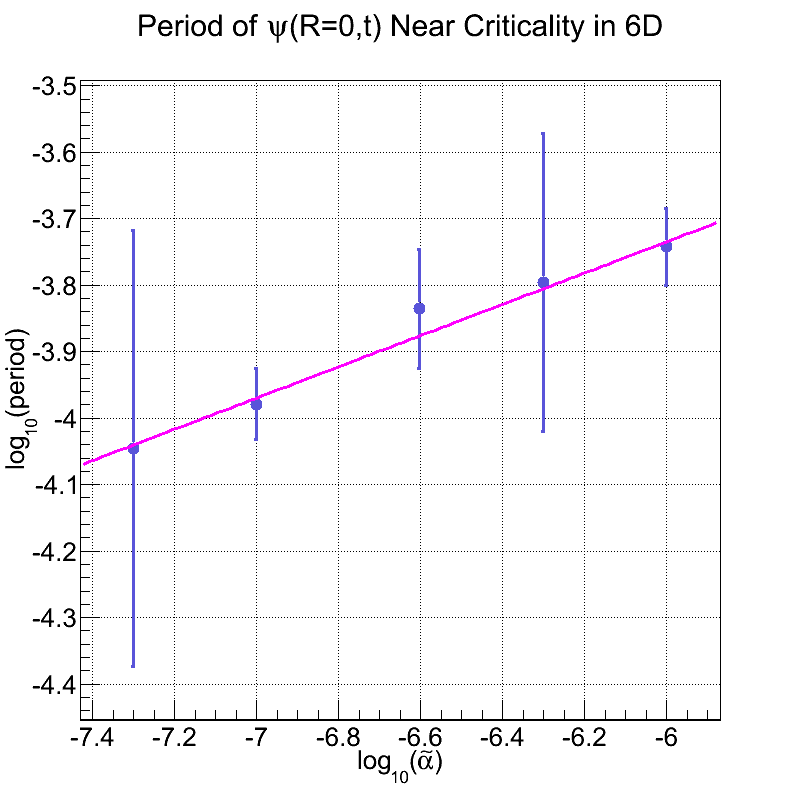}
\label{6Dscalar1c}
}
\caption{Scalar field oscillations}
\label{6Dscalar}
\end{figure}

In general relativity the discrete self-similarity of the critical solution results in oscillations of the scalar field at the origin with ever decreasing period. The presence of the dimensionful Gauss Bonnet parameter breaks the scale invariance which is thought to be the cause of discrete self similarity.  As shown by Golod and Piran \cite{Golod2012} this is indeed the case in 5 dimensions. The scalar field oscillations at the origin near criticality approach a constant period that depends on the value of the Gauss-Bonnet parameter. Since it was difficult to get close enough to criticality to guarantee that the period had converged, we plotted the values as a function of $\log(dA)$, where $dA\equiv |A-A^*|$. As seen in Figures \ref{5Dscalar1b} and \ref{6Dscalar1b} the convergence was exponential and we used a best fit to determine the value of the period $T$ and its corresponding error for each value of $\tilde{\alpha}$. The results are shown for 5 and 6 dimensions in Figures \ref{5Dscalar1c} and \ref{6Dscalar1c}. Our results are qualitatively similar to those in \cite{Golod2012}, namely
\bea
T_{(n)}&\propto& \tilde{\alpha}^\beta_{(n)} \label{eq:lambda}
\eea
Our exponents in five and six dimensions are:
\begin{align}
\beta_{(5)} &= 0.34\pm0.05\\
\beta_{(6)} &= 0.24\pm0.08
\end{align}
These both differ from the value of approximately 1/2 obtained in 5D by Golod and Piran \cite{Golod2012}, who argued that $\beta$ is one divided by the scaling dimension of the Gauss-Bonnet coupling coefficient. Intriguingly our results suggest a relationship of 
\be
\beta_{(n)} = 1/(n-2)
\ee
Note that the 6D plots show oscillations at late times which are likely due to the build up of numerical error.  It should also be mentioned here that in 6 dimensions the numerics did not let us get close enough to criticality to measure more than a few oscillations as can be seen in Figure \ref{6Dscalar1a}.  Although we calculated a period we can not say conclusively that the critical solution is periodic.  The oscillations may, in fact decrease in period as in the general relativistic case. 

We note also that the range of $\tilde{\alpha}$ that we considered was between $5 \times 10^{-8}$ and $10^{-6}$, which is outside the range $4\times10^{-6}$ to $4\times10^{-4}$ considered by \cite{Golod2012}, which may explain the discrepancy in our results. We were restricted to smaller values of the Gauss-Bonnet parameter because our PG co-ordinate code did not allow us to get close enough to criticality for large values of $\tilde{\alpha}$ in order to reliably measure the period of the scalar field.

\subsection{Critical Exponents}
In general relativity there exist universal scaling relations whose properties are determined in part by the critical solution. We now present two different sets of scaling plots in the Gauss-Bonnet case. The first is the value of the logarithm of the apparent horizon radius $R_{AH}$ on formation as a function of $\log(dA)$ as the critical parameter is approached from above (i.e. super-critical). The second is the log of the maximum value of the trace of the energy momentum tensor at the origin as a function of $\log(dA)$. We find as expected that if we are far enough from criticality that the curvatures stay small and the apparent horizon radius is large compared to the Gauss-Bonnet scale, we reproduce approximately the general relativity results: the curves are universal, with slope approximately equal to the general relativity critical exponent. The energy momentum tensor plots in this region are approximately straight lines with a small oscillation superimposed, whereas the radius plots show the large amplitude cusps observed in \cite{Ziprick2009c,Taves2011}.

\begin{figure}[ht!]
\centering
\subfigure[5D $\tilde{\alpha}=5\times10^{-7}$, amplitude and width separately]{
\includegraphics[width=0.4\linewidth]{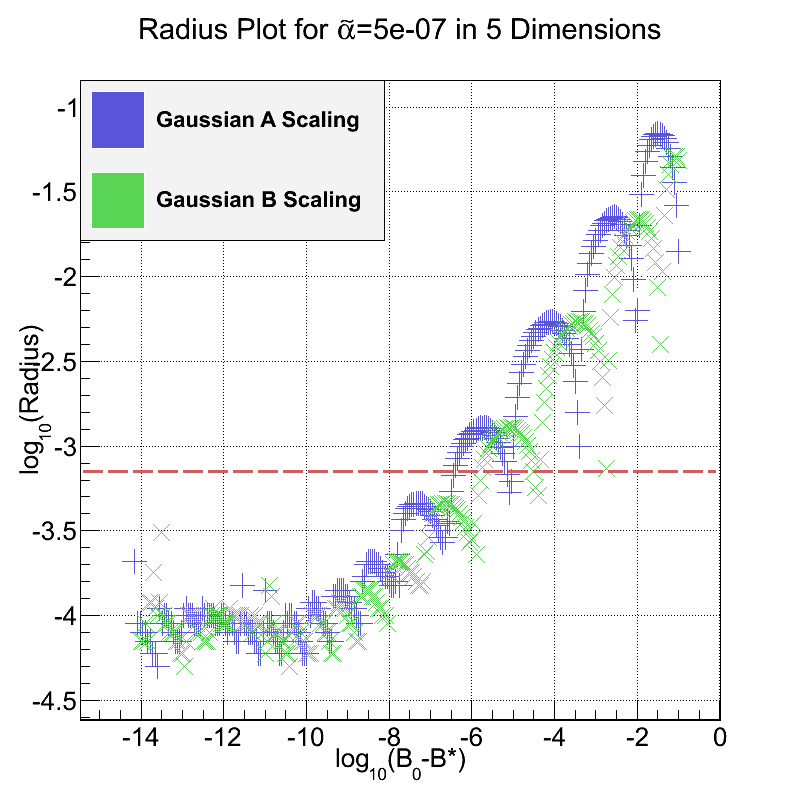}
\label{5DUniverality1}
}
\hspace{0.25in}
\subfigure[5D $\tilde{\alpha}=5\times10^{-7}$, amplitude and width shifted to lie on top of each other]{
\includegraphics[width=0.4\linewidth]{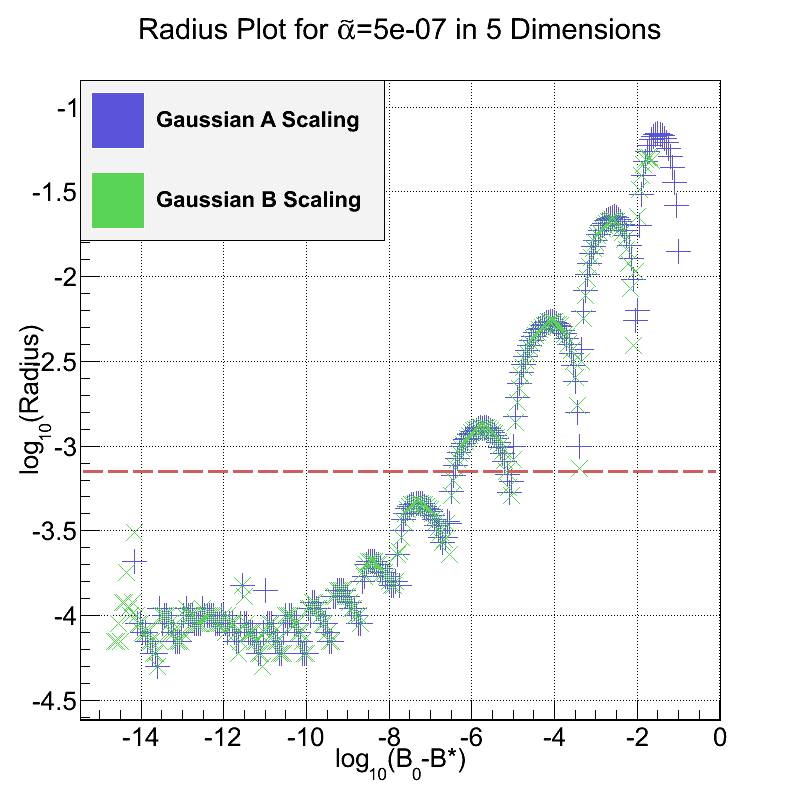}
\label{5DUniversality2}
}
\subfigure[5D $\tilde{\alpha}=5\times10^{-7}$, Radius Plots Superimposed]{
\includegraphics[width=0.4\linewidth]{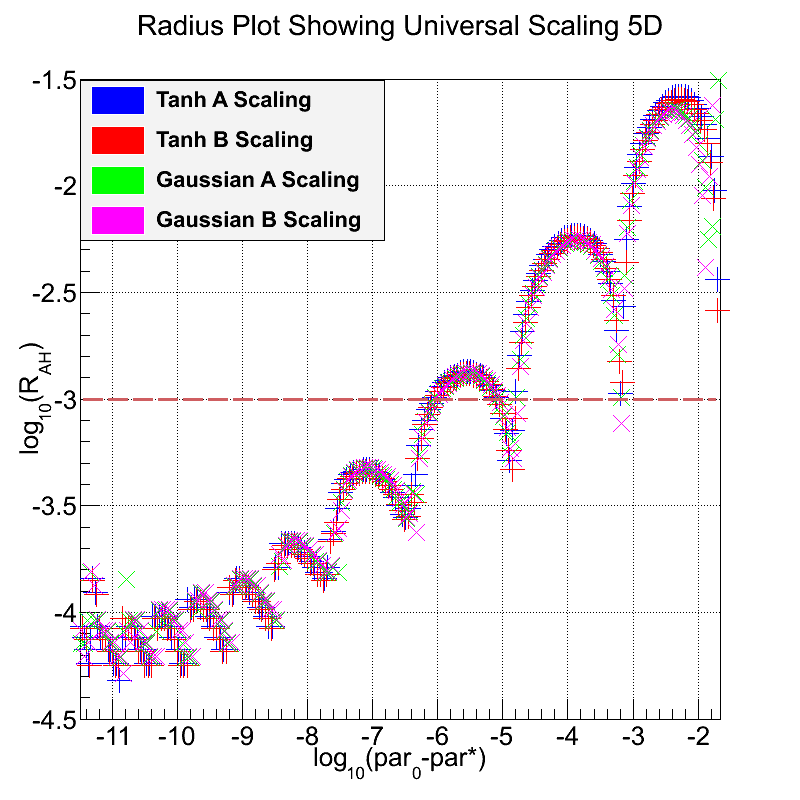}
\label{5DUniversality2b}
}
\hspace{0.25in}
\subfigure[5D $\tilde{\alpha}=5\times10^{-7}$, $T^\mu_\mu$ Plots Superimposed]{
\includegraphics[width=0.4\linewidth]{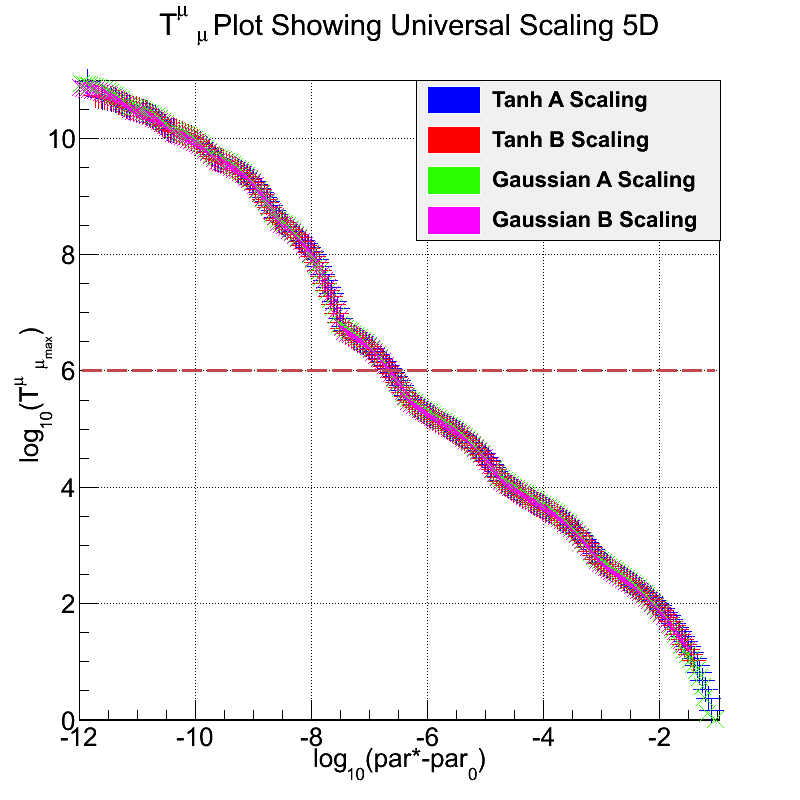}
\label{5DUniverality3}
}
\subfigure[6D, $\tilde{\alpha}=10^{-5}$, Radius Plots Superimposed]{
\includegraphics[width=0.4\linewidth]{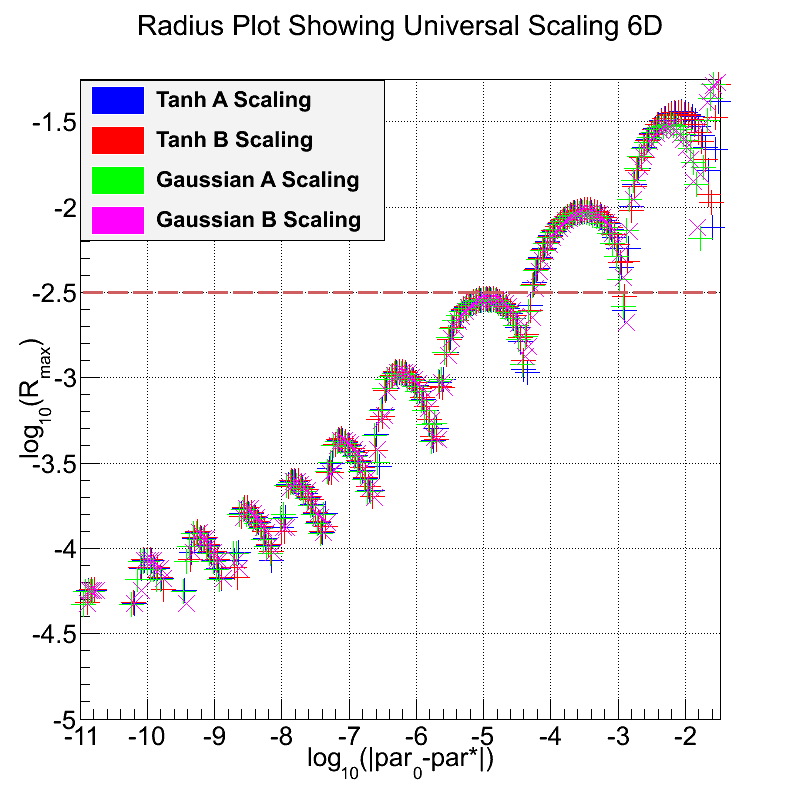}
\label{6DradiusUniversal}
}
\hspace{0.25in}
\subfigure[6D, $\tilde{\alpha}=10^{-5}$, $T^\mu_\mu$ Plots Superimposed]{
\includegraphics[width=0.4\linewidth]{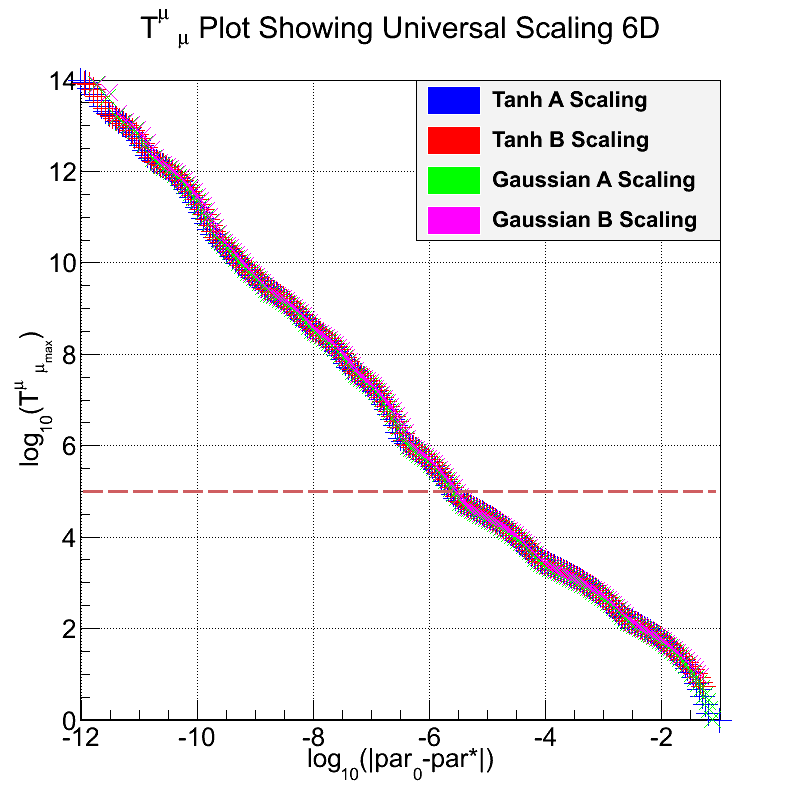}
\label{6DricciUniversal}
}
\caption{Universality in 5 and 6D}
\label{Universality}
\end{figure}

As the critical parameter is approached we enter into the Gauss-Bonnet region where the higher curvature terms dominate and things change. In the case of the $T^\mu_{\phantom{\mu}\mu}$ plots, the Gauss-Bonnet region occurs when $T^\mu_{\phantom{\mu}\mu} \tilde{\alpha} > 1$, whereas for the radius plots it can be defined by the relation $R_{AH}< \sqrt{\tilde{\alpha}}$. The boundary between the two regions is indicated in all the scaling plots by a horizontal dashed line.

\begin{figure}[ht!]
\centering
\subfigure[GR, slope$=0.413$]{
\includegraphics[width=0.4\linewidth]{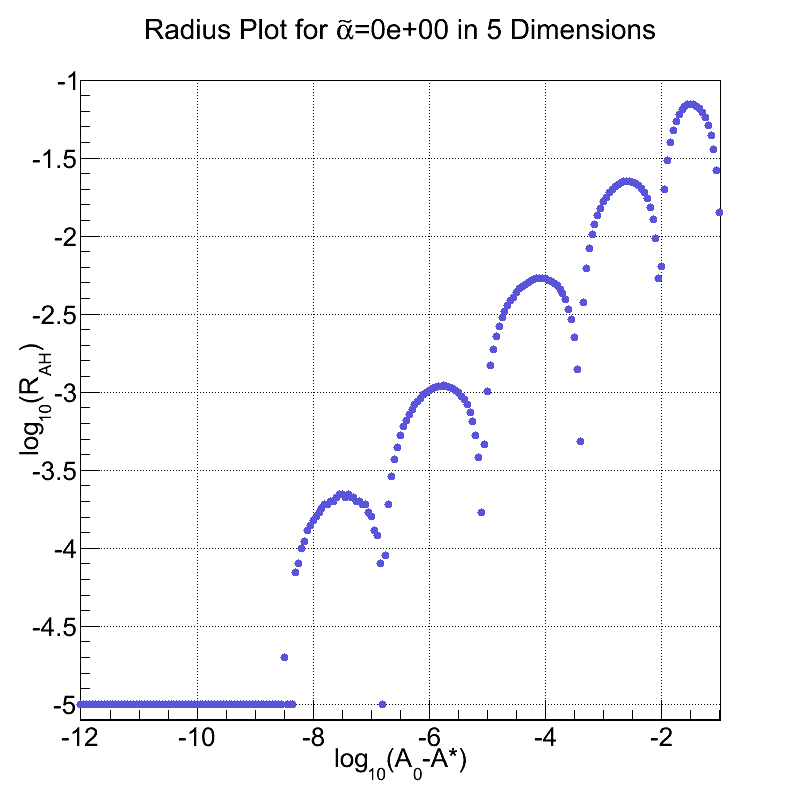}
\label{5DRadiusGR}
}
\hspace{0.25in}
\subfigure[$\tilde{\alpha}=10^{-8}$]{
\includegraphics[width=0.4\linewidth]{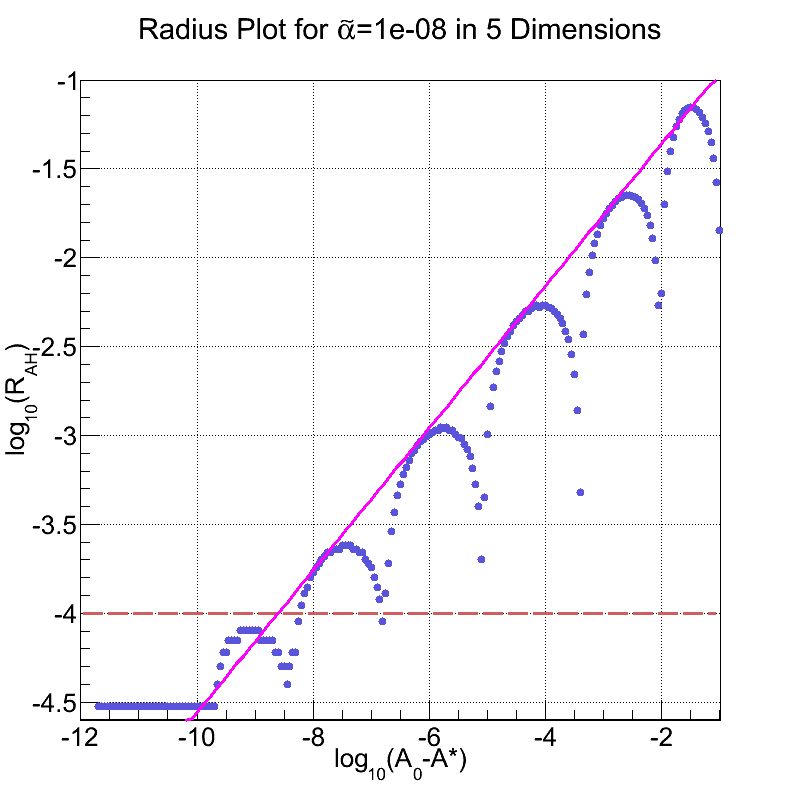}
\label{5DRadius-8}
}
\subfigure[$\tilde{\alpha}=10^{-7}$]{
\includegraphics[width=0.4\linewidth]{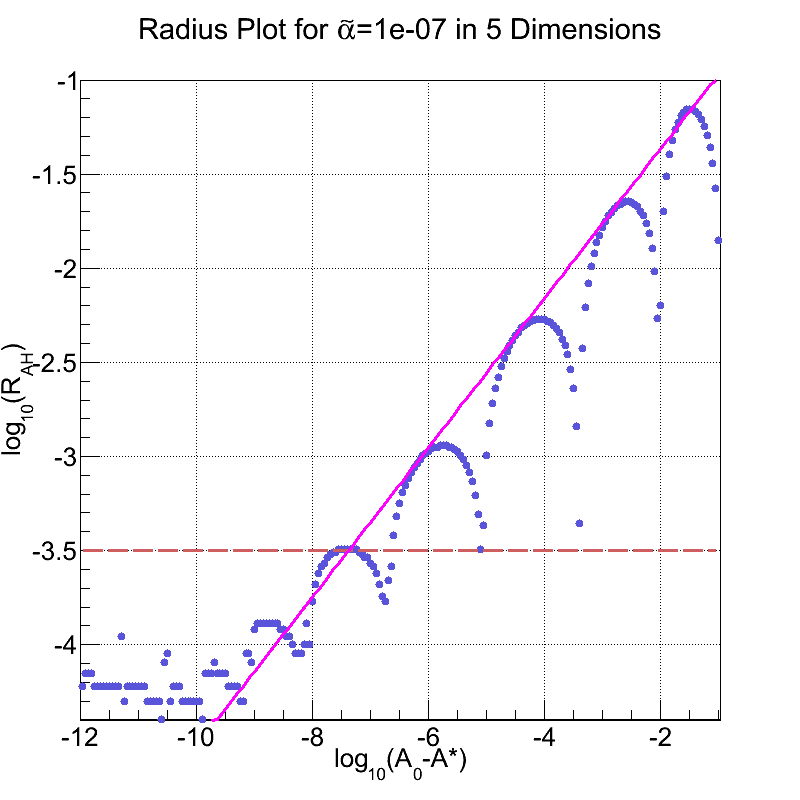}
\label{5DRadius-7}
}
\hspace{0.25in}
\subfigure[$\tilde{\alpha}=5\times10^{-7}$]{
\includegraphics[width=0.4\linewidth]{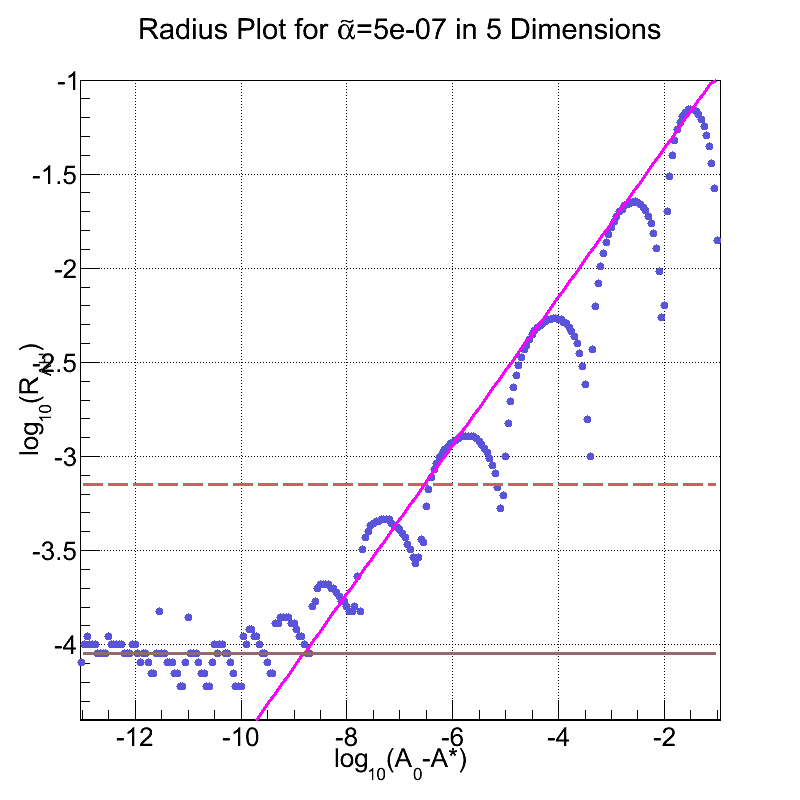}
\label{5DRadius5e-7}
}
\hspace{0.25in}
\subfigure[$\tilde{\alpha}=10^{-6}$]{
\includegraphics[width=0.4\linewidth]{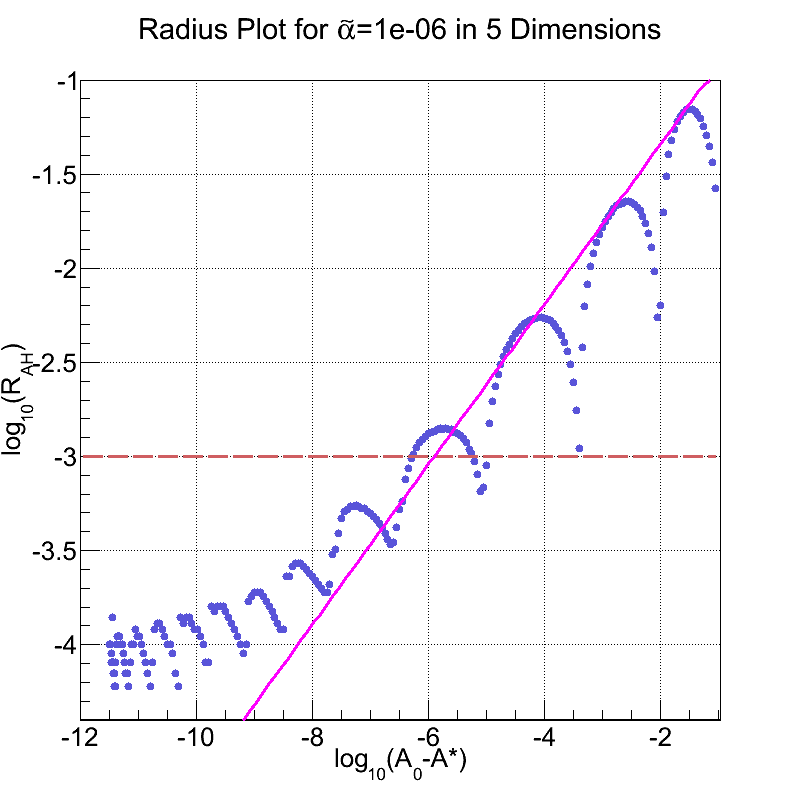}
\label{5DRadius-6}
}
\caption{Radius Scaling Plots - 5D. The lines represent the best-fit tangents to the curves in their respective regimes.}
\label{Radius5D}
\end{figure}

\begin{figure}[ht!]
\centering
\subfigure[GR, slope$=0.43$]{
\includegraphics[width=0.4\linewidth]{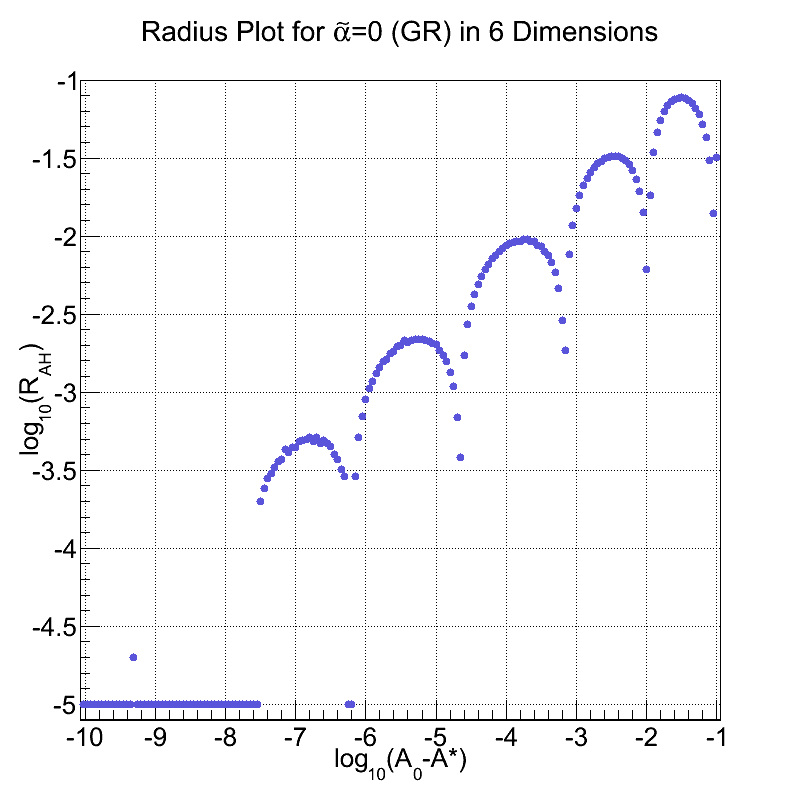}
\label{6DRadiusGR}
}
\hspace{0.25in}
\subfigure[$\tilde{\alpha}=10^{-7}$]{
\includegraphics[width=0.4\linewidth]{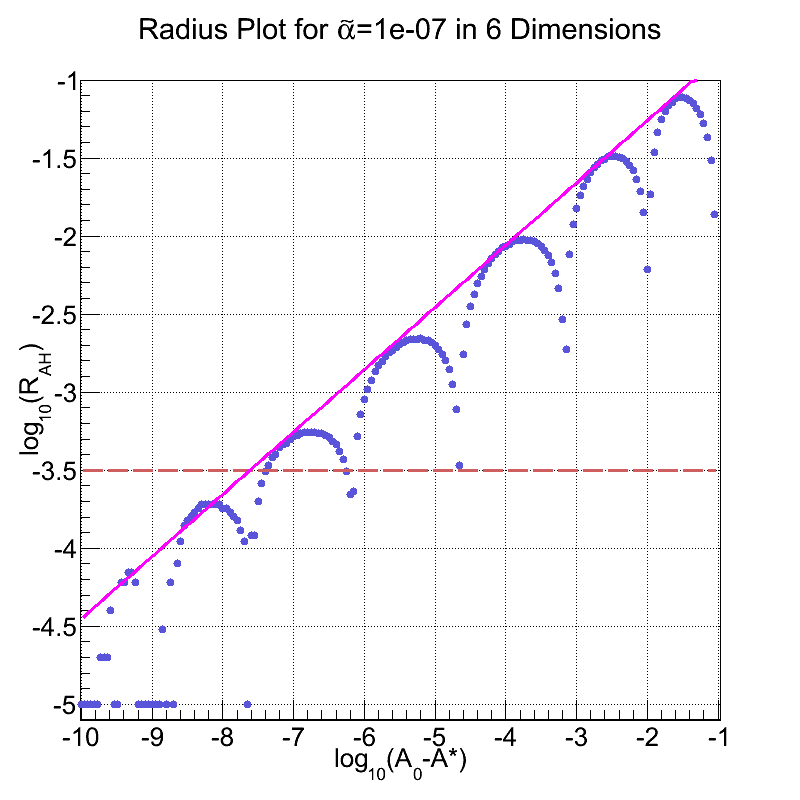}
\label{6DRadius-7}
}
\subfigure[$\tilde{\alpha}=5\times10^{-7}$]{
\includegraphics[width=0.4\linewidth]{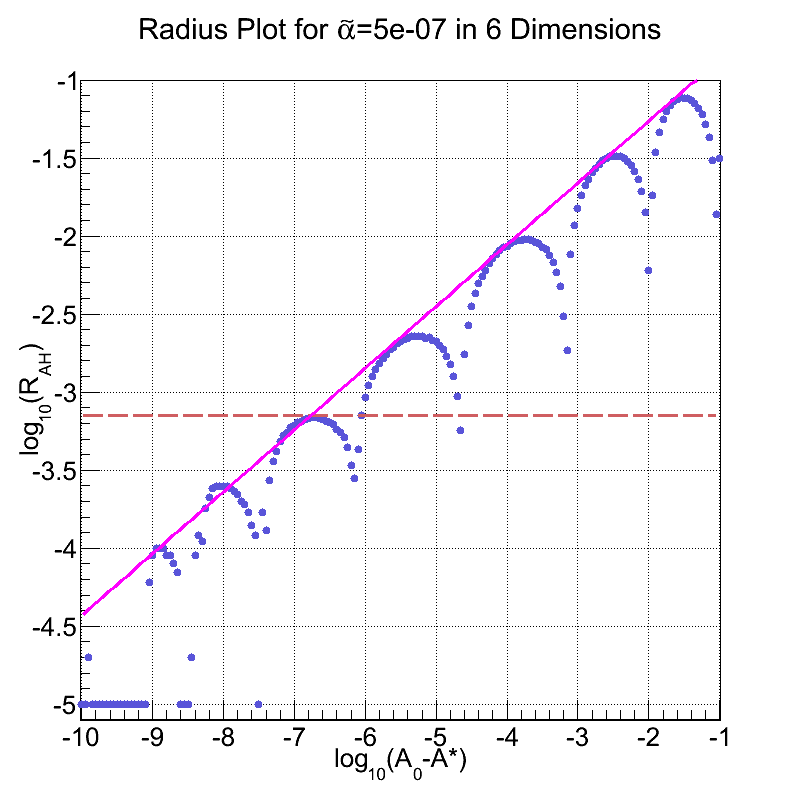}
\label{6DRadius5e-7}
}
\hspace{0.25in}
\subfigure[$\tilde{\alpha}=10^{-6}$]{
\includegraphics[width=0.4\linewidth]{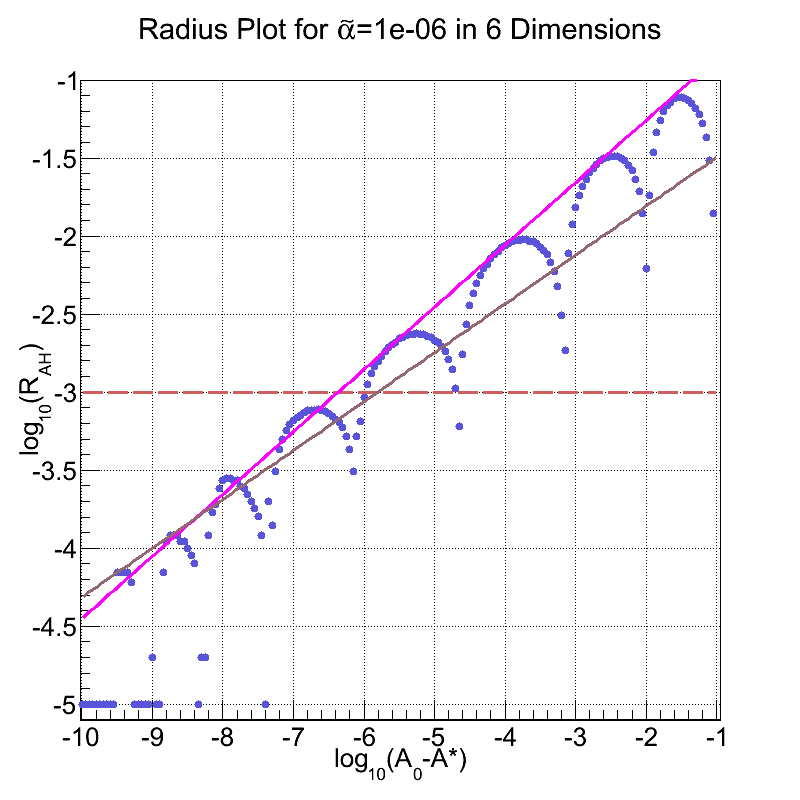}
\label{6DRadius-6}
}
\subfigure[$\tilde{\alpha}=10^{-5}$]{
\includegraphics[width=0.4\linewidth]{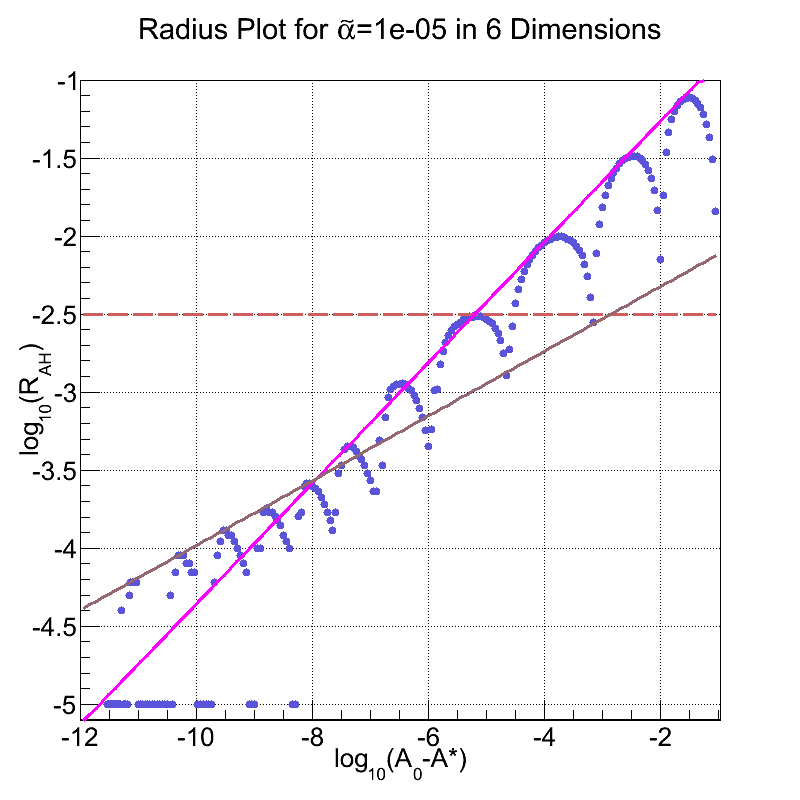}
\label{6DRadius-5}
}
\hspace{0.25in}
\subfigure[$\tilde{\alpha}=10^{-4}$]{
\includegraphics[width=0.4\linewidth]{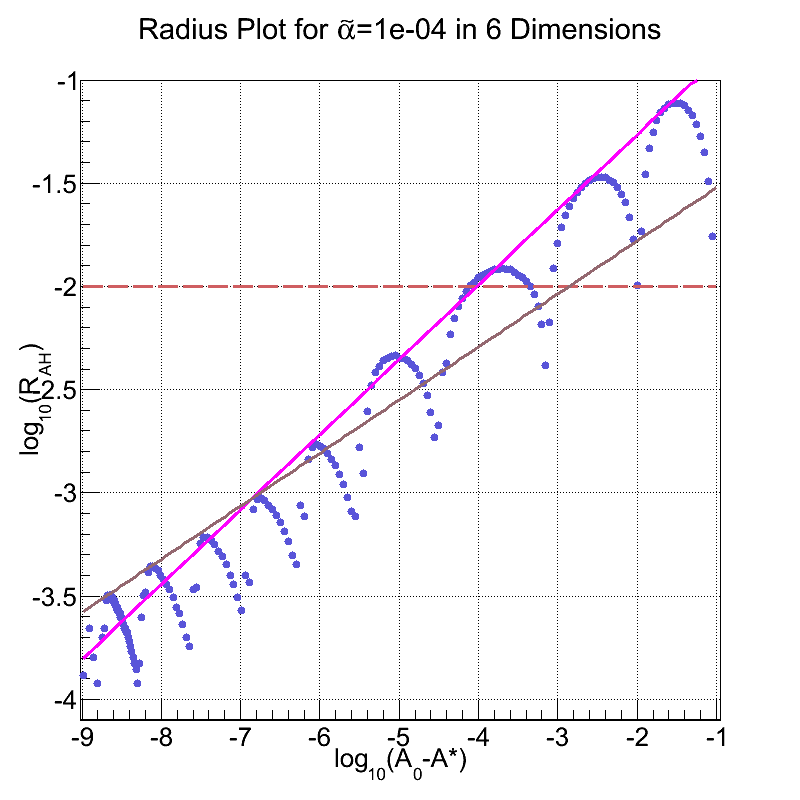}
\label{6DRadius-4}
}
\caption{Radius Scaling Plots - 6D. The lines represent the best-fit tangents to the curves in their respective regimes.}
\label{Radius6D}
\end{figure}

The radius plots of Figures \ref{Radius5D} and \ref{Radius6D} continue to exhibit cusps, but with a decreased period and slope. The $T^\mu_{\phantom{\mu}\mu}$ plots are also similar in the Gauss-Bonnet region to the general relativistic region in that they are approximately straight lines with oscillations superimposed. However, the slope changes quite suddenly when the transition from the general relativistic to Gauss-Bonnet region is made. The first important point is that the scaling plots are universal even in the Gauss-Bonnet region. This is illustrated for both 5D and 6D in Figure \ref{Universality}. There are qualitative differences in the scaling plots between 5D and 6D so we will now discuss the two cases separately.

In the 5D case there is evidence that the slope of the radius plot decreases continuously until a minimum radius is reached, i.e. that there is a radius gap. This is most evident in Figure \ref{5DRadius5e-7} but also appears to be the case in Figure \ref{5DRadius-6}. In the remaining 5D figures the numerics did not allow us to probe deeply enough into the Gauss-Bonnet region to fully observe this. 

A radius gap is not unexpected given the presence of the dimensionful Gauss-Bonnet parameter. Note that we focus on a radius gap instead of a mass gap because in 5D the former is trivial in light of Equation \ref{MAH}. 

\begin{figure}[ht!]
\centering
\subfigure[GR, slope$=0.826$]{
\includegraphics[width=0.4\linewidth]{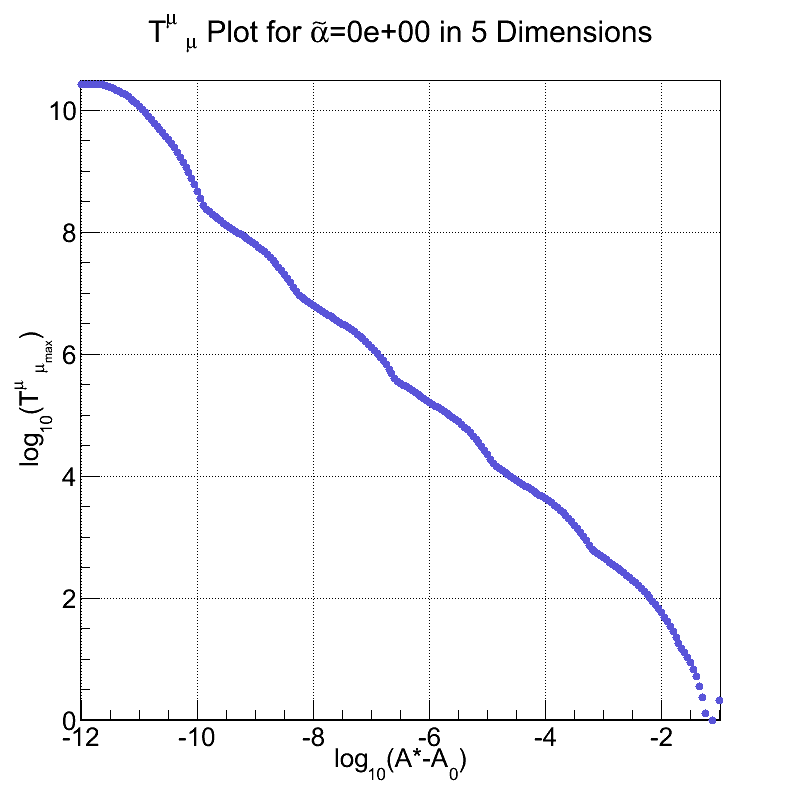}
\label{5DRicciGR}
}
\hspace{0.25in}
\subfigure[$\tilde{\alpha}=10^{-8}$]{
\includegraphics[width=0.4\linewidth]{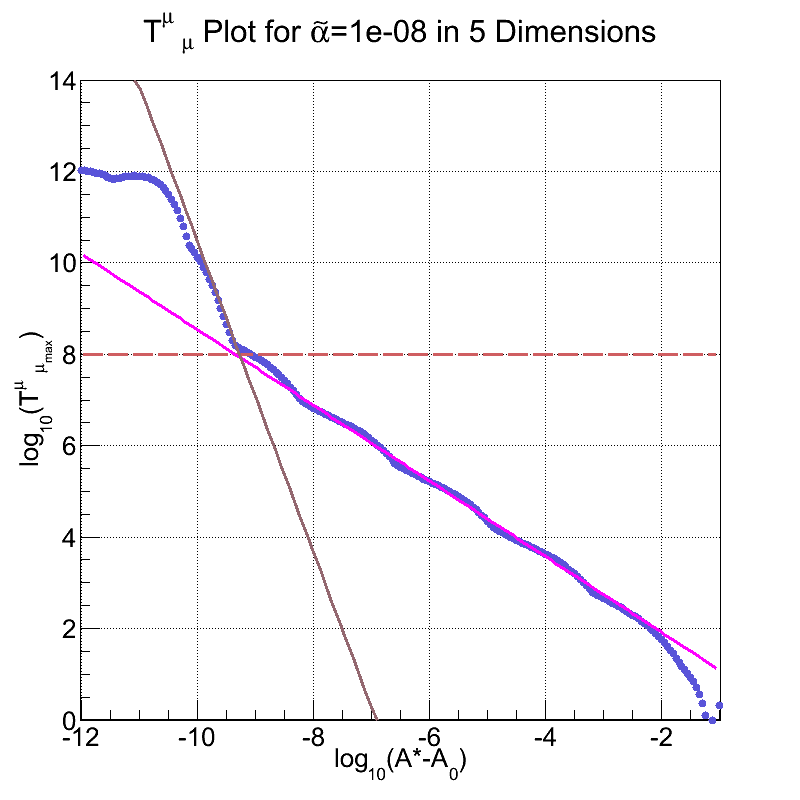}
\label{5DRicci-8}
}
\subfigure[$\tilde{\alpha}=10^{-7}$]{
\includegraphics[width=0.4\linewidth]{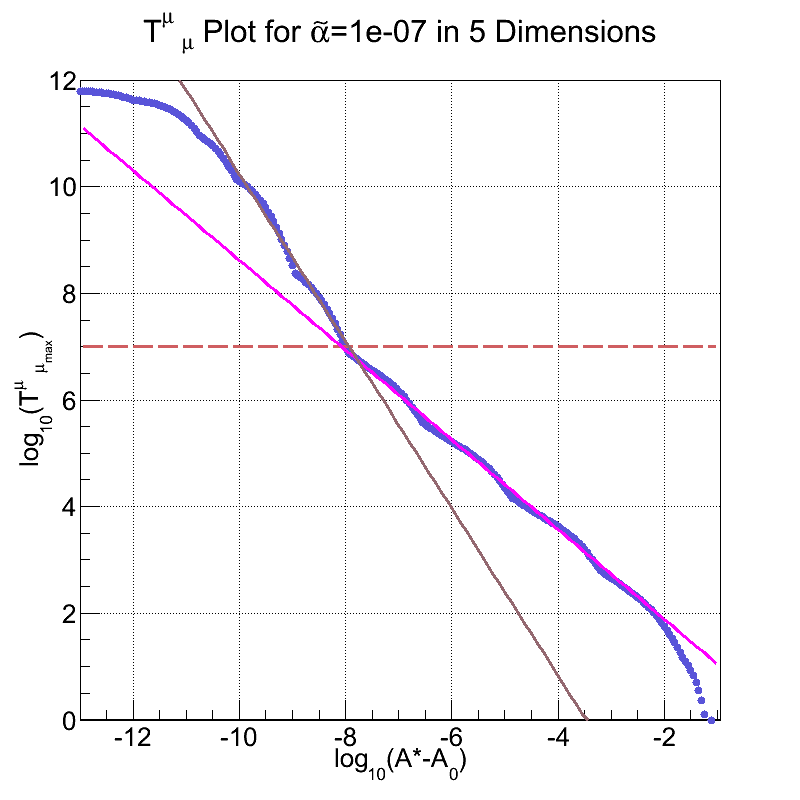}
\label{5DRicci-7}
}
\hspace{0.25in}
\subfigure[$\tilde{\alpha}=5\times10^{-7}$]{
\includegraphics[width=0.4\linewidth]{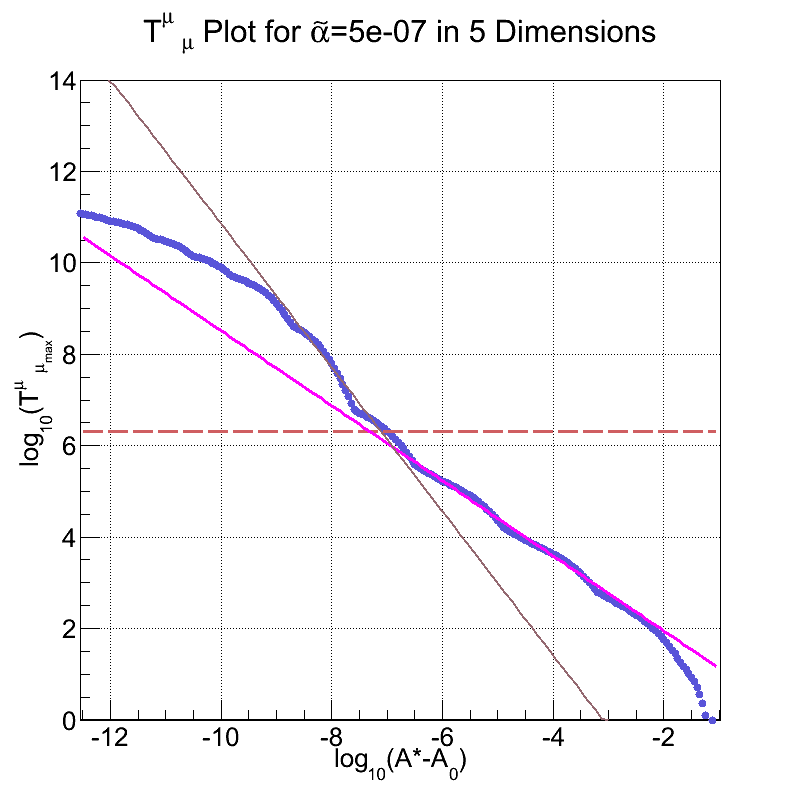}
\label{5DRicci5e-7}
}
\subfigure[$\tilde{\alpha}=10^{-6}$]{
\includegraphics[width=0.4\linewidth]{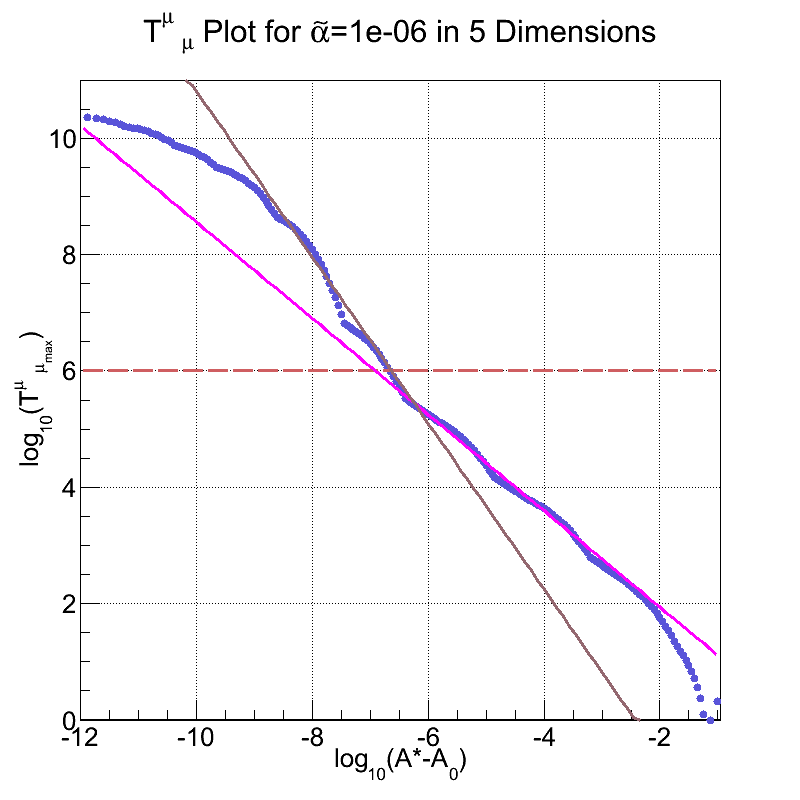}
\label{5DRicci-6}
}
\caption{$T^\mu_\mu$ Scaling Plots - 5D. The lines represent the best-fit mean slopes of the curves in their respective regimes.}
\label{Ricci5D}
\end{figure}

The $T^\mu_{\phantom{\mu}\mu}$ plots (Figure \ref{Ricci5D}), initially approximately straight, change slope quite suddenly as one moves from the general relativistic to the Gauss-Bonnet region, and then remain constant over a small range of $\log(dA)$. The slopes are given in Table \ref{Table 1}.  As criticality is approached the slope of the $T^\mu_{\phantom{\mu}\mu}$ plot gradually decreases, suggesting that there is a maximum value to $T^\mu_{\phantom{\mu}\mu}$ at the origin. This differs from general relativity, in which the critical solution is singular and $T^\mu_{\phantom{\mu}\mu}$ at the origin increases indefinitely as criticality is approached.  We emphasize again that these features are universal.

\begin{figure}[ht!]
\centering
\includegraphics[width=0.4\linewidth]{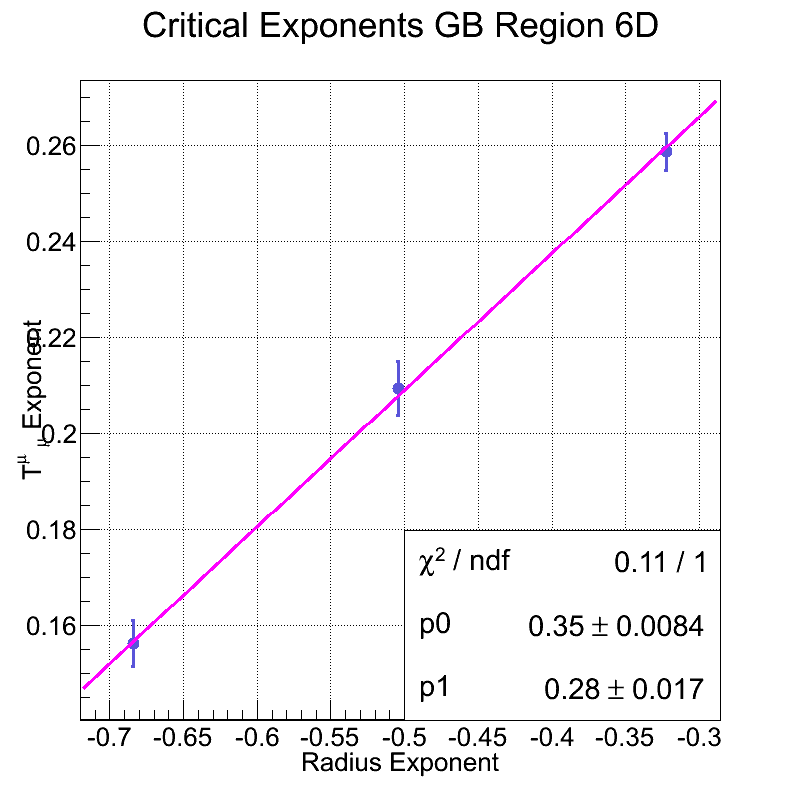}
\caption{Plot of Radius exponents vs $T^\mu_\mu$ 6D}
\label{Ricci vs Radius}
\end{figure}

In 6D things are different. There is no evidence of a radius gap in the radius scaling plots, and the slope of the $T^\mu_{\phantom{\mu}\mu}$ plots remains constant until we reach the limits of numerical accuracy. Thus it appears that there is a transition to a new set of scaling exponents, which are plotted in Table \ref{Table 2}.  Note that numerical uncertainties make the first and last entries in each column unreliable. The exponents are different for $T^\mu_{\phantom{\mu}\mu}$ and radius scaling, but the absolute value of both appear to increase with decreasing $\tilde{\alpha}$. Moreover a log-log plot of the three reliable Radius vs $T^\mu_{\phantom{\mu}\mu}$ exponents (Figure \ref{Ricci vs Radius}) reveals that they are related by:
\be
\gamma_{(T^\mu_{\phantom{\mu}\mu})} \approx -(2.24\pm0.04) \times \gamma_{(Radius)}^{0.28\pm0.02}
\label{eq:Ricci vs Radius}
\ee
This is to be compared to the general relativistic case in which the relation is determined purely by the dimension of the two quantities:
\be
\gamma_{(T^\mu_{\phantom{\mu}\mu})} = -2 \gamma_{(Radius)}
\ee

\begin{figure}[ht!]
\centering
\subfigure[GR, slope$=0.43$]{
\includegraphics[width=0.4\linewidth]{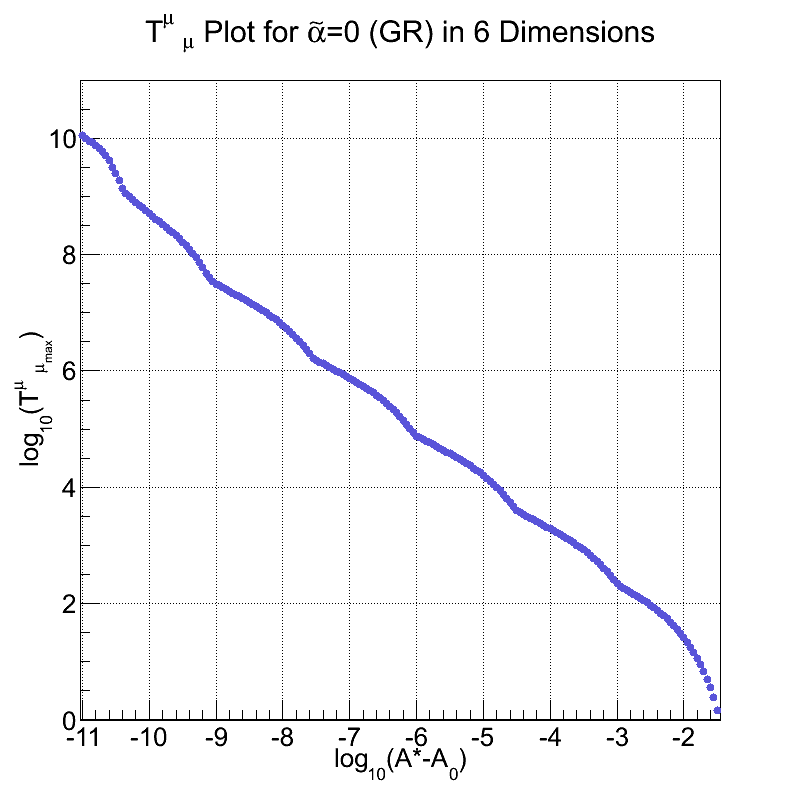}
\label{6DRicciGR}
}
\hspace{0.25in}
\subfigure[$\tilde{\alpha}=10^{-7}$]{
\includegraphics[width=0.4\linewidth]{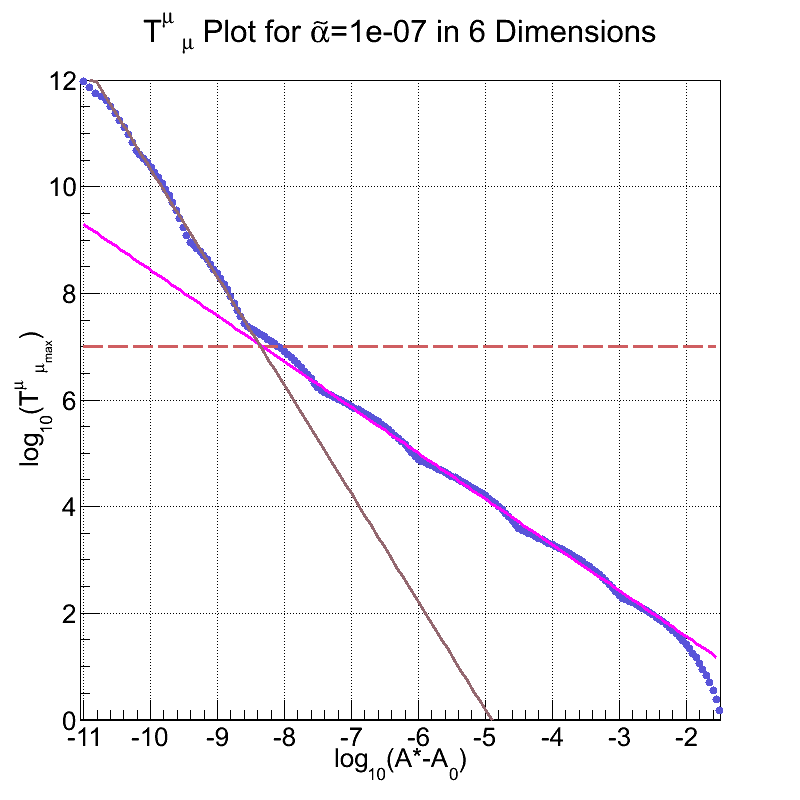}
\label{6DRicci-7}
}
\subfigure[$\tilde{\alpha}=5\times10^{-7}$]{
\includegraphics[width=0.4\linewidth]{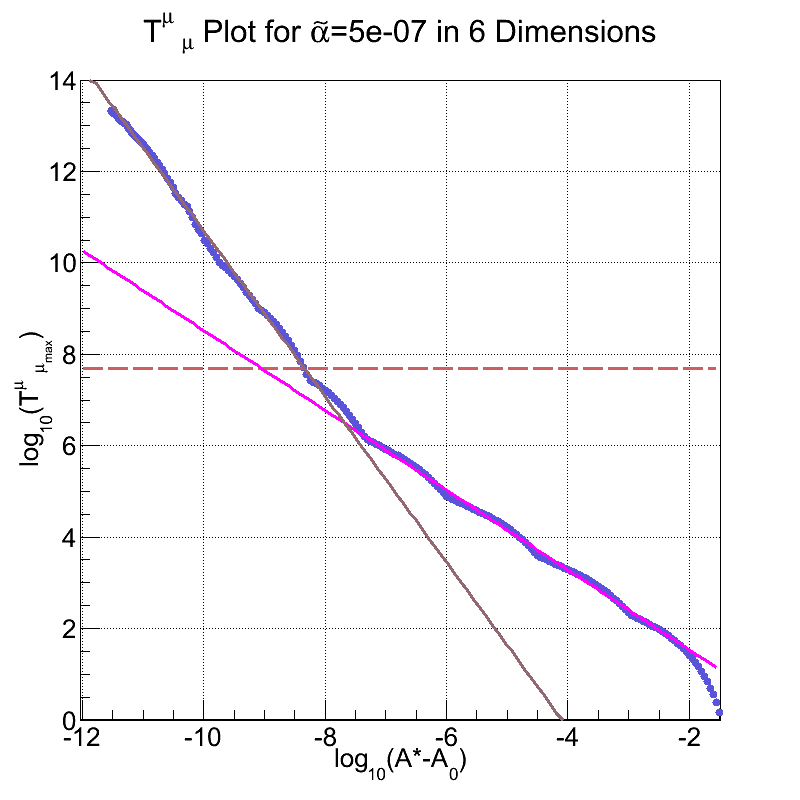}
\label{6DRicci5e-7}
}
\hspace{0.25in}
\subfigure[$\tilde{\alpha}=10^{-6}$]{
\includegraphics[width=0.4\linewidth]{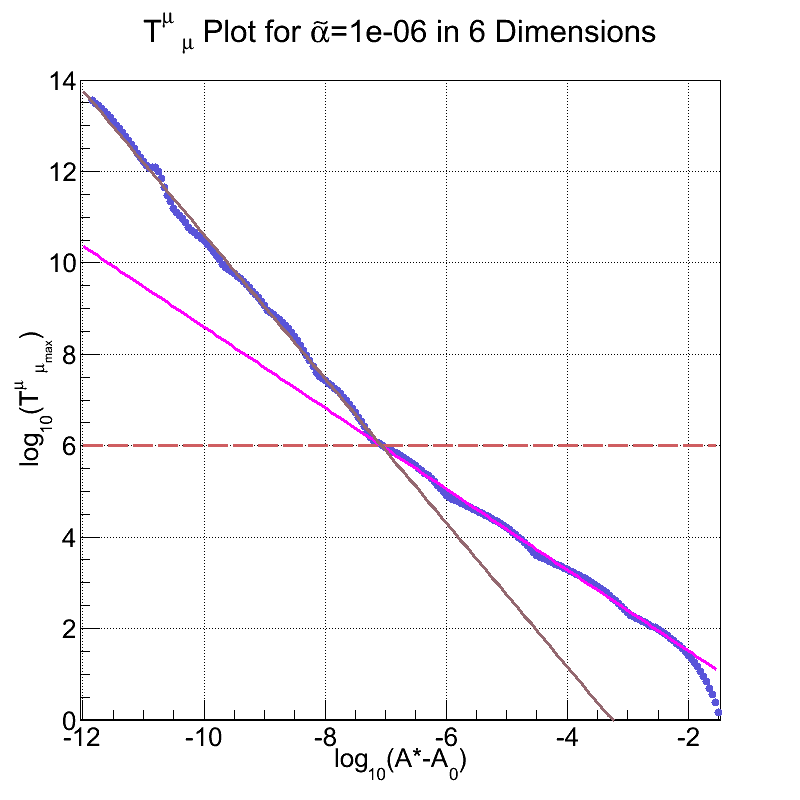}
\label{6DRicci-6}
}
\subfigure[$\tilde{\alpha}=10^{-5}$]{
\includegraphics[width=0.4\linewidth]{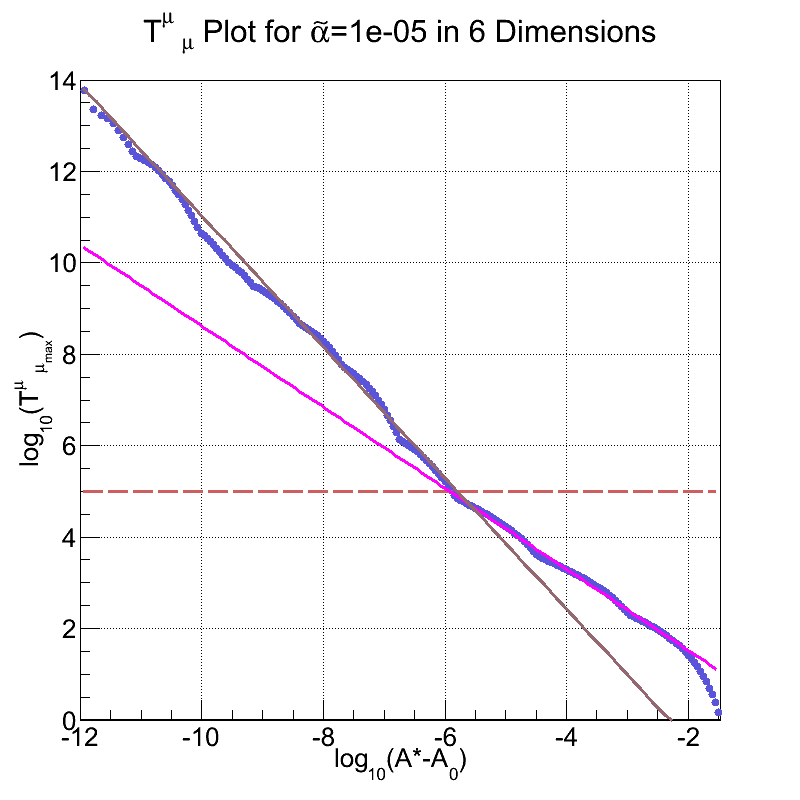}
\label{6DRicci-5}
}
\hspace{0.25in}
\subfigure[$\tilde{\alpha}=10^{-4}$]{
\includegraphics[width=0.4\linewidth]{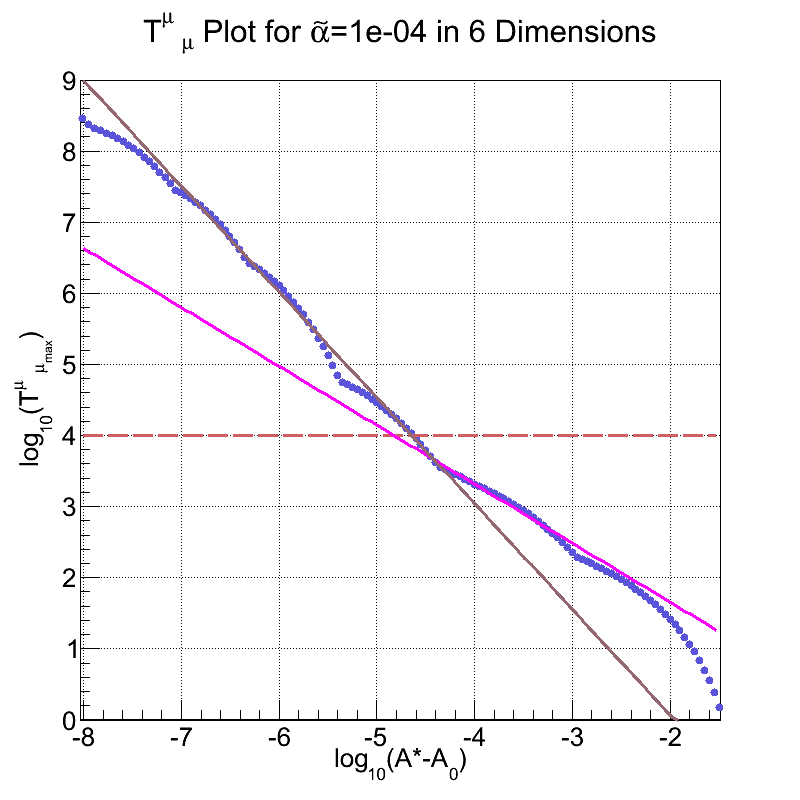}
\label{6DRicci-4}
}
\caption{$T^\mu_\mu$ Scaling Plots - 6D. The lines represent the best-fit mean slopes of the curves in their respective regimes.}
\label{Ricci6D}
\end{figure}

\begin{table}
\begin{center}
\begin{tabular}{|c|c|}
\hline
\centering $\tilde{\alpha}$ &  \centering 5D $T^\mu_\mu$ Scaling \tabularnewline
\hline
\centering $10^{-6}$   & \centering $-1.426\pm0.074$ \tabularnewline
\hline
\centering $5\times10^{-7}$   & \centering $-1.573\pm0.076$ \tabularnewline
\hline
\centering $10^{-7}$ &  \centering $-1.577\pm0.028$  \tabularnewline
\hline
\centering $10^{-8}$  & \centering $-3.397\pm0.049$  \tabularnewline
\hline
\end{tabular}  
\caption{5D $T^\mu_\mu$ scaling exponents in Gauss-Bonnet region.}
  \label{Table 1}
  \end{center}
\end{table}

\begin{table}
\begin{center}
\begin{tabular}{|c|c|c|}
\hline
\centering $\tilde{\alpha}$ &  \centering 6D $T^\mu_\mu$ Scaling& \centering  6D Radius Scaling \tabularnewline
\hline
\centering $10^{-4}$   & \centering $-1.488\pm0.128$ &  \centering $0.257\pm0.002$ \tabularnewline
\hline
\centering $10^{-5}$   & \centering $-1.433\pm0.016$ &  \centering $0.207\pm0.002$ \tabularnewline
\hline
\centering $10^{-6}$ &  \centering $-1.619\pm0.021$ &  \centering $0.313\pm0.002$ \tabularnewline
\hline
\centering $5\times10^{-7}$  & \centering $-1.814\pm0.016$ & \centering $0.476\pm0.002$ \tabularnewline
\hline
\centering $10^{-7}$ &  \centering $-2.029\pm0.027$ & \centering \centering $0.417\pm0.002$ \tabularnewline
\hline
\end{tabular}  
\caption{6D $T^\mu_\mu$ and AH radius scaling exponents in Gauss-Bonnet region.}
  \label{Table 2}
  \end{center}
\end{table}

\clearpage

\chapter{Conclusions}
\label{ch:Conclusions}

The first goal of this thesis was to perform the Hamiltonian analysis of spherically symmetric Lovelock gravity in terms of a mass function.  Although our analysis was classical our results may be useful for the study of quantum gravity.  Our second goal was to use the equations of motion for a massless scalar field, coupled to Lovelock gravity to numerically investigate the effects of higher dimensions and higher curvature terms on the formation of small black holes; specifically we were interested in the effect on Choptuik scaling.

In Chapter \ref{ch:LLADM}, \nameref{ch:LLADM}, we performed the Hamiltonian analysis for spherically symmetric spacetimes in general Lovelock gravity in arbitrary dimensions.  We showed that, as in general relativity, the areal radius and the generalized Misner-Sharp quasi-local mass $M$ are natural canonical variables that yield the remarkably simple, geometrical Lagrangian density of Equation \ref{eq:geometrodynamic L2}

\begin{align}
{\cal L}= P_M\dot{M} + P_S\dot{S}- N^M M'-N^S P_S \nonumber
\end{align}
for the generic theory. Using these variables also enabled us to rigorously derive the super-Hamiltonian and super-momentum constraints given in Equations \ref{eq:super H} and \ref{eq:super Hr}

\bea
H &=& \left(\frac{P_M y}{R'} -\frac{\Lambda}{R'}\right)M' + y P_s\, ,\nonumber \\
H_r &=& P_M M' +P_S S' \nonumber
\eea
for the most general theory, a task that would have be daunting at best, if not impossible, in terms of ADM variables.  Note that the super-Hamiltonian and the super-momentum constraints are in the same form as in general relativity.  All information specific to Lovelock theory is hidden in the mass function, $M$, $y$ and their relationships to the ADM variables.

These results are useful: the geometrodynamical variables allow the physical phase space of the vacuum theory to be explicitly parametrized in terms of the ADM mass and its conjugate momentum, as done for general relativity by Kucha\v{r} \cite{Kuchar1994}. This in turn provides a rigorous starting point for the quantization of Lovelock black holes using the techniques of \cite{Louko1996}. 

Finally, the simple form of the Hamiltonian allows us to gauge fix and derive the Hamiltonian equations of motion for the collapse of self gravitating matter in flat slice coordinates.  The equations of motion for a massless scalar field are given by Equations \ref{psidot} and \ref{Pisdot},

\begin{align}
\dot{\psi} =& N \left( \frac{\Pis}{{\cal A}_{n-2} R^{n-2}} + \psip\frac{N_r}{N}  \right),\nonumber \\
\dot{P}_\psi =& \left[ N \left( {\cal A}_{n-2} R^{n-2} \psip +  \Pis \frac{N_r}{N}  \right) \right]^\prime \nonumber
\end{align}
with the consistency conditions and Hamiltonian constraint given by Equations \ref{sigma1}, \ref{Nsig} and \ref{Cconstraint}: 
\be
\label{sigma1conc}
N^\prime \frac{\partial M}{\partial (N_r/N)} + N P_\psi \psip = 0, \nonumber
\ee

\begin{align}
\label{Nsigconc}
M= \frac{(n-2){\cal A}_{n-2}}{2\kappa_n^2}\sum_{p=0}^{[n/2]} \tilde{\alpha}_{(p)}R^{n-1-2p}\left(\frac{N_r}{N}\right)^{2p} \nonumber
\end{align}
and

\be
\label{Cconstraintconc}
-M^\prime + \frac{1}{2} \left( \frac{\Pis^2}{{\cal A}_{n-2}R^{n-2}} + {\cal A}_{n-2} R^{n-2} \psip^2 \right) + \Pis \psip  \frac{N_r}{N}= 0. \nonumber
\ee
Note that only the definition of $N_r/N$ in terms of $M$, from Equation \ref{Nsig} differentiates between general relativity and any other form of Lovelock gravity.  

The equations of motion plus the constraints put us in a position to study the dynamics of black hole formation in generic Lovelock gravity which was the subject of Chapter \ref{ch:CSEGBG}.

In Chapter \ref{ch:HDCS}, \nameref{ch:HDCS} we successfully confirmed the existence of cusps in the mass scaling function, Equation \ref{Chop}, in 4 to 8 dimensions as observed in 4 dimensions by Ziprick and Kunstatter \cite{Ziprick2009c,Ziprick2009a}.  This can be seen in Figures \ref{mass scaling plots} and \ref{8DChop} and are displayed below in 5 to 8 dimensions.

\begin{figure}[ht!]
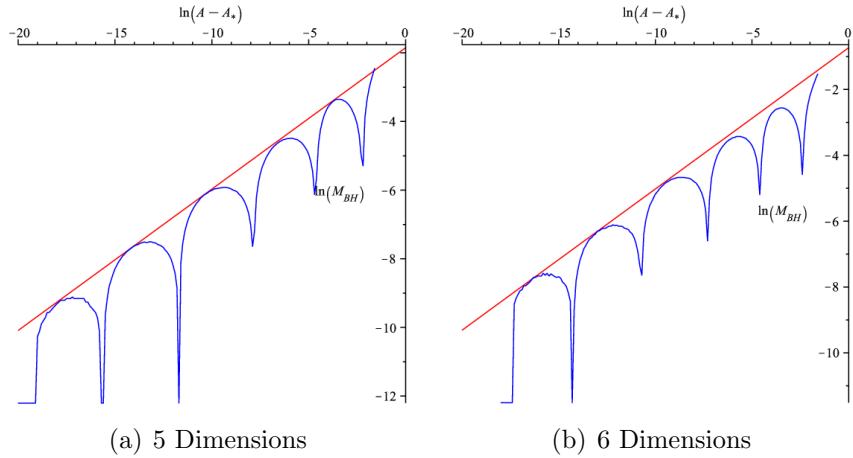
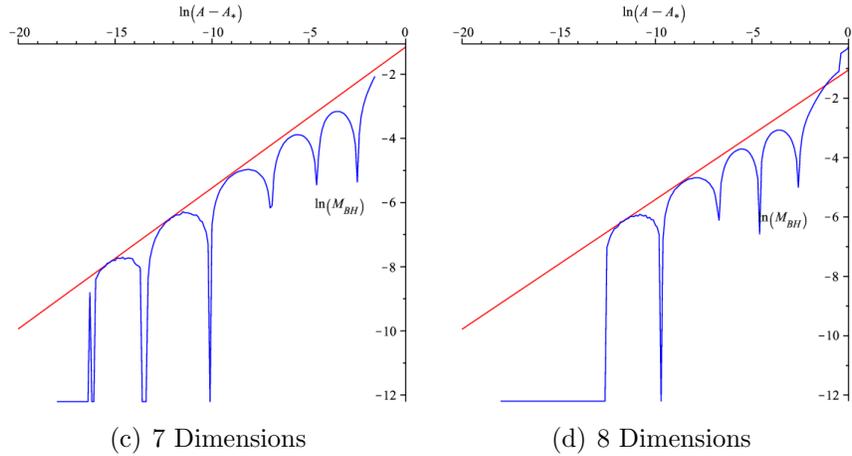

\centering
\hspace{0.25in}
\subfigure[5 Dimensions]{
\includegraphics[width=0.4\linewidth]{5Dchop.png}
}
\subfigure[6 Dimensions]{
\includegraphics[width=0.4\linewidth]{6Dchop.png}
}
\hspace{0.25in}
\subfigure[7 Dimensions]{
\includegraphics[width=0.4\linewidth]{7Dchop.png}
}
\subfigure[8 Dimensions]{
\includegraphics[width=0.4\linewidth]{8Dchop.png}
}
\caption*{Mass Scaling in 5, 6, 7 and 8 dimensions}
\end{figure}

In addition the mass scaling plots obtained using PG coordinates give critical data which agrees with previous results in 4, 5 and 6 dimensions.  However, in 7 dimensions, our critical exponent and echoing period agree with Bland {\it et al} \cite{Bland2005} but disagree with the values obtained by Sorkin and Oren \cite{Sorkin2005}. It is important to note that \cite{Bland2005} claimed that the critical exponent was a monotonic function of spacetime dimension that asymptotes to 1/2, whereas \cite{Sorkin2005} suggested that the critical exponent peaks near $n=10$. While both claims are intriguing, only one (at most) can be right. Since we were limited by numerics to $n\leq7$ we cannot make any definitive claims about the asymptotic behaviour. Our result in $n=7$ supports that of \cite{Bland2007}, as does our more tentative result in $n=8$. It is however impossible to make definitive claims about the asymptotic behaviour without pushing the PG calculation to higher dimensions.  This is the topic of ongoing work using advanced numerical techniques such as adaptive mesh refinement.

In addition to applying more advanced numerical techniques to the study of higher dimensional collapse it is intriguing to consider the possibility of generalizing the work of Gundlach\cite{Gundlach1997} as described in Section \ref{sec:Choptuik} to the higher dimensional case to get a more analytical understanding of the higher dimensional critical exponents.

In Chapter \ref{ch:CSEGBG}, \nameref{ch:CSEGBG} we studied the effects of the Gauss-Bonnet term on the dynamics of the collapse of a massless scalar field minimally coupled to gravity in five and six spacetime dimensions.  We found that the Gauss-Bonnet term destroys the self-similar behaviour in 5 dimensions, as demonstrated by the fact that near criticality the scalar field at the origin oscillates with a constant period as seen in Figure \ref{5Dscalar}.  In the 6 dimensional case numerics would not allow us to calculate enough oscillations of the solution to conclude that the discrete self-similar behaviour was broken by the Gauss-Bonnet parameter as seen in Figure \ref{6Dscalar}.  The relevant parts of Figures \ref{5Dscalar} and \ref{6Dscalar} are displayed below for convenience.
\begin{figure}[ht!]
\centering
\hspace{0.25in}
\subfigure[5D,$\psi(0,T_{PG})$ near criticality, $\tilde{\alpha}=10^{-6}$]{
\includegraphics[width=0.4\linewidth]{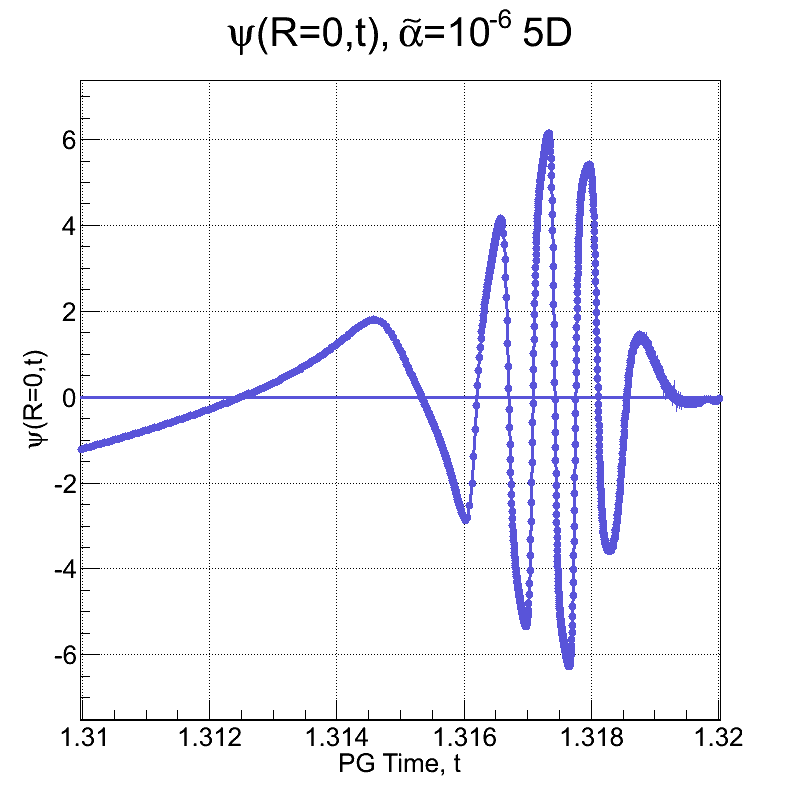}
}
\hspace{0.25in}
\subfigure[6D,$\psi(0,T_{PG})$ near criticality, $\tilde{\alpha}=10^{-6}$]{
\includegraphics[width=0.4\linewidth]{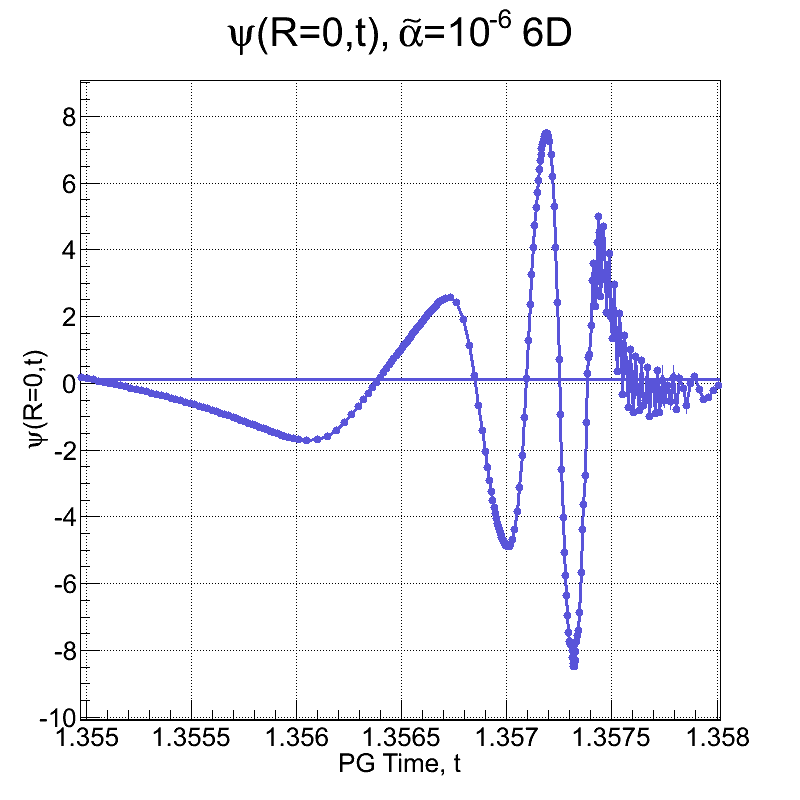}
}
\caption*{Scalar field oscillations}
\end{figure}
The period in five dimensions is proportional to roughly the cube root of the Gauss-Bonnet parameter. While these results differ from those in \cite{Golod2012} it must be emphasized that we have explored a different range of Gauss-Bonnet parameter, and this may account for the difference.

We also showed the existence of modified, but still universal, horizon and energy momentum scaling plots near criticality. We found evidence for the existence of a radius gap in five dimensions but not in six dimensions as can be seen in Figures \ref{Radius5D} and \ref{Radius6D}, excerpts of which are displayed below.
\begin{figure}[ht!]
\centering
\subfigure[5D, $\tilde{\alpha}=10^{-6}$]{
\includegraphics[width=0.4\linewidth]{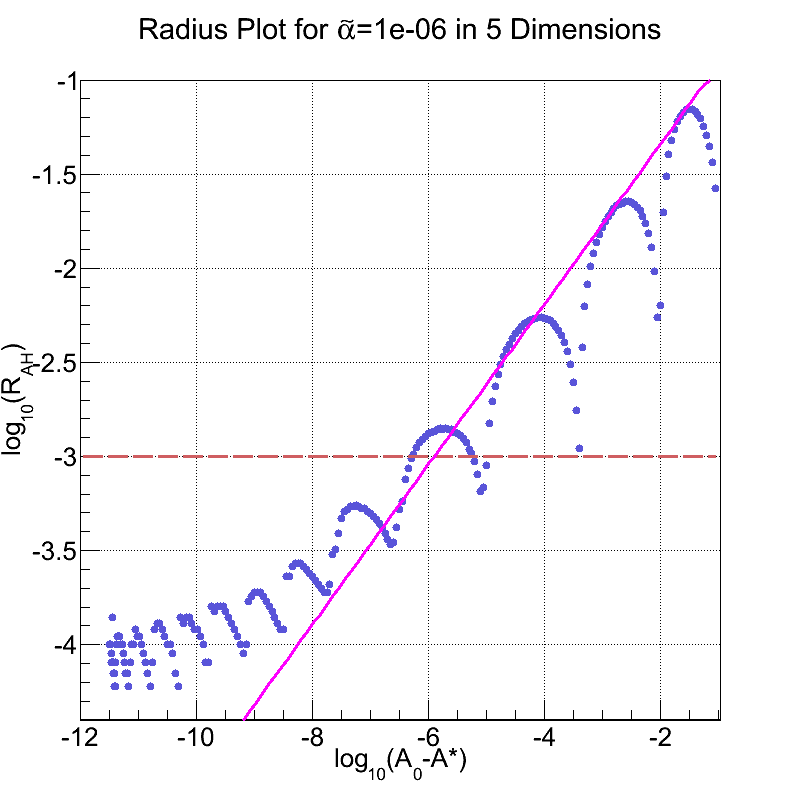}
}
\hspace{0.25in}
\subfigure[6D, $\tilde{\alpha}=10^{-4}$]{
\includegraphics[width=0.4\linewidth]{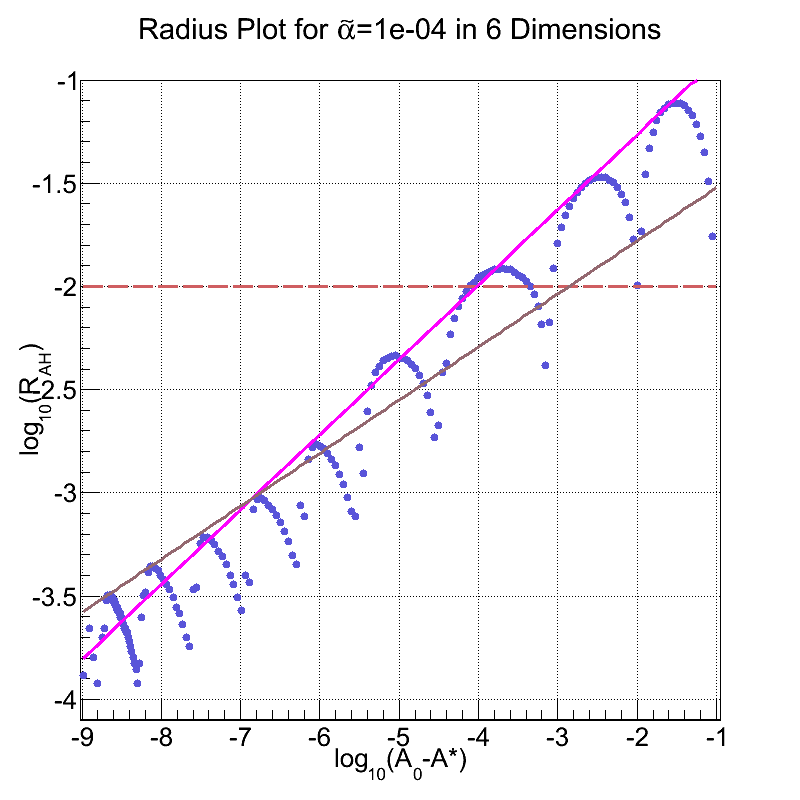}
}
\caption*{Radius Scaling Plots - 5D and 6D.}
\end{figure}
This qualitative difference is not completely unexpected. As mentioned below Equation \ref{Pisdot_code}, the time evolution equation in five dimensions is special, containing one less term than in the higher-dimensional cases. 


It is clearly of interest to confirm our results with further simulations and to try to understand analytically the source of the new scaling behaviour.  It would also be interesting to model the collapse of third order Lovelock gravity and Lovelock gravity coupled to other forms of matter, such as the electromagnetic field described in Section \ref{sec:CSF} This may be possible using the code with adaptive mesh refinement mentioned earlier.  These projects are currently underway.

\begin{appendix}

\chapter{Appendices}
\label{ap:Appendices}

\section{Derivations}
\label{app:derivations}
In this appendix, we present the details of several lengthy derivations of results in the main text.
The following relations will be useful for much of the following:
\begin{align}
y\Lambda \delta F=&y\Lambda \biggl(\frac{\delta({R'}^2)}{\Lambda^2}-2y\delta y\biggl)-2y(y^2+F)\delta \Lambda, \label{useful0}\\
y\Lambda \frac{\delta F}{F}=&2\delta(y\Lambda)-2y\delta \Lambda-R'\delta \biggl(\ln\biggl|\frac{R'+y\Lambda}{R'-y\Lambda}\biggl|\biggl). \label{useful1}
\end{align}
Equation \ref{useful1} is shown as
\begin{align}
y\Lambda \frac{\delta F}{F}=&y\Lambda \frac{\delta(({R'}^2-y^2\Lambda^2)/\Lambda^2)}{({R'}^2-y^2\Lambda^2)/\Lambda^2}  \nonumber \\
=&-2(y\Lambda)^2\frac{\delta(y\Lambda)}{{R'}^2-(y\Lambda)^2}-2y\delta \Lambda+y\Lambda \frac{2R'\delta R'}{{R'}^2-y^2\Lambda^2} \nonumber \\
=&2\delta(y\Lambda)-2y\delta \Lambda-2{R'}^2\frac{\delta(y\Lambda)}{{R'}^2-(y\Lambda)^2}+y\Lambda \frac{2R'\delta R'}{{R'}^2-y^2\Lambda^2} \nonumber \\
=&2\delta(y\Lambda)-2y\delta \Lambda-R'\delta \biggl(\ln\biggl|\frac{R'+y\Lambda}{R'-y\Lambda}\biggl|\biggl).
\end{align}
Equations \ref{useful0} and \ref{useful1} will also be used by replacing $\delta$ by $\partial_t$ or $\partial_x$.

\subsection{Lagrangian Density~(\ref{eq:lovelock simplified})}
\label{appendix1}
We now derive the Lagrangian density of Equation \ref{eq:lovelock simplified} from Equation \ref{L_p}.  Using the binomial expansion for the last two terms in Equation \ref{L_p} yields
\begin{align}
& 2p(p-1) \frac{(D^2R)^2 - (D^AD_BR)(D^BD_AR)}{R^2} \left(\frac{k-(DR)^2}{R^2}\right)^{p-2} \nonumber \\
& + p{}^{(2)}{\cal R} \left(\frac{k-(DR)^2}{R^2}\right)^{p-1} \nonumber \\
=&R^{-2(p-1)} \biggl[pk^{p-1}{}^{(2)}{\cal R}+ \nonumber \\
&\{(D^2R)^2 - (D^AD_BR)(D^BD_AR)\}\sum_{i=0}^{p-2}\frac{2(i+1)p!k^{p-2-i}(-1)^i(DR)^{2i} }{(i+1)!(p-2-i)!}\nonumber \\
& - (D^A R)\left(D^2D_AR - D_AD^2R\right)\sum_{i=0}^{p-2}\frac{2p!k^{p-2-i}(-1)^{i}(DR)^{2i} }{(i+1)!(p-2-i)!}\biggl],
\end{align}
where we used the following two-dimensional identity:
\begin{align}
(D R)^2 {}^{(2)}{\cal R}\equiv  2(D^A R)\left(D^2D_AR - D_AD^2R\right). 
\end{align}
This identity can be derived from Equation 2.10 of \cite{Taves2012}.

Equation \ref{L_p} now reduces to
\begin{align}
\label{eq:lp1}
{\cal L}_{(p)} =& \frac{(n-2)!}{(n-2p)!} \biggl[(n-2p)(n-2p-1)\left(\frac{k-(DR)^2}{R^2}\right)^{p} \nonumber \\
& -2p(n-2p)\frac{D^2R}{R}\left(\frac{k-(DR)^2}{R^2}\right)^{p-1} + pk^{p-1}R^{2-2p}{}^{(2)}{\cal R} \nonumber \\
&  + \sum_{i=0}^{p-2} \frac{2(-1)^ik^{p-2-i}p! (DR)^{2i}}{(i+1)!(p-2-i)!} \biggl\{(i+1)\frac{(D^2R)^2 - (D^AD_BR)(D^BD_AR)}{R^{2p-2}} \nonumber \\
&- \frac{D^AR(D^2D_AR - D_AD^2R)}{R^{2p-2}} \biggl\}\biggl].
\end{align}
Using integration by parts, we can rewrite the term in curly brackets in Equation \ref{eq:lp1} as
\begin{align}
&\sum_{i=0}^{p-2} \frac{2(-1)^ik^{p-2-i}p! (DR)^{2i}}{(i+1)!(p-2-i)!} R^{n-2p}\biggl\{(i+1)[(D^2R)^2 - (D^AD_BR)(D^BD_AR)] \nonumber \\
&- D^AR(D^2D_AR - D_AD^2R)\biggl\} \nonumber \\
=&\sum_{i=0}^{p-2} \frac{2(-1)^ik^{p-2-i}p! }{(i+1)!(p-2-i)!}(DR)^{2i}D_A(R^{n-2p}) \biggl\{ \frac12D^A((DR)^2)- (D^2R)(D^AR)\biggl\} \nonumber \\
&+\mbox{(t.d.)},
\end{align}
where we used the following identity:
\begin{align}
&(D^2R)^2 - (D^AD_BR)(D^BD_AR) \nonumber \\
&+ \frac{(D(DR)^2)^2}{2(DR)^2} - \frac{(D^AR)(D_A(DR)^2)(D^2R)}{(DR)^2} \equiv 0.
\end{align}
Using the above result together with integration by parts and the following identity;
\begin{align}
\sum_{i=0}^{p-2} \frac{2(-1)^ik^{p-2-i}p! (DR)^{2i}}{(i+1)!(p-2-i)!}=&-\sum_{w=1}^{p-1} \frac{2(-1)^{w}k^{p-1-w}p! (DR)^{2w}(DR)^{-2}}{w!(p-1-w)!} \nonumber \\
=&-\sum_{w=0}^{p-1} \frac{2(-1)^{w}k^{p-1-w}p! (DR)^{2w}(DR)^{-2}}{w!(p-1-w)!} \nonumber \\
&+2pk^{p-1} (DR)^{-2} \nonumber \\
=&\frac{2pk^{p-1}-2p(k-(DR)^2)^{p-1}}{(DR)^{2}},
\end{align}
we can rewrite Equation \ref{eq:lp1} in the form of Equation \ref{eq:lovelock simplified} up to a total derivative.

\subsection{Lagrangian Density (\ref{Lag-1})}
\label{appendix2}
In this appendix, we show the derivation of the Lagrangian density of Equation \ref{Lag-1} from the action given by Equation \ref{action-3}.  { While we consider only the spherically symmetric case $(k=1$) in the main text, here we derive the equations for general $k$.}

For this purpose, we separate Equation \ref{action-3} into two portions:
\begin{align}
I_{\rm M}=&I_1+I_2+\mbox{(t.d.)}, \label{I_M} \\
I_1:=& \frac{(n-2)V_{n-2}^{(k)}}{2\kappa_n^2}\sum^{[n/2]}_{p=0}\int d^2{\bar y}{\tilde \alpha}_{(p)} \nonumber \\
&\times\Biggl[2pk^{p-1}R^{n-2p-1}y(N_r\Lambda)'-\frac{2pk^{p-1}N}{n-2p}\biggl((R^{n-2p})'\Lambda^{-1}\biggl)'  \nonumber \\
&+ pR^{n-2p-1}\biggl\{k^{p-1}-(k-F)^{p-1}\biggl\}(\Lambda N_r y+\Lambda^{-1}NR')\frac{F'}{F}   \nonumber \\
& + (n-2p-1)\biggl\{\left(k-F\right)^{p} +2pk^{p-1}F\biggl\}N\Lambda R^{n-2-2p}\Biggr],\label{I1}\\
I_2:=&\frac{(n-2)V_{n-2}^{(k)}}{2\kappa_n^2}\sum^{[n/2]}_{p=0}\int d^2{\bar y}p{\tilde \alpha}_{(p)} R^{n-2p-1} \nonumber \\
&\times\biggl[-2k^{p-1}y {\dot \Lambda}-\{k^{p-1}-(k-F)^{p-1}\}\Lambda y\frac{\dot F}{F}\biggl].
\end{align}
In order to perform the variation of $I_M$, we have to deal with the terms containing $F'$ and ${\dot F}$.  Using the binomial expansion and integration by parts, we can rewrite the second line of Equation \ref{I1} as
\begin{align}
& \frac{(n-2)V_{n-2}^{(k)}}{2\kappa_n^2}\sum^{[n/2]}_{p=0}{\tilde \alpha}_{(p)} pR^{n-2p-1}\biggl\{k^{p-1}-(k-F)^{p-1}\biggl\}(\Lambda N_r y+\Lambda^{-1}NR')\frac{F'}{F}  \nonumber \\
=&-\frac{(n-2)V_{n-2}^{(k)}}{2\kappa_n^2}\sum^{[n/2]}_{p=2}{\tilde \alpha}_{(p)} pR^{n-2p-1}\times \nonumber \\
&\sum_{w=0}^{p-2}\frac{(p-1)!(-1)^{p-1-w}}{w!(p-1-w)!}k^{w}F^{p-1-w}(\Lambda N_r y+\Lambda^{-1}NR')\frac{F'}{F}   \nonumber \\
=&-\frac{(n-2)V_{n-2}^{(k)}}{2\kappa_n^2}\sum^{[n/2]}_{p=2}{\tilde \alpha}_{(p)} p\sum_{w=0}^{p-2}\frac{(p-1)!(-1)^{p-1-w}k^{w}}{w!(p-1-w)!}\nonumber \\
&\times \biggl\{ \sum_{j=0}^{p-2-w}\frac{2(p-2-w)!(-1)^{p-2-w-j}}{j!(p-2-w-j)!}\frac{(N_r R^{n-2p-1}\Lambda^{1-2j}{R'}^{2j})'y^{2(p-w-j)-1}}{2(p-w-j)-1} \nonumber \\
&+\Lambda N_r yR^{n-2p-1}F^{p-2-w}(\Lambda^{-2}{R'}^2)' -(\Lambda^{-1}NR'R^{n-2p-1})'\frac{F^{p-1-w}}{p-1-w}\biggl\}.\label{app2-1}
\end{align}
This expression does not contain $y'$ and can be used to obtain $P_R$.  An important observation is that the only part of the action $I_M$ in Equation \ref{I_M} that contributes to $P_\Lambda$ is $I_2$.  After a tedious but straightforward calculation, using binomial expansion and integration by parts yet again, we can rewrite the last term in $I_2$ as
\begin{align}
&-\frac{(n-2)V_{n-2}^{(k)}}{2\kappa_n^2}\sum^{[n/2]}_{p=0}\int d^2{\bar y}p{\tilde \alpha}_{(p)} R^{n-2p-1}\biggl\{k^{p-1}-(k-F)^{p-1}\biggl\}\Lambda y\frac{\dot F}{F} \nonumber \\
=&-\frac{(n-2)V_{n-2}^{(k)}}{\kappa_n^2} \nonumber \\
&\times\sum^{[n/2]}_{p=2}\int d^2{\bar y}p{\tilde \alpha}_{(p)} \sum_{w=1}^{p-1}\frac{(p-1)!(-1)^wk^{p-1-w}}{w!(p-1-w)!}R^{n-2p-1} F^{w-1} y\Lambda^{-2}{\dot \Lambda}{R'}^2 \nonumber \\
&+\frac{(n-2)V_{n-2}^{(k)}}{\kappa_n^2} \nonumber \\
&\times\sum^{[n/2]}_{p=2}\int d^2{\bar y}p{\tilde \alpha}_{(p)} \sum_{w=1}^{p-1}\frac{(p-1)!(-1)^wk^{p-1-w}}{w!(p-1-w)!}\sum_{j=0}^{w-1}\frac{(w-1)!(-1)^{w-1-j}}{j!(w-1-j)!} \nonumber \\
&\times \biggl[\frac{1}{2(w-j)+1}\partial_t(R^{n-2p-1}\Lambda^{1-2j}){R'}^{2j}y^{2(w-j)+1} \nonumber \\
&- \sum_{q=0}^{2(w-j)+1}\frac{(2w-2j)!(-1)^q}{q!(2w-2j+1-q)!}\frac{j{\dot R}^{2(w-j+1)}}{w-j+1} \nonumber \\
&\times(R^{n-2p-1}\Lambda^{1-2j}N^{-2(w-j)-1}{R'}^{2j-1+q}N_r^q)' \nonumber \\
&-\sum_{q=0}^{2(w-j)-1}\frac{(2w-2j-1)!(-1)^q}{q!(2w-2j-1-q)!}\frac{{\dot R}^{2w-2j-q}}{2w-2j-q} \nonumber \\
&\times(R^{n-2p-1}\Lambda^{-1-2j}N^{-2(w-j)+1}{R'}^{2j+1+q}N_r^q)'\biggl] \nonumber \\
&+\partial_t(\cdots)+\partial_x(\cdots).\label{app2-2}
\end{align}
Using Equations \ref{app2-1} and \ref{app2-2}, we obtain the Lagrangian density of Equation \ref{Lag-1}.

\subsection{Liouville Form (\ref{Liouville}) in Lovelock Gravity}
\label{app:Liouville}
In this appendix, we verify the Liouville form of Equation \ref{Liouville} in Lovelock gravity.  Note that the explicit form of $P_R$ is not used in the derivation.  In this appendix we take $\delta$ to be $\partial_t$.

Using Equations \ref{qlm-L} and \ref{eq:PM}, we write $P_M\delta M$ as
\begin{align}
P_M\delta M=&-\frac{(n-2){\cal A}_{n-2}}{2\kappa_n^2}\sum_{p=0}^{[n/2]}{\tilde \alpha}_{(p)}\frac{y\Lambda}{F}R^{n-2-2p} \nonumber \\ 
& \times\biggl[(n-1-2p)(1-F)^p\delta R-pR(1-F)^{p-1}\delta F\biggl] \nonumber \\
=&-\frac{(n-2){\cal A}_{n-2}}{2\kappa_n^2}\sum_{p=0}^{[n/2]}{\tilde \alpha}_{(p)}(n-1-2p)\frac{y\Lambda}{F}R^{n-2-2p}(1-F)^p\delta R \nonumber \\
&+\frac{(n-2){\cal A}_{n-2}}{2\kappa_n^2}\sum_{p=1}^{[n/2]}{\tilde \alpha}_{(p)}R^{n-1-2p}\frac{y\Lambda}{F}\delta F \nonumber \\
&+\frac{(n-2){\cal A}_{n-2}}{2\kappa_n^2}\sum_{p=2}^{[n/2]}{\tilde \alpha}_{(p)}pR^{n-1-2p}\sum_{w=1}^{p-1}\frac{(p-1)!(-1)^w}{w!(p-1-w)!}F^{w-1}y\Lambda\delta F.
\end{align} 
Using Equation \ref{useful1} for the second term and Equation \ref{useful0} for the last term together with the binomial expansion, we obtain
\begin{align}
P_M\delta M=&-\frac{(n-2){\cal A}_{n-2}}{2\kappa_n^2}\sum_{p=0}^{[n/2]}{\tilde \alpha}_{(p)}(n-1-2p)\frac{y\Lambda}{F}R^{n-2-2p}(1-F)^p\delta R \nonumber \\
& +\frac{(n-2){\cal A}_{n-2}}{2\kappa_n^2}\sum_{p=1}^{[n/2]}{\tilde \alpha}_{(p)}pR^{n-1-2p}
\nonumber \\
&\times \biggl\{2\delta(y\Lambda)-2y\delta \Lambda-R'\delta \ln\biggl|\frac{R'+y\Lambda}{R'-y\Lambda}\biggl|\biggl\} \nonumber \\
&+\frac{(n-2){\cal A}_{n-2}}{2\kappa_n^2}\sum_{p=2}^{[n/2]}{\tilde \alpha}_{(p)}\biggl[pR^{n-1-2p}\sum_{w=1}^{p-1}\frac{(p-1)!(-1)^w}{w!(p-1-w)!} \nonumber \\
&\times \sum_{j=0}^{w-1}\frac{(w-1)!(-1)^{w-1-j}}{j!(w-1-j)!}y^{2(w-1-j)}\biggl(\frac{{R'}^2}{\Lambda^2}\biggl)^{j} \nonumber \\
& \times \biggl\{y\Lambda \biggl(\frac{\delta({R'}^2)}{\Lambda^2}-2y\delta y\biggl)-2y(y^2+F)\delta \Lambda\biggl\}\biggl]. \label{PMdeltaM1}
\end{align} 
An important fact is that $P_M\delta M$ has the form of $P_M\delta M=P_\Lambda\delta \Lambda+(\cdots)\delta R+\delta \eta+\zeta'$, where $\delta \eta$ and $\zeta'$ directly appear in the Liouville form, of Equation \ref{Liouville}.  We can see this by comparing Equations \ref{PMdeltaM1} (and the last term of Equation \ref{PMdeltaM1b}) to \ref{PLambda3} and remembering that $S$ is defined by $S:=R$.  This means that all of the terms with $\delta \Lambda$ in Equation \ref{PMdeltaM1} are contained in the expression for $P_\Lambda$.
Using the binomial expansion and integration by parts, we can calculate the other terms in Equation \ref{PMdeltaM1} as follows
\begin{align}
&-\frac{(n-2){\cal A}_{n-2}}{2\kappa_n^2}\sum_{p=0}^{[n/2]}{\tilde \alpha}_{(p)}(n-1-2p)\frac{y\Lambda}{F}R^{n-2-2p}(1-F)^p\delta R \nonumber \\
& +\frac{(n-2){\cal A}_{n-2}}{2\kappa_n^2}\sum_{p=1}^{[n/2]}{\tilde \alpha}_{(p)}pR^{n-1-2p}\biggl\{2\delta(y\Lambda)-R'\delta \ln\biggl|\frac{R'+y\Lambda}{R'-y\Lambda}\biggl|\biggl\} \nonumber \\
&+\frac{(n-2){\cal A}_{n-2}}{2\kappa_n^2}\sum_{p=2}^{[n/2]}{\tilde \alpha}_{(p)}\biggl[pR^{n-1-2p}\sum_{w=1}^{p-1}\frac{(p-1)!(-1)^w}{w!(p-1-w)!} \nonumber \\
&\times \sum_{j=0}^{w-1}\frac{(w-1)!(-1)^{w-1-j}}{j!(w-1-j)!}y^{2(w-1-j)}\biggl(\frac{{R'}^2}{\Lambda^2}\biggl)^{j}y\Lambda \biggl(\frac{\delta({R'}^2)}{\Lambda^2}-2y\delta y\biggl)\biggl] \nonumber  \\
=&-\frac{(n-2){\cal A}_{n-2}}{2\kappa_n^2}\sum_{p=0}^{[n/2]}{\tilde \alpha}_{(p)}(n-1-2p)\frac{y\Lambda}{F}R^{n-2-2p}(1-F)^p\delta R \nonumber \\
& +\frac{(n-2){\cal A}_{n-2}}{2\kappa_n^2}\sum_{p=1}^{[n/2]}{\tilde \alpha}_{(p)}\delta \biggl[pR^{n-1-2p}\biggl\{2y\Lambda-R' \ln\biggl|\frac{R'+y\Lambda}{R'-y\Lambda}\biggl|\biggl\}\biggl] \nonumber \\
& -\frac{(n-2){\cal A}_{n-2}}{2\kappa_n^2}\sum_{p=1}^{[n/2]}{\tilde \alpha}_{(p)}p\biggl[2(n-1-2p)R^{n-2-2p}y\Lambda\delta R \nonumber \\
&-\biggl((n-1-2p)R^{n-2-2p}R'\delta R+R^{n-1-2p}\delta (R')\biggl)\ln\biggl|\frac{R'+y\Lambda}{R'-y\Lambda}\biggl|\biggl] \nonumber \\
&+\frac{(n-2){\cal A}_{n-2}}{\kappa_n^2}\sum_{p=2}^{[n/2]}{\tilde \alpha}_{(p)}pR^{n-1-2p}\sum_{w=1}^{p-1}\frac{(p-1)!(-1)^w}{w!(p-1-w)!}F^{w-1}\frac{yR'}{\Lambda}\delta(R') \nonumber\\
&-\frac{(n-2){\cal A}_{n-2}}{\kappa_n^2}\sum_{p=2}^{[n/2]}{\tilde \alpha}_{(p)}\delta \biggl[pR^{n-1-2p}\Lambda \sum_{w=1}^{p-1}\frac{(p-1)!(-1)^w}{w!(p-1-w)!} \nonumber \\
&\times \sum_{j=0}^{w-1}\frac{(w-1)!(-1)^{w-1-j}}{j!(w-1-j)!}\biggl(\frac{{R'}^2}{\Lambda^2}\biggl)^{j}\frac{y^{2(w-j)+1}}{2(w-j)+1}\biggl] \nonumber \\
&+\frac{(n-2){\cal A}_{n-2}}{\kappa_n^2}\sum_{p=2}^{[n/2]}{\tilde \alpha}_{(p)}\delta\biggl[R^{n-1-2p}\Lambda \sum_{w=1}^{p-1}\frac{(p-1)!(-1)^w}{w!(p-1-w)!} \nonumber \\
&\times \sum_{j=0}^{w-1}\frac{py^{2(w-j)+1}}{2(w-j)+1}\frac{(w-1)!(-1)^{w-1-j}}{j!(w-1-j)!}\biggl(\frac{{R'}^2}{\Lambda^2}\biggl)^{j}\biggl].
\label{PMdeltaM1b}
\end{align} 
Because there will not appear any more total variation terms, we can read off the total variation term $\eta$ as Equation \ref{deta}.

In order to derive $\zeta$, we write down the quantity $\Pi:= P_M\delta M-P_\Lambda\delta\Lambda-\delta \eta$:
\begin{align}
\Pi=&-\frac{(n-2){\cal A}_{n-2}}{2\kappa_n^2}\sum_{p=0}^{[n/2]}{\tilde \alpha}_{(p)}(n-1-2p)\frac{y\Lambda}{F}R^{n-2-2p}(1-F)^p\delta R \nonumber \\
& -\frac{(n-2){\cal A}_{n-2}}{2\kappa_n^2}\nonumber \\
& \times\sum_{p=1}^{[n/2]}{\tilde \alpha}_{(p)}p(n-1-2p)R^{n-2-2p}\biggl\{2y\Lambda -R'\ln\biggl|\frac{R'+y\Lambda}{R'-y\Lambda}\biggl|\biggl\}\delta R  \nonumber \\
& +\frac{(n-2){\cal A}_{n-2}}{2\kappa_n^2}\sum_{p=1}^{[n/2]}{\tilde \alpha}_{(p)}\biggl[pR^{n-1-2p}\ln\biggl|\frac{R'+y\Lambda}{R'-y\Lambda}\biggl|\delta R\biggl]' \nonumber \\
& -\frac{(n-2){\cal A}_{n-2}}{2\kappa_n^2}\sum_{p=1}^{[n/2]}{\tilde \alpha}_{(p)}\biggl[pR^{n-1-2p}\ln\biggl|\frac{R'+y\Lambda}{R'-y\Lambda}\biggl|\biggl]'\delta R \nonumber \\
&+\frac{(n-2){\cal A}_{n-2}}{2\kappa_n^2}\sum_{p=2}^{[n/2]}{\tilde \alpha}_{(p)}\biggl[2pR^{n-1-2p}\sum_{w=1}^{p-1}\frac{(p-1)!(-1)^w}{w!(p-1-w)!}F^{w-1}\frac{yR'}{\Lambda}\delta R \biggl]' \nonumber\\
&-\frac{(n-2){\cal A}_{n-2}}{2\kappa_n^2}\sum_{p=2}^{[n/2]}{\tilde \alpha}_{(p)}\biggl[2pR^{n-1-2p}\sum_{w=1}^{p-1}\frac{(p-1)!(-1)^w}{w!(p-1-w)!}F^{w-1}\frac{yR'}{\Lambda}\biggl]'\delta R  \nonumber\\
&+\frac{(n-2){\cal A}_{n-2}}{\kappa_n^2}\sum_{p=2}^{[n/2]}{\tilde \alpha}_{(p)}p(n-1-2p)R^{n-2-2p}\Lambda \sum_{w=1}^{p-1}\frac{(p-1)!(-1)^w}{w!(p-1-w)!} \nonumber \\
&\times \sum_{j=0}^{w-1}\frac{(w-1)!(-1)^{w-1-j}}{j!(w-1-j)!}\biggl(\frac{{R'}^2}{\Lambda^2}\biggl)^{j}\frac{y^{2(w-j)+1}}{2(w-j)+1}\delta R \nonumber \\
&+\frac{(n-2){\cal A}_{n-2}}{\kappa_n^2}\sum_{p=2}^{[n/2]}{\tilde \alpha}_{(p)}pR^{n-1-2p} \sum_{w=1}^{p-1}\frac{(p-1)!(-1)^w}{w!(p-1-w)!} \nonumber \\
&\times \sum_{j=0}^{w-1}\frac{(w-1)!(-1)^{w-1-j}}{j!(w-1-j)!}\frac{2j{R'}^{2j-1}\Lambda^{1-2j}}{2(w-j)+1}y^{2(w-j)+1}\delta (R').
\end{align} 
The last term generates a total derivative term by integration by parts.  Now we see all the total derivative terms and can read off the total variation term $\zeta$ to be Equation \ref{zeta'}.

In order to prove the Liouville form of Equation \ref{Liouville}, we write down the quantity $\Xi\delta R:= P_M\delta M-P_\Lambda\delta\Lambda-\delta \eta-\zeta'$ as
\begin{align}
\Xi\delta R=&-\frac{(n-2){\cal A}_{n-2}}{2\kappa_n^2}\sum_{p=0}^{[n/2]}{\tilde \alpha}_{(p)}(n-1-2p)\frac{y\Lambda}{F}R^{n-2-2p}(1-F)^p\delta R \nonumber \\
& -\frac{(n-2){\cal A}_{n-2}}{2\kappa_n^2}\nonumber \\
&\times\sum_{p=1}^{[n/2]}{\tilde \alpha}_{(p)}pR^{n-2-2p}\biggl[2(n-1-2p)y\Lambda+R\biggl(\ln\biggl|\frac{R'+y\Lambda}{R'-y\Lambda}\biggl|\biggl)'\biggl] \delta R\nonumber \\
&+\frac{(n-2){\cal A}_{n-2}}{\kappa_n^2}\sum_{p=2}^{[n/2]}{\tilde \alpha}_{(p)}p(n-1-2p)R^{n-2-2p}\Lambda \sum_{w=1}^{p-1}\frac{(p-1)!(-1)^w}{w!(p-1-w)!} \nonumber \\
&\times \sum_{j=0}^{w-1}\frac{(w-1)!(-1)^{w-1-j}}{j!(w-1-j)!}\biggl(\frac{{R'}^2}{\Lambda^2}\biggl)^{j}\frac{y^{2(w-j)+1}}{2(w-j)+1}\delta R \nonumber \\
&-\frac{(n-2){\cal A}_{n-2}}{\kappa_n^2}\sum_{p=2}^{[n/2]}{\tilde \alpha}_{(p)}\biggl[pR^{n-1-2p}\sum_{w=1}^{p-1}\frac{(p-1)!(-1)^w}{w!(p-1-w)!} \nonumber \\
&\times \biggl\{F^{w-1}\frac{yR'}{\Lambda} + \sum_{j=0}^{w-1}\frac{(w-1)!(-1)^{w-1-j}}{j!(w-1-j)!}\frac{2j{R'}^{2j-1}\Lambda^{1-2j}}{2(w-j)+1}y^{2(w-j)+1}\biggl\}\biggl]'\delta R.
\end{align}

We now show that   
\begin{align}
\Xi=\frac{1}{R'}(\Lambda P_\Lambda'+P_MM'),
\end{align}
which is sufficient to verify the Liouville form, Equation \ref{Liouville}.  From Equation \ref{PLambda3}, we obtain
\begin{align}
\Lambda P_\Lambda'=&-\frac{(n-2){\cal A}_{n-2}}{\kappa_n^2}\sum_{p=1}^{[n/2]}{\tilde \alpha}_{(p)}p\Lambda \biggl\{(n-1-2p)R^{n-2-2p}yR'+R^{n-1-2p}y'\biggl\} \nonumber \\
&+\frac{(n-2){\cal A}_{n-2}}{2\kappa_n^2}\sum_{p=2}^{[n/2]}{\tilde \alpha}_{(p)}p(n-1-2p)R^{n-2-2p}R'\Lambda \nonumber \\
&\times  \biggl\{-2y(y^2+F)\sum_{w=1}^{p-1}\frac{(p-1)!(-1)^w}{w!(p-1-w)!}F^{w-1} \nonumber \\
&+\sum_{w=1}^{p-1}\frac{(p-1)!(-1)^w}{w!(p-1-w)!}\times \nonumber \\
&\sum_{j=0}^{w-1}\frac{(w-1)!(-1)^{w-1-j}}{j!(w-1-j)!}\frac{2(1-2j)}{2(w-j)+1}y^{2(w-j)+1}(y^2+F)^{j}\biggl\} \nonumber \\
&+\frac{(n-2){\cal A}_{n-2}}{2\kappa_n^2}\times \nonumber \\
&\sum_{p=2}^{[n/2]}{\tilde \alpha}_{(p)}pR^{n-1-2p}\Lambda \biggl\{-2y(y^2+F)\sum_{w=1}^{p-1}\frac{(p-1)!(-1)^w}{w!(p-1-w)!}F^{w-1} \nonumber \\
&+\sum_{w=1}^{p-1}\frac{(p-1)!(-1)^w}{w!(p-1-w)!}\times \nonumber \\
&\sum_{j=0}^{w-1}\frac{(w-1)!(-1)^{w-1-j}}{j!(w-1-j)!}\frac{2(1-2j)}{2(w-j)+1}y^{2(w-j)+1}(y^2+F)^{j}\biggl\}'.
\end{align}  
Using 
\begin{align}
R'\biggl\{\ln\biggl|\frac{R'+y\Lambda}{R'-y\Lambda}\biggl|\biggl\}'=&2(y\Lambda)'-2y\Lambda'-y\Lambda \frac{F'}{F}
\end{align}
for the logarithmic term, we finally obtain
\begin{align}
\Xi\delta R=&-\frac{(n-2){\cal A}_{n-2}}{2\kappa_n^2}\sum_{p=0}^{[n/2]}{\tilde \alpha}_{(p)}(n-1-2p)\frac{y\Lambda}{F}R^{n-2-2p}(1-F)^p\delta R \nonumber \\
& -\frac{(n-2){\cal A}_{n-2}}{2\kappa_n^2} \sum_{p=1}^{[n/2]}{\tilde \alpha}_{(p)}\biggl[2p(n-1-2p)R^{n-2-2p}y\Lambda\nonumber \\
&+pR^{n-1-2p}\frac{1}{R'}\biggl(2y'\Lambda-y\Lambda \frac{F'}{F}\biggl)\biggl] \delta R\nonumber \\
&+\frac{(n-2){\cal A}_{n-2}}{\kappa_n^2}\sum_{p=2}^{[n/2]}{\tilde \alpha}_{(p)}p(n-1-2p)R^{n-2-2p}\Lambda \sum_{w=1}^{p-1}\frac{(p-1)!(-1)^w}{w!(p-1-w)!} \nonumber \\
&\times \sum_{j=0}^{w-1}\frac{(w-1)!(-1)^{w-1-j}}{j!(w-1-j)!}\biggl(\frac{{R'}^2}{\Lambda^2}\biggl)^{j}\frac{y^{2(w-j)+1}}{2(w-j)+1}\delta R \nonumber \\
&-\frac{(n-2){\cal A}_{n-2}}{\kappa_n^2}\sum_{p=2}^{[n/2]}{\tilde \alpha}_{(p)}\biggl[pR^{n-1-2p}\sum_{w=1}^{p-1}\frac{(p-1)!(-1)^w}{w!(p-1-w)!} \nonumber \\
&\times \biggl\{F^{w-1}\frac{yR'}{\Lambda} \nonumber \\
&+\sum_{j=0}^{w-1}\frac{(w-1)!(-1)^{w-1-j}}{j!(w-1-j)!}\frac{2j{R'}^{2j-1}\Lambda^{1-2j}}{2(w-j)+1}y^{2(w-j)+1}\biggl\}\biggl]'\delta R.
\end{align}
On the other hand, using Equations \ref{qlm-L} and \ref{eq:PM}, we obtain
\begin{align}
P_MM' =&-\frac{(n-2){\cal A}_{n-2}}{2\kappa_n^2}\sum_{p=0}^{[n/2]}{\tilde \alpha}_{(p)}(n-1-2p)R^{n-2-2p}(1-F)^pR'\frac{y\Lambda}{F} \nonumber \\
&+\frac{(n-2){\cal A}_{n-2}}{2\kappa_n^2}\sum_{p=1}^{[n/2]}{\tilde \alpha}_{(p)}pR^{n-1-2p}(1-F)^{p-1}F'\frac{y\Lambda}{F}.
\end{align}  
Some useful cancellations allow us to derive
\begin{align}
\Xi-&\frac{\Lambda P_\Lambda+P_MM'}{R'}=\nonumber \\
& -\frac{(n-2){\cal A}_{n-2}}{2\kappa_n^2}\sum_{p=2}^{[n/2]}{\tilde \alpha}_{(p)}pR^{n-1-2p}\biggl[\frac{F'}{R'}\frac{y\Lambda}{F}\biggl\{(1-F)^{p-1}-1\biggl\} \nonumber \\
&+\sum_{w=1}^{p-1}\frac{(p-1)!(-1)^w}{w!(p-1-w)!} \biggl\{2yF^{w-1}(y^2+F)\biggl(\frac{\Lambda}{R'}\biggl)'+2F^{w-1}y^2y'\frac{\Lambda}{R'}\biggl\}  \nonumber \\
&+\sum_{w=1}^{p-1}\frac{(p-1)!(-1)^w}{w!(p-1-w)!} \sum_{j=0}^{w-1}\frac{(w-1)!(-1)^{w-1-j}}{j!(w-1-j)!}\frac{2jy^{2(w-j)+1}(y^2+F)^{j-1}}{2(w-j)+1}\nonumber \\
&\times \biggl\{2(y^2+F)\biggl(\frac{\Lambda}{R'}\biggl)' +(2yy'+F')\frac{\Lambda}{R'} \biggl\}\biggl].
\end{align}
Expanding the first term, we finally obtain
\begin{align}
&\Xi-\frac{\Lambda P_\Lambda+P_MM'}{R'} \nonumber \\
=& -\frac{(n-2){\cal A}_{n-2}}{2\kappa_n^2}\biggl\{2(y^2+F)\biggl(\frac{\Lambda}{R'}\biggl)' +(2yy'+F')\frac{\Lambda}{R'} \biggl\}\sum_{p=2}^{[n/2]}{\tilde \alpha}_{(p)}pR^{n-1-2p}\nonumber \\
&\times \sum_{w=1}^{p-1}\frac{(p-1)!(-1)^w}{w!(p-1-w)!}\nonumber \\
&\times\biggl[F^{w-1}y+\sum_{j=0}^{w-1}\frac{(w-1)!(-1)^{w-1-j}}{j!(w-1-j)!}\frac{2jy^{2(w-j)+1}(y^2+F)^{j-1}}{2(w-j)+1}\biggl].
\end{align}
By direct calculations, we can show
\begin{align}
2(y^2+F)\biggl(\frac{\Lambda}{R'}\biggl)' +(2yy'+F')\frac{\Lambda}{R'}  =0
\end{align}  
and complete the proof.

\subsection{Equation~(\ref{eq:GRfinal}) in Lovelock Gravity}
\label{appendix3}
In this appendix, we derive Equation \ref{eq:GRfinal} in Lovelock gravity.  { While we consider the spherically symmetric case $(k=1$) in the main text, we derive the equations for general $k$ in this appendix.}

Let us start from the action in the form of Equation \ref{I_M}:
\begin{align}
I_M=& \frac{(n-2)V_{n-2}^{(k)}}{2\kappa_n^2}\sum^{[n/2]}_{p=0}\int d^2{\bar y}{\tilde \alpha}_{(p)}\Biggl[2pk^{p-1}R^{n-2p-1}y(N_r\Lambda)' \nonumber \\
&-\frac{2pk^{p-1}N}{n-2p}\biggl((R^{n-2p})''\Lambda^{-1}+(R^{n-2p})'(\Lambda^{-1})'\biggl)  \nonumber \\
&+ pR^{n-2p-1}\biggl\{k^{p-1}-(k-F)^{p-1}\biggl\}\biggl\{(\Lambda N_r y+\Lambda^{-1}NR')\frac{F'}{F}-\Lambda y\frac{\dot F}{F}\biggl\}   \nonumber \\
& +(n-2p-1)\biggl\{\left(k-F\right)^{p} +2pk^{p-1}F\biggl\}N\Lambda R^{n-2-2p}\nonumber \\
&-2pk^{p-1}R^{n-2p-1}y {\dot \Lambda}\biggl].
\end{align}
Using  
\begin{align}
(\Lambda^{-1})'=\frac{\Lambda(F'+2yy')}{2{R'}^2}-\frac{R''}{R'\Lambda}
\end{align}
for the second line and integration by parts for the first line, we obtain
\begin{align}
I_M=& \frac{(n-2)V_{n-2}^{(k)}}{2\kappa_n^2}\sum^{[n/2]}_{p=0}\int d^2{\bar y}{\tilde \alpha}_{(p)}\Biggl[-2pk^{p-1}\Lambda\biggl(N_r+\frac{Ny}{R'}\biggl)(R^{n-2p-1}y)' \nonumber \\
&+ pR^{n-2p-1}\biggl\{k^{p-1}-(k-F)^{p-1}\biggl\}\biggl\{\biggl(N_r+\frac{Ny}{R'}\biggl)\frac{\Lambda y}{F}F'-\Lambda y\frac{\dot F}{F}\biggl\}  \nonumber \\
& - pR^{n-2p-1}\frac{N\Lambda}{R'}(k-F)^{p-1}F' +(n-2p-1)\left(k-F\right)^{p}N\Lambda R^{n-2-2p} \nonumber \\
&-2pk^{p-1}R^{n-2p-1}y {\dot \Lambda}\biggl]+\mbox{(t.d.)},
\end{align}
where we also used ${R'}^2\Lambda^{-1}=F\Lambda+y^2\Lambda$.
Using 
\begin{align}
\frac{N\Lambda}{R'}M' =&\frac{(n-2)V_{n-2}^{(k)}}{2\kappa_n^2}\sum_{p=0}^{[n/2]}{\tilde\alpha}_{(p)}\frac{N\Lambda}{R'}R^{n-1-2p}\nonumber \\
&\times\biggl[-p(k-F)^{p-1}F'+(n-1-2p)(k-F)^p\frac{R'}{R}\biggl],\\
P_M{\dot M} =&\frac{(n-2)V_{n-2}^{(k)}}{2\kappa_n^2}\sum_{p=0}^{[n/2]}{\tilde \alpha}_{(p)}\frac{y\Lambda}{F}R^{n-1-2p}\nonumber \\
&\times\biggl[p(k-F)^{p-1}{\dot F}-(n-1-2p)(k-F)^p\frac{\dot R}{R}\biggl]
\end{align}  
which can be obtained from Equations \ref{qlm-L} and \ref{eq:PM}, we obtain
\begin{align}
&I_M-\int d^2{\bar y}\biggl(P_M{\dot M}+\frac{N\Lambda}{R'}M'\biggl) \nonumber \\
=& \frac{(n-2)V_{n-2}^{(k)}}{2\kappa_n^2}\sum^{[n/2]}_{p=0}\int d^2{\bar y}{\tilde \alpha}_{(p)}\Biggl[-2pk^{p-1}\Lambda\biggl(N_r+\frac{Ny}{R'}\biggl)(R^{n-2p-1}y)' \nonumber \\
&+ pR^{n-2p-1}\biggl\{k^{p-1}-(k-F)^{p-1}\biggl\}\biggl(N_r+\frac{Ny}{R'}\biggl)\frac{\Lambda y}{F}F'-2pk^{p-1}R^{n-2p-1}y {\dot \Lambda}  \nonumber \\
&-pk^{p-1}R^{n-2p-1}\Lambda y\frac{\dot F}{F} +\frac{y\Lambda}{F}(k-F)^p\partial_t(R^{n-1-2p})\biggl]+\mbox{(t.d.)}.
\end{align}
Using Equation \ref{useful1} for $y\Lambda{\dot F}/F$ together with integration by parts and ${\dot R}/R'=Ny/R'+N_r$, we rewrite the above expression as
\begin{align}
&I_M-\int d^2{\bar y}\biggl(P_M{\dot M}+\frac{N\Lambda}{R'}M'\biggl) \nonumber \\
=& \frac{(n-2)V_{n-2}^{(k)}}{2\kappa_n^2}\sum^{[n/2]}_{p=0}\int d^2{\bar y}{\tilde \alpha}_{(p)}\Biggl[-2pk^{p-1}\Lambda \frac{{\dot R}}{R'}(R^{n-2p-1}y)' \nonumber \\
&+ pR^{n-2p-1}\biggl\{k^{p-1}-(k-F)^{p-1}\biggl\}\frac{{\dot R}}{R'}\frac{\Lambda y}{F}F' +\frac{y\Lambda}{F}(k-F)^p\partial_t(R^{n-1-2p}) \nonumber \\
&+pk^{p-1}\biggl\{2y\Lambda\partial_t(R^{n-2p-1})+\frac{1}{n-2p}\partial_t(R^{n-2p})\partial_x\biggl(\ln\biggl|\frac{R'+y\Lambda}{R'-y\Lambda}\biggl|\biggl)\biggl\}\biggl]+\mbox{(t.d.)}.
\end{align}
Replacing the logarithmic term by 
\begin{align}
\partial_x\biggl(\ln\biggl|\frac{R'+y\Lambda}{R'-y\Lambda}\biggl|\biggl)=&\frac{-2R''y\Lambda+2(y\Lambda)'R'}{F\Lambda^2} \nonumber \\
=&\frac{y\Lambda}{FR'}\biggl(-F'+\frac{2y'}{y}F\biggl),
\end{align}
we finally obtain
\begin{align}
I_M-&\int d^2{\bar y}\biggl(P_M{\dot M}+\frac{N\Lambda}{R'}M'\biggl) \nonumber \\
=& \frac{(n-2)V_{n-2}^{(k)}}{2\kappa_n^2}\sum^{[n/2]}_{p=0}\int d^2{\bar y}{\tilde \alpha}_{(p)}\frac{{\dot R}\Lambda yR^{n-2-2p}(k-F)^{p-1}}{R'F} \nonumber \\
&\times \Biggl[(n-2p-1)(k-F)R'- pRF'\biggl]+\mbox{(t.d.)}\nonumber \\
=& \int d^2{\bar y}\frac{{\dot R}\Lambda y}{R'F}M'+\mbox{(t.d.)}
\end{align}
which completes the derivation.

\section[Boundary Condition at Spacelike Infinity]{\texorpdfstring{Boundary Condition at \\ Spacelike Infinity}{Boundary Condition at Spacelike Infinity}}
\label{app:boundary}
In this appendix, we confirm that under the transformation from $\{\Lambda,P_\Lambda;R, P_R\}$ to $\{M, P_M;S, P_S\}$ the boundary term and total variation in Equation \ref{Liouville} are finite with suitable asymptotic fall-off rates of the ADM variables consistent with asymptotic flatness.  We also consider the transformation to slicings which approach the PG gauge at spatial infinity, $\{\Lambda,P_\Lambda;R, P_R\}$ to $\{M, \tilde{P}_M;S, \tilde{P}_S\}$ and show that it is also valid canonical transformation.

We consider the following behaviour of the ADM variables near spacelike infinity:
\begin{align}
&N\simeq N_\infty(t)+\mathcal{O}(x^{-\epsilon_1}),\\
&N_r\simeq N_r^{\infty}(t) x^{-(n-3)/2-\epsilon_2},\\
&\Lambda \simeq 1+\Lambda_1(t) x^{-(n-3)-\epsilon_3},\\
&R\simeq x+R_1(t) x^{-(n-4)-\epsilon_4},
\end{align}
where $\epsilon_1$ is positive, $\epsilon_2$, $\epsilon_3$ and $\epsilon_4$ are non-negative numbers, and we require the following three conditions:\\
 (I) the canonical transformation from $\{\Lambda,P_\Lambda;R, P_R\}$ to $\{M, P_M;S, P_S\}$ is well-defined, \\
 (II) the Hamiltonian is finite in terms both of $\{\Lambda,P_\Lambda;R, P_R\}$ and $\{M, P_M;S, P_S\}$, and \\
 (III) the Misner-Sharp mass is non-zero and finite at spacelike infinity $M\simeq M^\infty(t)$.
 
We will see below that these requirements are fulfilled for $\epsilon_1>0$, $\epsilon_2>\max[0,-(n-5)/2]$, $\epsilon_3= 0$, and $\epsilon_4>\max[0,-(n-5)]$.  Note that to approach PG slicings we must have $\epsilon_2=0$ and $\epsilon_3 \to \infty$ (or $\Lambda=1$ everywhere) for all $n$.  This does not greatly alter the analysis below and will be addressed when appropriate.

First we look at the following integrated Liouville form, Equation \ref{Liouville}:
\begin{align}
\int_{-\infty}^{\infty} dx(P_\Lambda\delta \Lambda+P_R\delta R)-\int_{-\infty}^{\infty} dx(P_M\delta M+P_S\delta S)=\delta\int_{-\infty}^{\infty}\eta dx +[~\zeta~]^{x=\infty}_{x=-\infty},\label{intLiouv}
\end{align} 
where $P_\Lambda$, $P_R$, $S$, $P_S$, $M$, $P_M$, $\eta$, and $\zeta$ are defined by Equations \ref{PLambdaGR2}, \ref{PR}, \ref{eq:R}, \ref{eq:PM}, \ref{MS}, \ref{eq:Pr1}, \ref{deta-gr}, and \ref{zeta'-gr}, respectively.  For the well-definedness of the canonical transformation, two conditions must hold at spacelike infinity; (i) $\zeta$ vanishes, and (ii) the integrands in the left-hand side and $\eta$ converge to zero faster than $\mathcal{O}(x^{-1})$.  The second requirement ensures the finiteness of the integrals.  For the case of slicings which approach the PG gauge Equation \ref{intLiouv} becomes

\begin{align}
&\int_{-\infty}^{\infty} dx(P_\Lambda\delta \Lambda+P_R\delta R)-\int_{-\infty}^{\infty} dx((P_M+\Delta P_M)\delta M+(P_S + \Delta P_S )\delta S) \nonumber \\
&=\delta\int_{-\infty}^{\infty}(\eta + \Delta \eta) dx +[~\zeta + \Delta \zeta ~]^{x=\infty}_{x=-\infty}, \label{intLiouvPG}
\end{align} 
where we define $\eta_{PG}:=\Delta\eta + \eta$ and $\zeta_{PG}:=\Delta\zeta + \zeta$ as the total derivative terms appropriate for slicings which approach PG gauge at spacelike infinity and $\Delta P_M$ and $\Delta P_S$ are defined as $\Delta P_M := \tilde{P}_M-P_M$ and $\Delta P_M := \tilde{P}_M-P_M$.  We know from Equation \ref{eq:PM PG} that $\Delta P_M=S^\prime \sqrt{1-F}/F$.  In order to preserve the form of the diffeomorphism constraint, $H_r$ we must insist that $\Delta P_M M^\prime + \Delta P_S S^\prime = 0$ which gives $\Delta P_S = -M^\prime \sqrt{1-F}/F$.   Since $\Delta P_M$ and $\Delta P_S$ are not zero it will be important to consider them (in any equations containing $P_M$ and $P_S$), as well as $\Delta \eta$ and $\Delta \zeta$ when we consider PG boundary conditions.

It will be important later to notice that we can use the fact that Equation \ref{M-F} tells us that $M=M(F,S)$ to write

\begin{align}
&\int_{-\infty}^{\infty} dx(\Delta P_M\delta M+\Delta P_S\delta S) =\delta\int_{-\infty}^{\infty}\Delta \eta~ dx +[~\Delta \zeta ~]^{x=\infty}_{x=-\infty} \nonumber \\
&=\delta\int_{-\infty}^{\infty} X^\prime(S)Y(F) ~ dx -[~ \dot{X}(S)Y(F) ~]^{x=\infty}_{x=-\infty}, \label{intLiouvPG2}
\end{align} 
where

\be
X(S):=-\frac{(n-2)V_{n-2}^{(k)}}{2\kappa_n^2}\frac{S^{n-2}}{n-2}
\label{eq:XS}
\ee
and
\be
Y(F):= \int{dF\frac{\sqrt{1-F}}{F}}.
\label{eq:YF}
\ee

Using the PG boundary conditions one can easily show that $\Delta \eta = X^{\prime}Y$ diverges and $\Delta \zeta = \dot{X}Y$ goes to zero at spacelike infinity.  We will see later that the $\Delta \eta$ will cancel other diverging terms when the PG boundary conditions are used.

Let us now look at the asymptotic behaviour of the momentum conjugates for the case of slices which approach the Schwarzschild form at spacelike infinity.  Near spacelike infinity, $P_\Lambda$ and $P_R$ behave as
\begin{align}
P_\Lambda\simeq& -\frac{(n-2)\ma A_{n-2}}{\kappa_n^2}N_\infty^{-1}\biggl({\dot R_1} x^{1-\epsilon_4}-N_r^{\infty} x^{(n-3)/2-\epsilon_2}\biggl), \\
P_R\simeq & -\frac{(n-2)\ma A_{n-2} }{\kappa_n^2} N_\infty^{-1}\nonumber \\
&\times\biggl[N_r^{\infty}(t) \biggl(-\frac{n-3}{2}+\epsilon_2\biggl)x^{(n-5)/2-\epsilon_2}+{\dot \Lambda}_1(t) x^{-\epsilon_3}+(n-3){\dot R_1} x^{-\epsilon_4}\biggl],
\end{align}
with which we obtain
\begin{align}
P_\Lambda\delta\Lambda\simeq& \mathcal{O}(x^{-(n-4)-\epsilon_3-\epsilon_4})+\mathcal{O}(x^{-(n-3)/2-\epsilon_2-\epsilon_3}), \\
P_R\delta R\simeq & \mathcal{O}(x^{-(n-3)/2-\epsilon_2-\epsilon_4})+\mathcal{O}(x^{-(n-4)-\epsilon_3-\epsilon_4})+\mathcal{O}(x^{-(n-4)-2\epsilon_4}).
\end{align}
Hence, $\epsilon_4>0$ and $\epsilon_2+\epsilon_3>0$ are required for $n=5$.  For $n=4$, the requirement is $\epsilon_4>1/2$, $\epsilon_3+\epsilon_4>1$, $\epsilon_2+\epsilon_4>1/2$, and $\epsilon_2+\epsilon_3>1/2$.

For the next check, we see the behaviour of $F$, defined by Equation \ref{defF}.  Using 
\begin{align}
y \simeq&N_\infty^{-1}({\dot R}_1x^{-(n-4)-\epsilon_4}-N_r^{\infty} x^{-(n-3)/2-\epsilon_2}),\label{eval-y}
\end{align}
where $y$ is defined by Equation \ref{defy}, we obtain
\begin{align}
F\simeq &N_\infty^{-2}{\dot R}_1x^{-(n-4)}(-{\dot R}_1x^{-(n-4)-2\epsilon_4}+2N_r^{\infty}x^{-(n-3)/2-\epsilon_2-\epsilon_4}) \nonumber \\
&+1-2\Lambda_1 x^{-(n-3)-\epsilon_3}-N_\infty^{-2}{N_r^{\infty}}^2x^{-(n-3)-2\epsilon_2}\nonumber \\
&-2(n-4+\epsilon_4)R_1x^{-(n-3)-\epsilon_4}, \label{asympF}
\end{align} 
which converges to $1$.  To satisfy condition (III) $F$ must behave near infinity as 
\begin{align}
F\simeq 1-F_1(t)x^{-(n-3)}
\end{align} 
and then, for $M\simeq M^\infty(t)$,  $F_1$ is identified as
\begin{align}
F_1(t)\equiv \frac{2\kappa_n^2M^\infty(t)}{(n-2)\ma A_{n-2}}.
\end{align} 
Since we have already required $(n-4)+2\epsilon_4>1$ and $(n-3)/2+\epsilon_2+\epsilon_4>1$ in the previous argument, the first line in Equation \ref{asympF} converges to zero faster than $\mathcal{O}(x^{-(n-3)})$.  Therefore the condition (III) requires $\epsilon_2\epsilon_3\epsilon_4=0$, where $F_1(t)$ is determined depending on the cases.  It will be important for later to note that for PG slicings, where $\epsilon_2=0$, $F_1(t)$ is given by

\be
F_1(t)=-N_\infty^{-2}N_r^{\infty 2}
\label{eq:F1PG}
\ee

Using Equation \ref{eval-y} and $M\simeq M^\infty(t)$, we obtain the asymptotic behaviour of $P_M\delta M$ as
\begin{align}
P_M\delta M\simeq -N_\infty^{-1}\delta M^\infty({\dot R}_1x^{-(n-4)-\epsilon_4}-N_r^{\infty} x^{-(n-3)/2-\epsilon_2}).
\label{eq:PM2}
\end{align}
This provides additional requirements; $\epsilon_2>0$ for $n=5$ and $\epsilon_4>1$ and $\epsilon_2>1/2$ for $n=4$.
{ Note however that if one uses the momentum variable $\tilde{P}_M$ appropriate for asymptotically PG slices, then there is an extra term on the right hand side of Equation \ref{eq:PM2}, $\Delta P_M \delta M$, that cancels the second term. This in turn allows the choice $\epsilon_2=0$ for all spacetime dimensions as required by the PG slicing.}

The conditions for $\epsilon_2$ and $\epsilon_4$ obtained up to here are summarized as $\epsilon_2>\max[0,-(n-5)/2]$ and $\epsilon_4>\max[0,-(n-5)]$ for slicings which approach the Schwarschild slicings at spatial infinity.
We will see below that the requirements (I) and (II) are fulfilled under these conditions.

Using the following asymptotic expansion of $P_S$:
\begin{align}
P_S\simeq& -\frac{(n-2)\ma A_{n-2} }{\kappa_n^2} N_\infty^{-1}\biggl({\dot \Lambda}_1 x^{-\epsilon_3}+(n-4+\epsilon_4){\dot R_1} x^{-\epsilon_4}\biggl)\nonumber \\
&+\mathcal{O}(x^{-(n-1)/2-\epsilon_2-\epsilon_M}), \label{asympPS} 
\end{align}
where $\epsilon_M$ is some positive number defined by the next leading-order of $M$ as $M\simeq M^\infty+\mathcal{O}(x^{-\epsilon_M})$, we obtain
\begin{align}
P_S\delta S \simeq & -\delta R_1\frac{(n-2)\ma A_{n-2} }{\kappa_n^2} N_\infty^{-1}\nonumber \\
&\times\biggl({\dot \Lambda}_1 x^{-(n-4)-\epsilon_3-\epsilon_4}+(n-4+\epsilon_4){\dot R_1} x^{-(n-4)-2\epsilon_4}\biggl) \nonumber \\
&+\mathcal{O}(x^{-3(n-3)/2-\epsilon_2-\epsilon_4-\epsilon_M}).
\label{eq:eta2}
\end{align}
Under the present conditions this converges to zero faster than $\mathcal{O}(x^{-1})$.

Next let us evaluate $\zeta$ and $\eta$.  We write the logarithmic term in Equations \ref{deta-gr} and \ref{zeta'-gr} as
\begin{align}
\ln\biggl(\frac{R'+y\Lambda}{R'-y\Lambda}\biggl)=&\ln\biggl(\frac{1-W}{1+W}\biggl),
\end{align} 
where
\begin{align} 
W:=&\frac{\kappa_n^2\Lambda P_\Lambda}{(n-2)\ma A_{n-2}R^{n-3}R' }.
\end{align} 
Using the fact that $W$ converges to zero as
\begin{align}
W\simeq -N_\infty^{-1}\biggl({\dot R_1} x^{-(n-4)-\epsilon_4}-N_r^{\infty} x^{-(n-3)/2-\epsilon_2}\biggl),
\end{align}
we can evaluate the logarithmic term as
\begin{align}
\ln\biggl(\frac{R'+y\Lambda}{R'-y\Lambda}\biggl)\simeq &-2W-\frac23 W^3.
\end{align} 
Using this, we can show that $\zeta$ converges to zero under the present conditions as
\begin{align}
\zeta\simeq &\mathcal{O}(x^{-(n-5)-2\epsilon_4})+\mathcal{O}(x^{-(n-5)/2-\epsilon_2-\epsilon_4}).
\end{align} 
We also see that $\eta$ is evaluated as
\begin{align}
\eta=&\frac{(n-2)\ma A_{n-2}R^{n-3}R' }{2\kappa_n^2}\biggl[2W+\ln\biggl(\frac{1-W}{1+W}\biggl)\biggl] \nonumber \\
\simeq &-\frac{(n-2)\ma A_{n-2}x^{n-3}}{3\kappa_n^2}W^3 \nonumber \\
\simeq &\frac{-(n-2)A_{n-2}}{3\kappa_n^2}\left(\frac{\dot{R}_1}{N_\infty}\right)^3x^{-2(n-4)+1-3\epsilon_4} \nonumber \\
&+ \frac{(n-2)A_{n-2}}{3\kappa_n^2}\left(\frac{N_r^\infty}{N_\infty}\right)^3x^{-(n-3)/2-3\epsilon_2}
\end{align} 
and converges to zero faster than $\mathcal{O}(x^{-1})$ under the present conditions.  Notice that for the PG case, where $\epsilon_2=0$, the second term does not go to zero fast enough for $n \le 5$.  Applying the PG boundary conditions to Equations \ref{eq:XS}, \ref{eq:YF} and \ref{eq:F1PG} we see that this term is cancelled by $\Delta \eta$  

Lastly, let us check the well-definedness of the Hamiltonian $\int^\infty_{-\infty}dx(NH+N_rH_r)$, where $H_r$ and $H$ are defined by Equations \ref{Hr} and \ref{H}, respectively.  It requires that the fall-off rates of $NH$ and $N_rH_r$ are faster than $\mathcal{O}(x^{-1})$.
From the following asymptotic expressions;
\begin{align}
\Lambda P_\Lambda'\simeq& -\frac{(n-2)\ma A_{n-2}}{\kappa_n^2}N_\infty^{-1}\biggl(1+\Lambda_1 x^{-(n-3)-\epsilon_3}\biggl) \nonumber \\
&\times \biggl[(1-\epsilon_4){\dot R_1} x^{-\epsilon_4}-\biggl(\frac{n-3}{2}-\epsilon_2\biggl)N_r^{\infty} x^{(n-5)/2-\epsilon_2}\biggl], \\
R'P_R\simeq & -\frac{(n-2)\ma A_{n-2} }{\kappa_n^2} N_\infty^{-1}\biggl(1-(n-4+\epsilon_4)x^{-(n-3)-\epsilon_4}\biggl) \nonumber \\
&\times \biggl[N_r^{\infty} \biggl(-\frac{n-3}{2}+\epsilon_2\biggl)x^{(n-5)/2-\epsilon_2}+{\dot \Lambda}_1 x^{-\epsilon_3}+(n-3){\dot R_1} x^{-\epsilon_4}\biggl], 
\end{align}
we see that the dangerous terms of the order $x^{(n-5)/2-\epsilon_2}$ in $H_r=-\Lambda P_\Lambda'+R'P_R$ are cancelled out and obtain
\begin{align}
N_rH_r\simeq&  \mathcal{O}(x^{-(n-3)/2-\epsilon_2-\epsilon_3})+\mathcal{O}(x^{-(n-3)/2-\epsilon_2-\epsilon_4}).
\end{align}
This is faster than $\mathcal{O}(x^{-1})$ under the present conditions.  Next let us see the behaviour of $NH$.  Using the followings asymptotic expansions:
\begin{align}
\Lambda P_\Lambda \simeq& -\frac{(n-2)\ma A_{n-2}}{\kappa_n^2}N_\infty^{-1}\biggl(1+\Lambda_1x^{-(n-3)-\epsilon_3}\biggl)\biggl({\dot R_1} x^{1-\epsilon_4}-N_r^{\infty} x^{(n-3)/2-\epsilon_2}\biggl), \\
RP_R\simeq & -\frac{(n-2)\ma A_{n-2} }{\kappa_n^2} N_\infty^{-1}\biggl(x+R_1x^{-(n-4)-\epsilon_4}\biggl) \nonumber \\
&\times \biggl[N_r^{\infty}(t) \biggl(-\frac{n-3}{2}+\epsilon_2\biggl)x^{(n-5)/2-\epsilon_2}+{\dot \Lambda}_1(t) x^{-\epsilon_3}+(n-3){\dot R_1} x^{-\epsilon_4}\biggl], 
\end{align}
we obtain 
\begin{align}
\frac{P_{\Lambda}}{R^{n-2}}\biggl(RP_{R}-\frac{n-3}{2}\Lambda P_{\Lambda}\biggl) \simeq &\mathcal{O}(x^{-(n-4)-\epsilon_3-\epsilon_4})+ \mathcal{O}(x^{-(n-3)/2-\epsilon_2-\epsilon_4})\nonumber \\
&+\mathcal{O}(x^{-(n-3)/2-\epsilon_2-\epsilon_3}) + 
\mathcal{O}(x^{-(n-4)-2\epsilon_4}).
\end{align} 
This shows that the first line of the following super-Hamiltonian;
\begin{align}
H=&-\frac{\kappa_n^2P_{\Lambda}}{(n-2)\ma A_{n-2}R^{n-2}}\biggl(RP_{R}-\frac{n-3}{2}\Lambda P_{\Lambda}\biggl) \nonumber \\
&-\frac{(n-2)\ma A_{n-2}}{\kappa_n^2}\biggl\{-R^{n-3}(R' \Lambda^{-1})'+\frac{n-3}{2} \Lambda R^{n-4}(1-\Lambda^{-2}{R'}^2)\biggl\}
\end{align} 
converges to zero faster than $\mathcal{O}(x^{-1})$ under the present conditions.  On the other hand, the second line is evaluated as
\begin{align}
-R^{n-3}(R' \Lambda^{-1})'+\frac{n-3}{2} \Lambda R^{n-4}(1-\Lambda^{-2}{R'}^2) \simeq \epsilon_3\mathcal{O}(x^{-1-\epsilon_3})+\epsilon_4\mathcal{O}(x^{-1-\epsilon_4}),
\end{align} 
which also converges faster than $\mathcal{O}(x^{-1})$.  As a result, $NH$ converge to zero faster than $\mathcal{O}(x^{-1})$ and hence the Hamiltonian is well-defined.  In the PG case, where $\Lambda=1$ everywhere this still converges to zero faster than $\mathcal{O}(x^{-1})$.

We also check the well-definedness of the Hamiltonian with a new set of variables: $\int^\infty_{-\infty}dx(N^M M'+N^S P_S)$, where $N^M$ and $N^S$ are defined by Equations \ref{NM} and \ref{NS}, respectively.  Using Equation \ref{eval-y}, we obtain $N^M\simeq N_\infty$.  Combining this with $M'\simeq \mathcal{O}(x^{-1-\epsilon_M})$, it is shown that $N^MM'$ converges to zero faster than $\mathcal{O}(x^{-1})$ under the present conditions.  On the other hand, using $N^S\simeq \mathcal{O}(x^{-(n-4)-\epsilon_4})$ and Equation \ref{asympPS}, we obtain
\begin{align}
N^SP_S\simeq& \mathcal{O}(x^{-(n-4)-\epsilon_3-\epsilon_4})+\mathcal{O}(x^{-(n-4)-2\epsilon_4})+\mathcal{O}(x^{-3(n-3)/2-\epsilon_2-\epsilon_4-\epsilon_M}).
\end{align}
This also converges to zero faster than $\mathcal{O}(x^{-1})$ under the present conditions.

\end{appendix}

\bibliography{library}
\bibliographystyle{/home/tim/unsrturl} 

\end{document}